\documentclass[
 reprint,
 superscriptaddress,
 amsmath,amssymb,
 aps,
 prb,twocolumn,
 showkeys
]{revtex4-2}
\usepackage{graphicx}
\usepackage{siunitx}
\usepackage{bm}
\usepackage{graphicx}
\usepackage{textcomp,gensymb}
\usepackage{multirow}
\usepackage[export]{adjustbox}
\usepackage{stackengine}
\usepackage{natbib}
\usepackage{mathtools}
\usepackage{mathrsfs}
\usepackage{float}
\usepackage{placeins}
\usepackage{glossaries}
\let\g\gls

\usepackage{hyperref}
\hypersetup{
	colorlinks=true,
	linkcolor=blue,
	urlcolor=blue,
	citecolor=blue
}
\newacronym[plural=magnetoresistances]{mr}{MR}{magnetoresistance}
\newacronym{nlmr}{NLMR}{negative longitudinal magnetoresistance}
\newacronym{omr}{OMR}{orbital magnetoresistance}
\newacronym[plural=transition-metal dichalcogenides]{tmd}{TMD}{transition-metal dichalcogenide}
\newacronym[plural=Weyl points]{wp}{WP}{Weyl point}
\newacronym[plural=Dirac points]{dp}{DP}{Dirac point}
\newacronym{vdw}{vdW}{van-der-Waals}
\newacronym{cme}{CME}{chiral magnetic effect}
\newacronym{trs}{TRS}{time-reversal symmetry}
\newacronym{is}{IS}{inversion symmetry}
\newacronym{xps}{XPS}{x-ray photoemission spectroscopy}
\newacronym{pdms}{PDMS}{Polydimethylsiloxane}
\newacronym{ipa}{IPA}{isopropyl alcohol}
\newacronym{rt}{\textit{R-T}}{resistance-over-temperature-curve}
\newacronym[plural=field-coolings]{fc}{FC}{field-cooling}
\newacronym[plural=zero-field-coolings]{zfc}{ZFC}{zero-field-cooling}
\newacronym{fw}{FW}{field-warming}
\newacronym{zfw}{ZFW}{zero-field-warming}
\newacronym{dft}{DFT}{density functional theory}
\newacronym{2d}{2D}{two-dimensional}
\newacronym{dos}{DOS}{density of states}
\newacronym{wal}{WAL}{weak antilocalization}
\newacronym{lowess}{LOWESS}{locally weighted scatterplot smoothing}
\newacronym{sh}{SH}{sample holder}
\newacronym{hbn}{hBN}{hexagonal boron nitride}
\newacronym{fwhm}{FWHM}{full width half maximum}
\newacronym{afm}{AFM}{atomic force microscopy}
\newacronym{ohe}{OHE}{orbital Hall effect}
\newacronym{lk}{L-K}{Lifshitz-Kosevich}
\newacronym{fft}{FFT}{fast Fourier transform}
\newacronym{sdh}{SdH}{Shubnikov-de Haas}
\newacronym{soc}{SOC}{spin orbit coupling}
\newacronym[plural=Landau levels]{ll}{LL}{Landau level}
\newacronym{hln}{HLN}{Hikami-Larkin-Nagaoka}
\newacronym{cps}{CPS}{counts per second}
\newacronym{be}{BE}{binding energy}
\newcommand{\titleText}{2D Shubnikov-de Haas Oscillations in \pt: A fermiological Charge Carrier Investigation}
\usepackage{xspace} 
\newcommand{\pt}{$\text{PtSe}_2$\xspace}

\newcommand{\fig}{Fig.~}
\newcommand{\figs}{Figs.~}
\newcommand{\eq}{Eq.~}

\newcommand{\secText}{Sec.~}
\newcommand{\tab}{Table~}
\newcommand{\refText}{Ref.~}
\newcommand{\refsText}{Refs.~}
\newcommand{\freq}{\mathfrak{F}}
\newcommand{\supp}{Supplemental Material \cite{Supplemental}\xspace}
\renewcommand{\t}[1]{\,\text{{#1}}}
\newcommand{\E}[1]{\times10^{{#1}}}
\newcommand{\insertArrow}{\mathrel{\rotatebox[origin=c]{180}{$\Lsh\,$}}}
\usepackage{tikz}
\usetikzlibrary{decorations.pathmorphing,arrows.meta}
\newcommand{\longrightsquigarrow}[1]{
\raisebox{0.5ex}{
  \tikz[baseline={(A.base)}]{
    \node (A) at (0,0) {};
    \draw[->,
      decorate,
      decoration={snake, amplitude=0.3mm, segment length=1.5mm},
      line width=0.3pt,
      shorten >=-5pt,
      >=Latex
    ] (-0.1,0) -- (1,0);
    \node[anchor=south] at (0.5,0) {\scriptsize#1};
}}}
\newcommand{\figureS}{\setcounter{figure}{0}\renewcommand{\thefigure}{S\arabic{figure}}}
\newcommand{\tableS}{\setcounter{table}{0}\renewcommand{\thetable}{S\arabic{table}}}
\newcommand{\equationS}{\setcounter{equation}{0}\renewcommand{\theequation}{S\arabic{equation}}}

\usepackage[normalem]{ulem}

\newcommand{\corr}[2]{#2}
\newcommand{\new}[1]{#1}

\begin{document}
\preprint{APS/123-QED}
\title{\titleText}
\author{Julian Max Salchegger}
\email{julian.salchegger@jku.at}
\affiliation{Institut f\"ur Halbleiter-und-Festk\"orperphysik, Johannes Kepler University, Altenbergerstr. 69, A-4040 Linz, Austria}
\author{Rajdeep Adhikari}
\email{rajdeep.adhikari@jku.at}
\affiliation{Institut f\"ur Halbleiter-und-Festk\"orperphysik, Johannes Kepler University, Altenbergerstr. 69, A-4040 Linz, Austria}
\affiliation{Linz institute of Technology, Johannes Kepler University, Altenbergerstr. 69, A-4040 Linz, Austria}
\author{Bogdan Faina}
\affiliation{Institut f\"ur Halbleiter-und-Festk\"orperphysik, Johannes Kepler University, Altenbergerstr. 69, A-4040 Linz, Austria}
\author{Alberta Bonanni}
\email{alberta.bonanni@jku.at}
\affiliation{Institut f\"ur Halbleiter-und-Festk\"orperphysik, Johannes Kepler University, Altenbergerstr. 69, A-4040 Linz, Austria}
\date{\today}
\begin{abstract}
High magnetic field and low temperature transport measurements are carried out in order to gain insight into the properties of the charge carriers of \pt. In particular, the Shubnikov-de Haas oscillations arising at applied magnetic field strengths $\gtrsim 4.5\,\text{T}$ are found to occur exclusively in plane. An analysis of the oscillations \textit{via} the Lifshitz-Kosevich formalism allows determining the charge carrier's cyclotron mass, quantum transport time, Berry phase, Fermi surface cross section and Dingle temperature. The oscillations emerge at a layer thickness of $\approx 18\,\text{nm}$ and decrease in frequency for thinner \pt flakes. Further, \g{wal} is observed despite the presence of magnetic moments from Pt vacancies, which typically inhibit such effects. An explanation is provided on how \g{wal} and the Kondo effect can be observed within the same material.
\end{abstract}
\maketitle
\section{Introduction}

The \g{tmd} \pt is a \g{vdw}-layered semimetal with type-II Dirac cones \cite{HuangPtSe2BANDDFTandARPES,KenanPtSe2BANDDFTandARPES}, which has been suggested for applications in electronic \cite{YimPtSe2Electric,YimPtSe2Sens,ZhaoPtSe2FET} and spintronic \cite{Jo2022Rashba} devices, and has recently been considered as a platform for the detection of the \g{ohe} \cite{IntrinsicOHE, SahuOrbitalHall}. Unlike the conventional Hall effect, the \g{ohe} does not require an applied magnetic field to emerge: The orbital momentum of the charge carriers couples to an applied electric field, resulting in the generation of a transversal orbital current. By injecting the orbital current into a material with high \g{soc}, or \textit{e.g.} through Rashba \g{soc} \cite{DingOHERashba}, the orbital current can be converted to a spin current. Harnessing orbital currents promises applications in spintronic \cite{SOTMRAM} and orbitronic \cite{WangOrbitalTHz,WangTopoOrbital} devices. Additionally, the \g{ohe} may account for the discrepancy between the observed and theoretical magnitudes of the spin Hall effect in materials such as Pt \cite{TanakaOHE}. \pt is predicted to host significant orbital currents \cite{SahuOrbitalHall} and offers the advantage of inherent \g{soc}, due to the presence of heavy Pt atoms. Understanding the nature of the charge carriers present in \pt, as well as its electronic structure, is a fundamental step towards the experimental realization of an orbital Hall device.

Here, high magnetic field and low temperature transport studies are carried out to investigate key aspects of the charge carriers of \pt flakes, such as the cyclotron mass, the quantum scattering time, the mean free path and the Berry phase. In a previous work \cite{Ouroboros}, we demonstrated that \pt flakes, exfoliated from a bulk crystal, contain Pt vacancies, resulting in a Kondo effect. Here, a further analysis reveals that the vacancy concentration is not uniform, since the Kondo effect varies across samples. A reduced number of Pt vacancies leads to the emergence of \g{wal} in the \new{4-terminal} resistance measurements, heralding \g{soc} in the samples. To investigate the above mentioned charge carrier properties, \g{sdh} oscillations, which are closely linked to the orbital motion of the charge carriers and to their Fermi surface, are analyzed \cite{FuPtTe2SdH, XingzeZrP2, BuschBi2Se3SdH}. The oscillations arise in both the longitudinal and transverse resistance when the applied magnetic field exceeds $\sim 4\,\t{T}$.

The spectral analysis of the oscillations points at \corr{the charge transport being dominated by a single electron-like band.}{only one dominant type of charge-carrier orbit.} The quantum scattering time $\tau$ and the cyclotron mass $m_c$ are determined together with the Berry phase $\Phi_B$. The thickness-dependence of the extremal Fermi surface cross-sectional area indicates that the bulk limit of the electronic structure establishes only for thicknesses $t>20\t{nm}$. Using the angular dispersion of the \g{sdh} oscillations, insights into the Fermi surface shape are gained. 

\section{Experimental}
\begin{figure}[htb]
    \centering
    \includegraphics[width=0.45\linewidth]{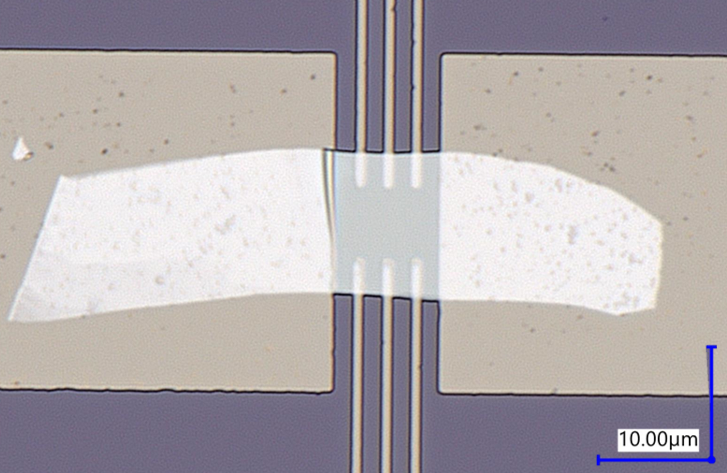}
    \includegraphics[width=0.4\linewidth]{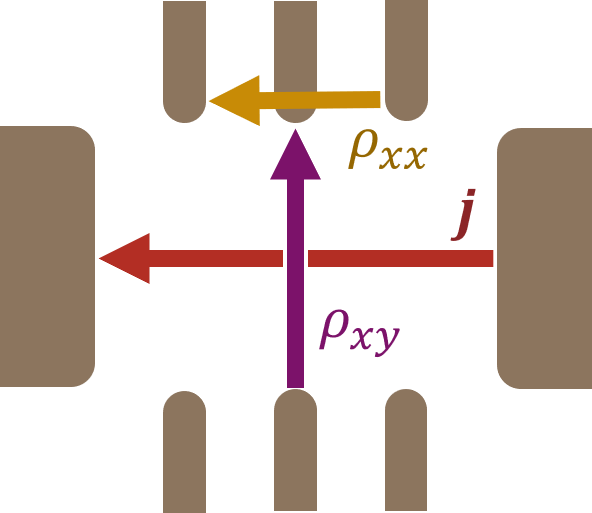}
\stackinset{c}{-0.9\linewidth}{c}{0.3\linewidth}{a)}{\hspace{0pt}}
\stackinset{c}{-0.4\linewidth}{c}{0.3\linewidth}{b)}{\hspace{0pt}}
    \caption{a) Optical image of sample X1. b) Schematic of the Hall bar with the direction of the current density $\boldsymbol{j}$. Measuring the voltage difference between two terminals which lie on line parallel to $\boldsymbol{j}$, results in $\rho_{xx}$ while the voltage difference between terminals, which lies on a line normal to $\boldsymbol{j}$, yields $\rho_{xy}$.}
    \label{fig_optical}
\end{figure}

The \pt flakes considered in this work are mechanically exfoliated from bulk crystals grown by HQ Graphene \cite{HQGraphene}. Optical microscopy (\fig \ref{fig_optical}) shows that the angles between flake edges are multiples of $60\degree$, as expected from the $C_3$ symmetry of the crystal. Atomic force microscopy and \g{xps} are performed to characterize the flatness, crystallinity and chemical constituents of the flakes, as depicted in \figs S3 and S4 of the \supp. The samples are listed in \tab \ref{tab_samples}: Samples A - F are obtained from bulk crystal Batch 1, while samples X1 and X2 are from Batch 2. The samples cover a range of thicknesses from $8\,\text{nm}$ to $26\,\text{nm}$, resulting in a semimetallic band structure \cite{KandemirThinPtSe2}, while also providing variation in the extent of 2D confinement of the charge carriers and in the surface-to-bulk ratio. A cover-layer of \g{hbn} is included to improve the signal-to-noise ratio for the thinnest samples and to study the interface effects originating from the interaction between hBN and the Pt vacancy magnetic moments.

Pre-fabricated Hall bars are patterned \textit{via} electron-beam lithography and $10\,\t{nm}$ of Pt are deposited \textit{via} sputtering. The flakes are then placed onto the Pt-contacts. An optical image of sample X1 is shown in \fig \ref{fig_optical} a). In \fig \ref{fig_optical} b), the direction of the current density $\boldsymbol{j}$ and exemplary voltage terminal selections for the measurement of the longitudinal resistance and the Hall voltage are marked.

Au wires are bonded to the Hall bar using an In solder agent. The low $T$/high $\mu_0 H$ transport measurements are performed in a Janis Super Variable Temperature 7TM-SVM cryostat equipped with a $7\,\t{T}$ superconducting magnet and a homemade rotatory \gls{sh} with two angular degrees of freedom. A lock-in amplifier \textit{ac} technique at $277\,\text{Hz}$ is used for measuring the magnetotransport properties of the \pt flakes.
\begin{table}[htb]
\centering
\begin{tabular}{c|S[table-format=2.0, table-number-alignment = center]|c|c}
sample& {$t$ (nm)}& cover& bulk crystal Batch \\ \hline
A      & 18                 & -     & 1\\
B      & 20                 & -     & 1\\
C      & 26                 & -     & 1\\
D      & 8                  & hBN   & 1\\
E      & 10                 & hBN   & 1\\
F      & 11                 & -     & 1\\
X1     & 17                 & -     & 2\\
X2     & 12                 & -     & 2\\
\end{tabular}
\caption{Considered samples: thickness $t$, cover-layer and respective batch of bulk crystals.}
\label{tab_samples}
\end{table}

\section{Results and Discussion}
\subsection{Hall voltage}
The Hall voltage $V_\text{H}$ is obtained when choosing voltage probing terminals orthogonal to the source-drain direction. The slope of $V_\text{H}(H)$ is used to determine the charge carrier density $n = \frac{\mu_0 H}{V_\text{H}} \frac{I}{t e}$, with $I$ being the source-drain current. The samples show $n(2\,\text{K}) \sim (2\times 10^{20})\, \text{cm}^{-3}$. Exemplary plots of $V_\text{H}(H)$ are presented in the \supp \fig S7 for samples F and X1.

\subsection{Shubnikov-de Haas oscillations}
\begin{figure}[htb]
    \centering
    \includegraphics[width=0.82\linewidth]{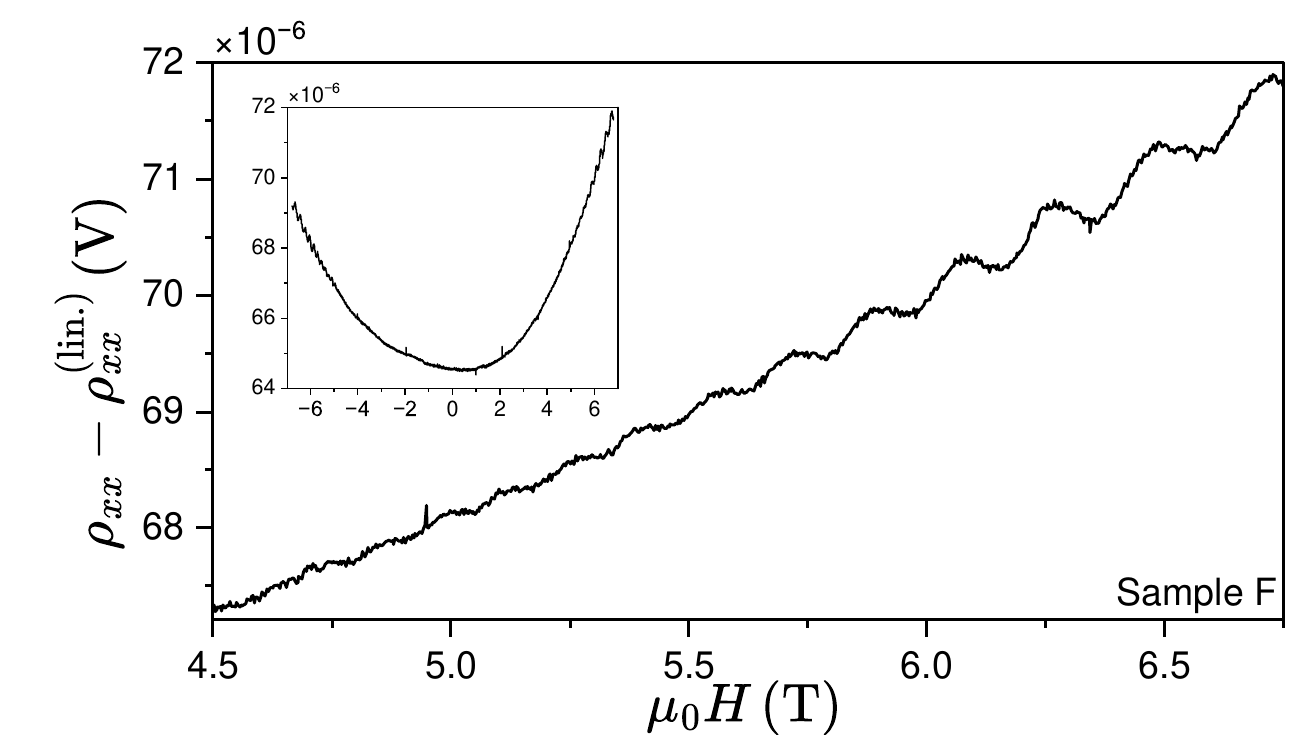}
    \includegraphics[width=0.8\linewidth]{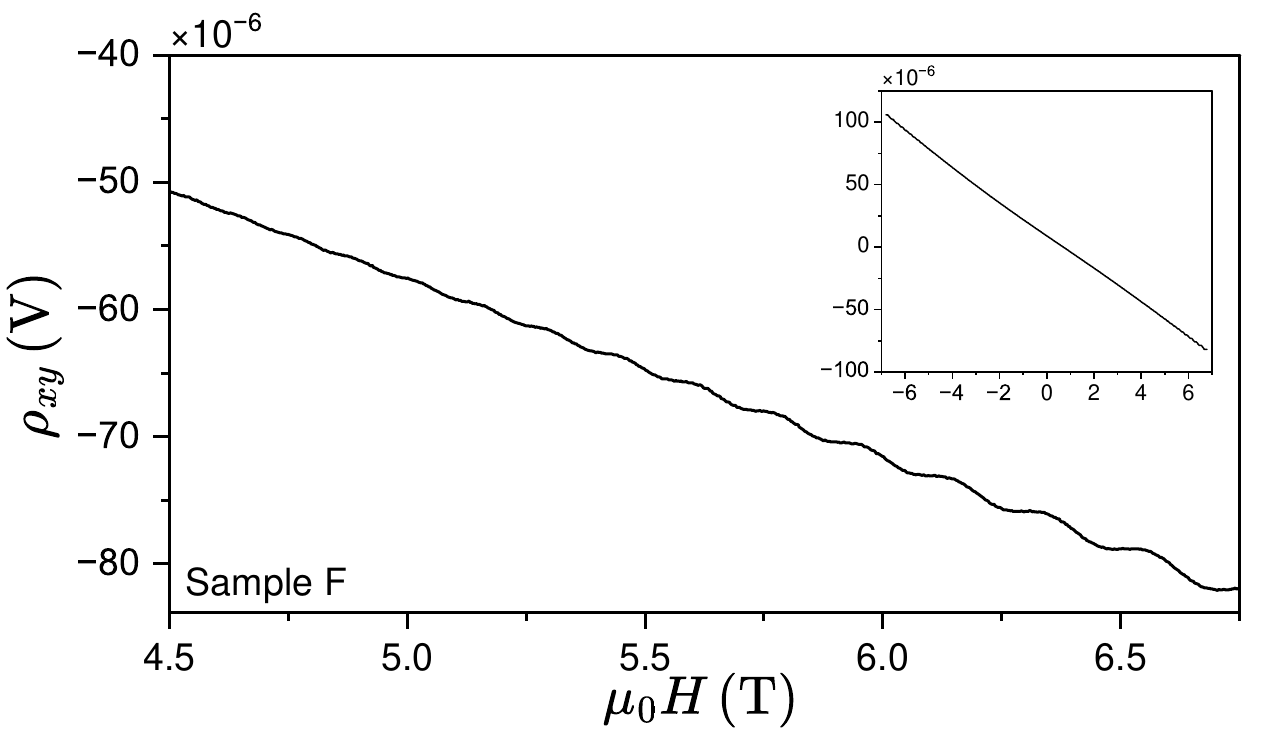}
    \caption{Upper panel: longitudinal \new{4-terminal} resistance $\rho_{xx}-\rho_{xx}^{(\text{lin.})}$ of sample F over applied magnetic field for $\mu_0 H\geq 4.5\t{T}$ at $2\t{K}$. Inset: increased range of $|\mu_0H|\leq 6.8\,\text{T}$. Lower panel: equivalent plot for $\rho_{xy}$.}
    \label{fig_Rasa_MR_2K}
\end{figure}

The current density $\boldsymbol{j}$ is induced between the source-drain contacts and the \new{4-terminal} longitudinal resistivity $\rho_{xx}$ is determined by measuring the voltage difference between two terminals which lie on a line parallel to $\boldsymbol{j}$. The transversal resistivity $\rho_{xy}$ (cognate with $V_\text{H}$) is measured between two terminals which lie on a line normal to $\boldsymbol{j}$. The longitudinal resistance $\rho_{xx}(H)$ of sample F at $2\t{K}$ is reported in the inset to the upper panel of \fig \ref{fig_Rasa_MR_2K}, with $H$ being the scalar value of the applied magnetic field ($H = \boldsymbol{H}\cdot \hat{e}_H$ and $\hat{e}_H$ is the axis along which the magnetic field is applied). Since the flake edges are, in general, not exactly parallel to $\boldsymbol{j}$, the longitudinal resistivity contains a linear component \corr{$\rho_{xx}^{(\text{lin.})}$}{$\rho_{xx}^{(\text{lin.})} \approx (2.62\E{-6})\,\text{V/T}$} which originates from the Hall voltage. The longitudinal resistance $\rho_{xx}-\rho_{xx}^{(\text{lin.})}$ shows a dependence which is approximately quadratic in $H$.

For $|\mu_0 H|\gtrsim 4.5\t{T}$, \gls{sdh} oscillations emerge, as displayed in the upper panel of \fig \ref{fig_Rasa_MR_2K}. The Hall voltage as a function of applied magnetic field captured at $2\t{K}$ shows a quasi-linear dependence $\rho_{xy}\propto -H$, as depicted in the inset to the lower panel of \fig \ref{fig_Rasa_MR_2K}, until, for $|\mu_0 H|\gtrsim 4.5\t{T}$, \gls{sdh} oscillations emerge, as reported in the lower panel of \fig \ref{fig_Rasa_MR_2K}.

The origin of the \gls{sdh} oscillations is the Landau-quantization of the electronic states, which modulates the \gls{dos} at the Fermi level when varying $\boldsymbol{H}$. A local maximum in $\rho_{xx}(H)$ corresponds to a \g{ll} coinciding with the Fermi level, as the increased number of available states (compared to no \g{ll} coinciding with the Fermi level) enhances the electron-electron scattering. A similar mechanism also affects $\rho_{xy}$, in which case the modulation of the \gls{dos} at the Fermi level alters the charge carrier density $n$ and thus the Hall voltage $\rho_{xy}\cong V_\text{H} \propto \frac{1}{n}$ shows a minimum when a \g{ll} coincides with the Fermi level.

\begin{figure}[htb]
    \centering
    \includegraphics[width=0.8\linewidth]{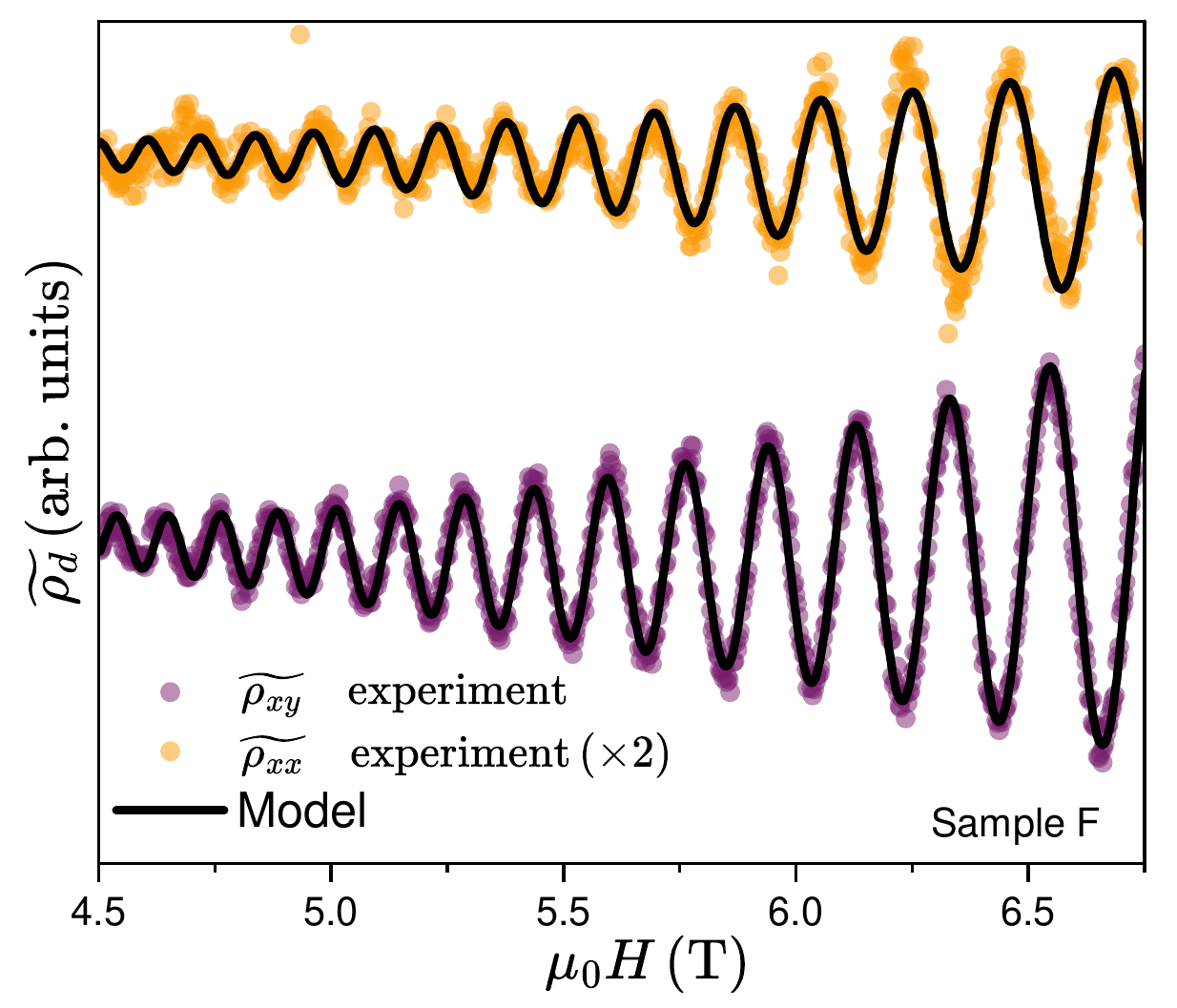}
    \caption{$\widetilde{\rho_d}$ over applied magnetic field for sample F at $2\t{K}$. Solid curves: \gls{lk} fits. The obtained values are given in \tab \corr{S1 of the \supp.}{\ref{tab_LK_parameters}}}
    \label{fig_Rasa_LKS_fit_both_2K}
\end{figure}

Let $d\in \{xx,xy\}\equiv\{\text{longitudinal},\text{transversal}\}$ denote the measurement direction and $\widetilde{\rho_d}$ be the oscillatory part of $\rho_d$, such that $\rho_d = \rho_d^\text{(bg.)}+\widetilde{\rho_d}$, where $\rho_d^\text{(bg.)}$ is a smooth background. Then, $\widetilde{\rho_d}(H,T)$ follows from the \gls{lk} relation \cite{LK, Shoenberg}:
\begin{equation}
\begin{aligned}
    \widetilde{\rho_d} = &A T \left(\frac{5}{2}\left(\frac{H}{2 \freq}\right)^\frac{1}{2}+\frac{3}{2}\left(\frac{H}{2\freq}\right)\right) \\
    \times&\frac{\exp(-x \frac{NM}{H}) \cos(\pi M)}{\sinh(T \frac{NM}{H})}\cos\left(2\pi\left( \frac{\freq}{H}+\phi\right)\right),
\end{aligned}
\label{eq_LK_rho_inMain}
\end{equation}
where $A$ is an amplitude prefactor, $\phi$ the phase of the oscillations, $x$ the Dingle temperature ($x$ is not a physical temperature, but is related to defects and impurities in the sample), $M=\frac{m_c}{m_0}$ the relative electron cyclotron mass ($m_c$ is the electron cyclotron mass and $m_0$ the free electron rest mass) and $N = 2\pi^2 \frac{m_0 k_\text{B}}{e \hbar}$ a constant ($\hbar$ is the reduced Planck constant, $e$ the elementary charge and $k_\text{B}$ the Boltzmann constant). Finally, $\freq$ is the frequency of the oscillations (periodic in $\frac{1}{H}$), which is linked to $S_\text{F}$, \textit{i.e.} the extremal cross-sectional area of the Fermi surface normal to $\boldsymbol{H}$, \textit{via} $\freq = \frac{\hbar}{2 \pi e} S_\text{F}$. The adaptation of the \gls{lk} formula given in \refText \cite{Shoenberg} is described in Appendix B. 
The experimental value of $\widetilde{\rho_d}$ is obtained by subtracting a polynomial background from $\rho_d$. \eq \ref{eq_LK_rho_inMain} is augmented with a second-order polynomial,
\begin{equation}
    \widetilde{\rho_d} \rightarrow \widetilde{\rho_d}+A^{(0)}+A^{\text({lin.})} H+A^{\text({quad.})} H^2,
    \label{eq_LK_and_quadBg}
\end{equation}
to account for any residual background, and the fitting of $\widetilde{\rho_d}$ as a function of the applied magnetic field of sample F at $2\,\text{K}$ with \eq \ref{eq_LK_and_quadBg} is reported in \fig \ref{fig_Rasa_LKS_fit_both_2K}.

\begin{figure}[htb]
    \centering
    \includegraphics[width=0.88\linewidth]{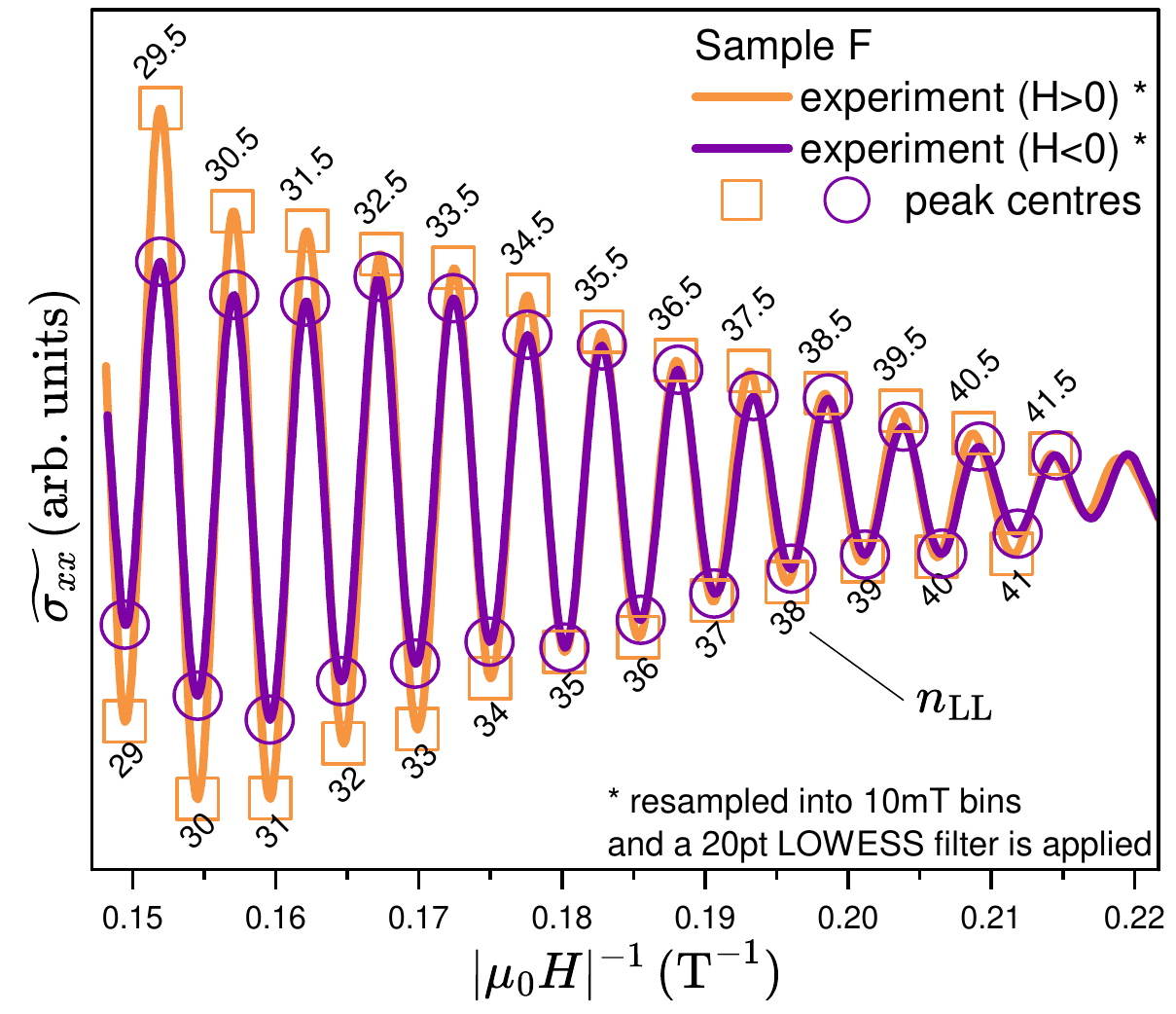}
    \includegraphics[width=0.9\linewidth]{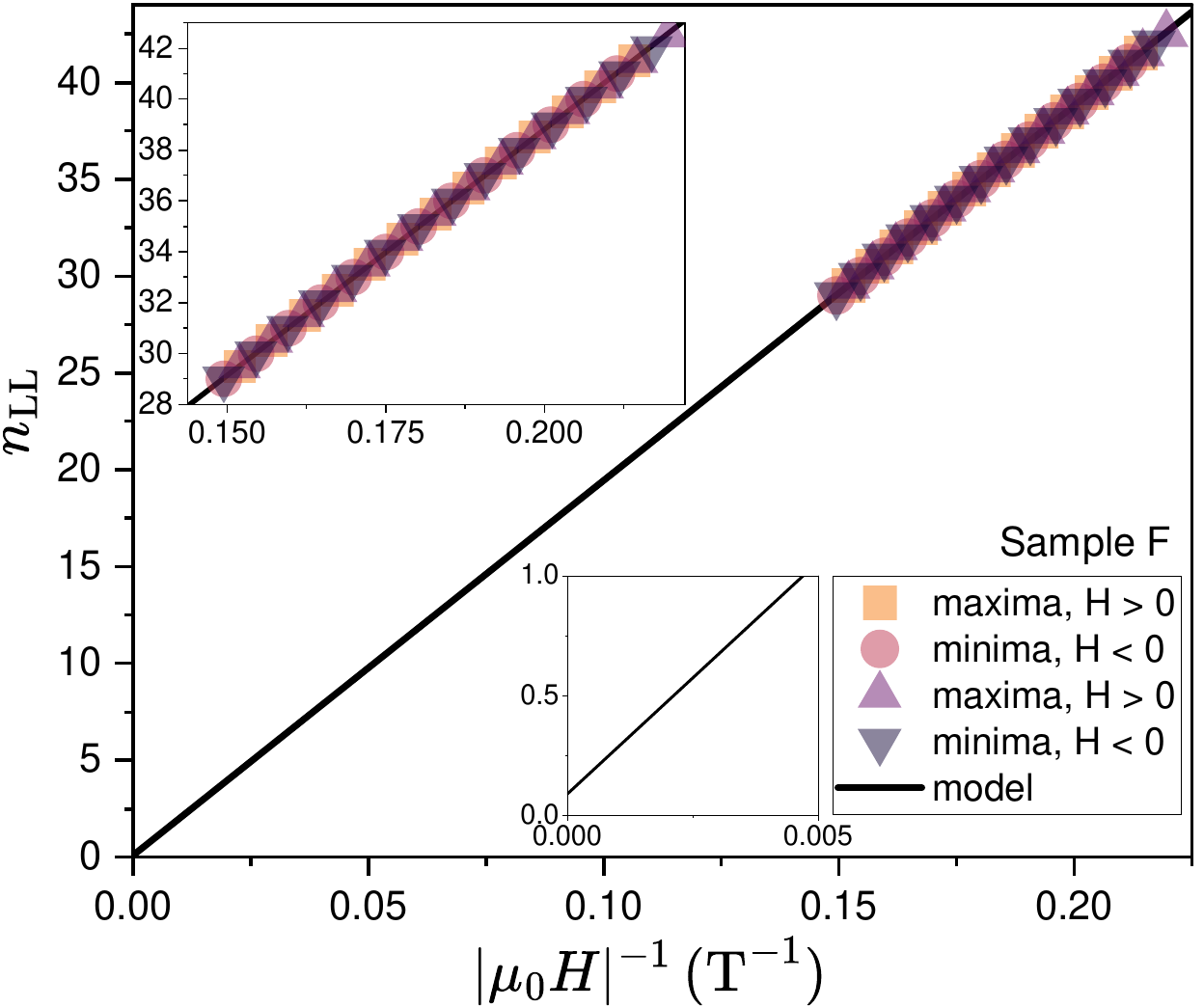}
    \caption{Upper panel: $\widetilde{\sigma_{xx}}$ of sample F at $2\,\text{K}$ over inverse applied magnetic field magnitude and extracted peak center positions marked as squares/circles. Lower panel: Landau-fan diagram showing the \g{ll} indices over the peak positions from the upper panel and a linear fit. Insets: magnification of the identified peak positions and of the axis cutoff region. The obtained parameters are provided in \tab S1 of the \supp.}
    \label{fig_F_LandauFan}
\end{figure}

The maxima in $\widetilde{\rho_{xx}}(H)$ occur when the scattering rate is maximal and thus correspond to (whole) integer \g{ll} numbers $n_\text{LL}$ (and minima to half-integers). Therefore, in the conductivity $\sigma_{xx} = \frac{\rho_{xx}}{\rho_{xx}^2+\rho_{xy}^2}$, minima of $\widetilde{\sigma_{xx}}(H)$ correspond to integer \g{ll} numbers. \corr{The identified levels can be found in the upper panel of \fig \ref{fig_F_LandauFan}, which shows $\widetilde{\sigma_{xx}}(|H|^{-1})$ for sample F at $2\,\text{K}$, while the resulting Landau-fan diagram $n_\text{LL}(|H|^{-1})$ is depicted in the lower panel of \fig \ref{fig_F_LandauFan}.}{The identified levels can be found in the upper panel of \fig \ref{fig_F_LandauFan}, which shows $\widetilde{\sigma_{xx}}(|H|^{-1})$ for sample F at $2\,\text{K}$ for both $H>0$ and $H<0$. The oscillations match in frequency and phase, which points at the underlying Landau quantization being symmetric under inversion of the applied magnetic field. The differing amplitudes are attributed to an asymmetry in the flake placement with respect to the source-drain axis, resulting in asymmetric contribution of the edge conductance channels.} The index-axis cutoff of the Landau-fan diagram is the oscillation phase $\phi$, which relates to the Berry phase \cite{MikitikBerryPhase} according to:
\begin{align}
    \Phi_\text{B} = 2\pi\left(\frac{1}{2}-\phi\pm\delta\right) \mod 2\pi.
\end{align}
The phase shift $\delta$ is taken to be zero \cite{BuschBi2Se3SdH} in the case of a 2D \g{sdh} effect, which is reflected in the fact that only the out-of-plane components of $\boldsymbol{H}$ contribute to the \g{sdh} oscillations. From the Landau-fan diagram, $\Phi_\text{B} = (0.82\pm 0.09)\pi$ is determined, suggesting non-trivial charge carriers. The physical origin of this non-trivial value is unclear, since the Dirac cone is located $\approx 1.5\,\text{eV}$ below the Fermi level, the dispersion of which forms a hole pocket at the Fermi level \cite{PtSe2ARPES}. The transport of the investigated system is, however, dominated by electrons, as evidenced by the sign of the Hall resistance and by the value of $S_\text{F}^{(11\,\text{nm})}\approx 0.02 \,\text{Å}^{-2}$ being smaller than the observed size of the hole pocket \cite{PtSe2ARPES}.  It has to be mentioned, that the extrapolation from $n_\text{LL} = 29$ to $0$ is volatile with regard to the exact manner in which the extrema of $\widetilde{\sigma_{xx}}(|H|^{-1})$ are detected and at which value of $|H|^{-1}$ the cutoff is set, since the amplitude of the oscillations damps significantly as $|H|^{-1} \rightarrow \infty$. To substantiate the value of the Berry phase, a direct \g{lk} fit of \corr{$\rho_{xx}^{(\text{corr.})} = \sigma_{xx}^{-1}$}{$\sigma_{xx}^{-1}$} is performed (provided  \fig S1 of the \supp) which yields $\Phi_\text{B}^{(\text{LK})} = (1.01\pm 0.05)\pi$.

\begin{figure}[htb]
    \centering
    \includegraphics[width=1.0\linewidth]{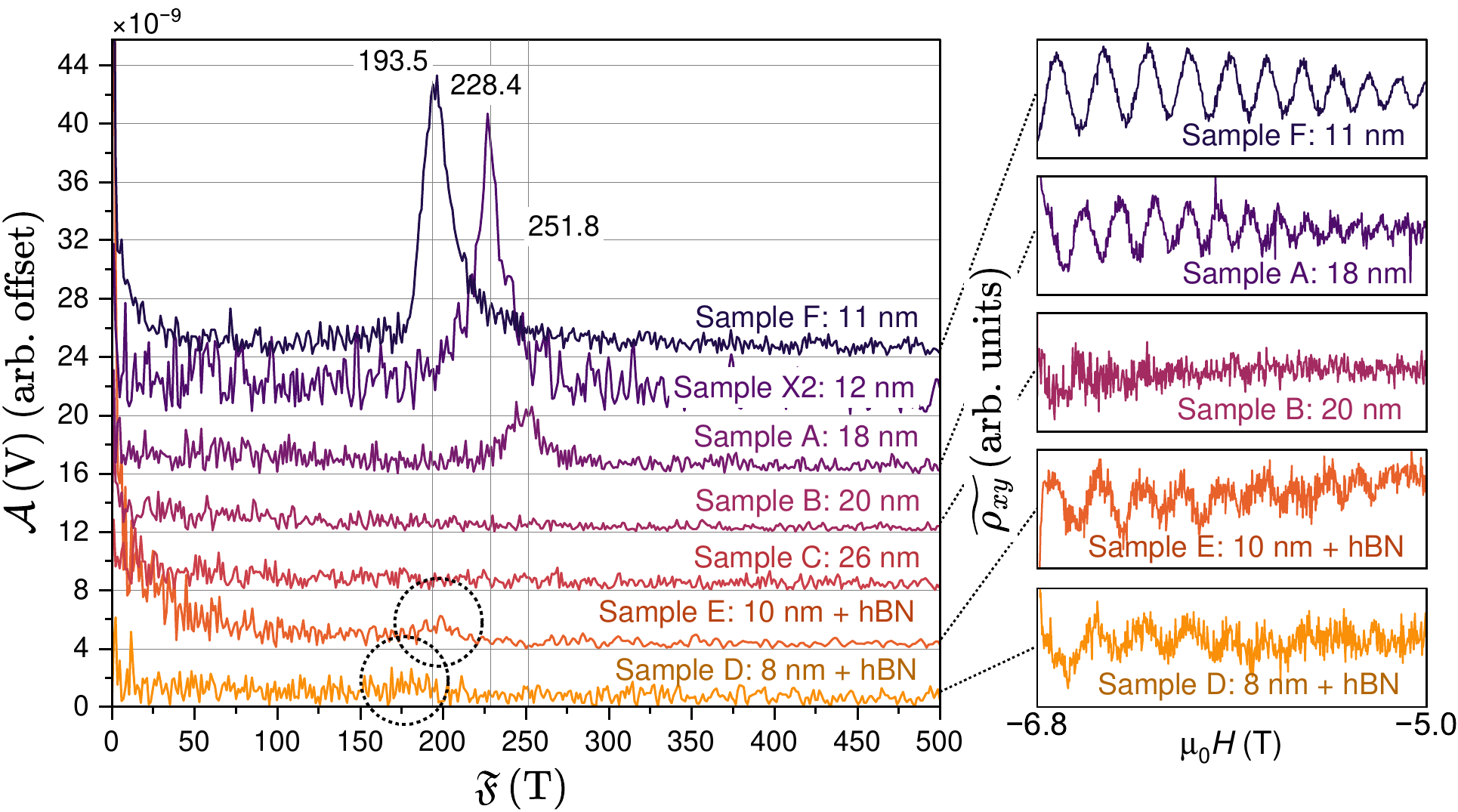}
    \caption{Left panel: amplitude $\mathcal{A}$ resulting from the \gls{fft} spectra of $\widetilde{\rho_{xy}}$ over frequency $\freq$ for specific flake thicknesses. The dotted circles are a guide to the eye. Right panel: $\widetilde{\rho_{xy}}$ as a function of applied magnetic field for specific flake thicknesses.}
    \label{fig_FFT}
\end{figure}

A \gls{fft} of the \g{sdh} oscillations is performed to gain further insight into the charge carrier properties. The data treatment is detailed in\secText B of the \supp. The resulting spectra are shown in the left panel of \fig \ref{fig_FFT}, where the \gls{fft} amplitude $\mathcal{A}$ is given as a function of the oscillation frequency $\freq$: For sample F, the position of the local maximum at $\freq_\text{FFT} \approx 195 \t{T}$ is in agreement with the frequency $\freq_\text{LK} = 193.6 \t{T}$ from the \gls{lk} fit. Similarly, the frequencies obtained for sample A are \corr{$\freq_\text{FFT} \approx 250 \t{T} \approx \freq_\text{LK} = 251.8 \t{T}$}{$\freq_\text{FFT} \approx 250 \t{T} \approx \freq_\text{LK} = 254.9 \t{T}$}. The shift in the value of $\freq$ with increasing the sample thickness $t$ from $11\t{nm}$ to $18\t{nm}$, points at $t=11\t{nm}$ not having a bulk bandstructure, because $\freq \propto S_\text{F}$. For $t \geq 20\t{nm}$ (samples B and C) no local maximum at $\freq>0$ is discernible, and indeed, no \gls{sdh} oscillations can be gleaned in $\widetilde{\rho_{xy}}$, as evidenced in the right panel of \fig \ref{fig_FFT}. A similar effect is observed  For sample E, which has a thickness comparable to the one of sample F, but is covered with \g{hbn}, a peak at $\freq\approx 200\t{T}$ can be found and $\widetilde{\rho_{xy}}$ does show \gls{sdh} oscillations. However, the peak height of $\freq_\mathcal{F}$ and the signal-to-noise ratio observed in $\widetilde{\rho_{xy}}$ are substantially lower than those of sample F. The \g{hbn} coating has a similar effect on sample D, with $t=8\t{nm}$: The maximum at $\freq\approx 180\t{T}$ is barely detectable, and the signal-to-noise ratio in $\widetilde{\rho_{xy}}$ is reduced in comparison to sample F.

\begin{figure}[htb]
    \centering
    \includegraphics[width=1\linewidth]{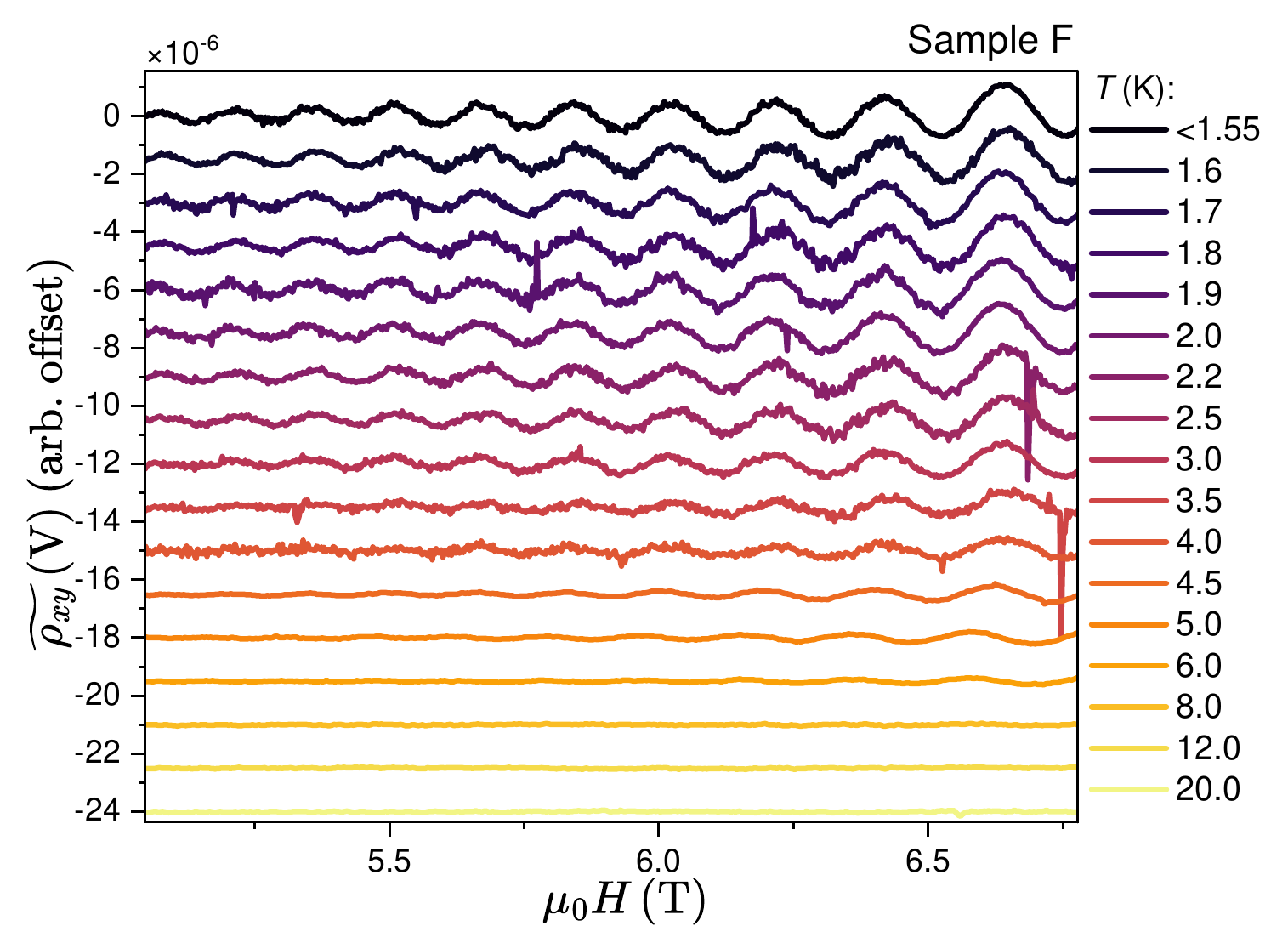}
    \caption{Oscillations in the Hall voltage $\widetilde{\rho_{xy}}$ of sample F as a function of applied magnetic field at various temperatures.}
    \label{fig_F_osc}
\end{figure}

The \g{sdh} oscillations reduce in magnitude with increasing temperature due to the broadening of the Fermi edge. The oscillations $\widetilde{\rho_{xy}}(H,T)$ as a function of applied magnetic field for various temperatures are given in \fig \ref{fig_F_osc} for sample F.  An equivalent plot for sample X2 is reported in \fig S9 of the \supp.

\begin{figure}[htb]
    \centering
    \includegraphics[width=0.9\linewidth]{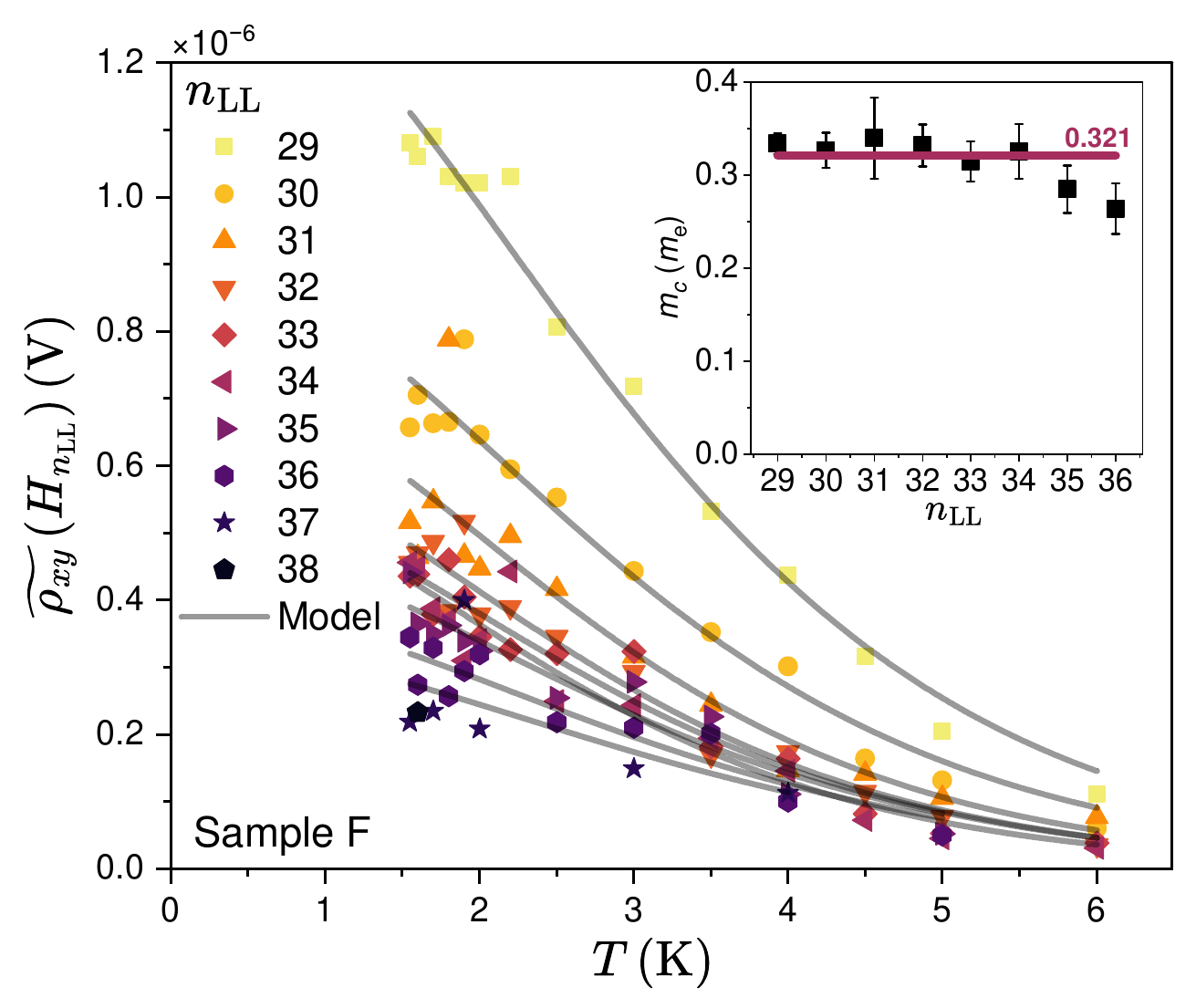}
    \caption{Amplitudes of $\widetilde{\rho_{xy}}$ for sample F at applied magnetic fields corresponding to $n_\text{LL}\in[29;38]$ over temperature and fitting with \eq \ref{eq_LK_T_dependence}. Inset: $m_c$ for each fitted $n_\text{LL}$ and average $m_c$.}
    \label{fig_lambda_peak_heights}
\end{figure}

By fitting the amplitude of a specific \g{ll} index $n_\text{LL}$ as a function of $T$ with the temperature dependence following \eq \ref{eq_LK_rho_inMain}, the reduction of the oscillation amplitude over temperature can be employed to extract the cyclotron mass $m_c$ according to:
\begin{equation}
    \widetilde{\rho_d}(H=\text{const.},T) = A_T \frac{T}{\sinh(T \frac{NpM}{H})},
    \label{eq_LK_T_dependence}
\end{equation}
with $A_T\in\mathbb{R}_+$ representing the temperature-independent terms. In \fig \ref{fig_lambda_peak_heights}, the amplitudes of $\widetilde{\rho_{xy}}$ corresponding to specific $n_\text{LL}\in[29;38]$ are given over a temperature range $T\in[1.55;6.00]\,\text{K}$. Now, using \eq \ref{eq_LK_T_dependence}, the $\widetilde{\rho_{xy}}$ amplitude at each $n_\text{LL}$ as a function of temperature is fitted (solid lines in \fig \ref{fig_lambda_peak_heights}). The resulting values of the cyclotron mass $m^{(n_\text{LL},\,xy)}_{c}$ as a function of $n_\text{LL}$ are plotted in the inset to \fig \ref{fig_lambda_peak_heights}. The errors of $m^{(n_\text{LL},\,xy)}_{c}$ are not statistically independent, since they originate from the same dataset and the weighted mean value is found to be $m^{(xy)}_c = (0.32\pm0.05)\,m_e$. Similarly, the amplitude values extracted from the $\widetilde{\rho_{xx}}$ data of sample F yield $m^{(xx)}_c = (0.32\pm0.02)\,m_e$ and a value of $m^{(xy)}_c = (0.39\pm0.05)\,m_e$ is obtained from the $\widetilde{\rho_{xy}}$ data of sample X2. The amplitude over temperature dependence and the fitting are provided in \figs S10 and S11, respectively, in the \supp.

\begin{figure}[htb]
    \centering
    \includegraphics[width=0.8\linewidth]{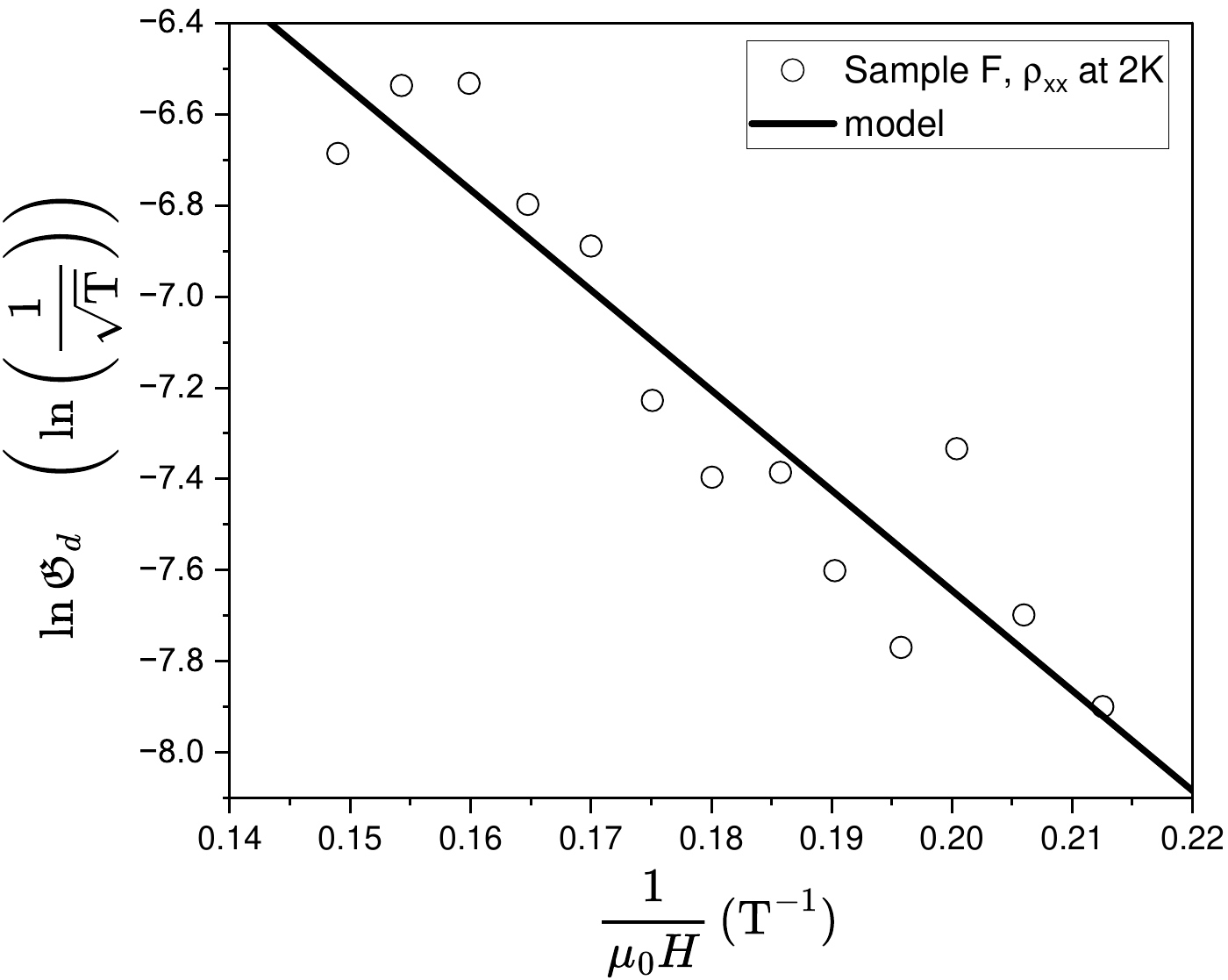}
    \includegraphics[width=0.8\linewidth]{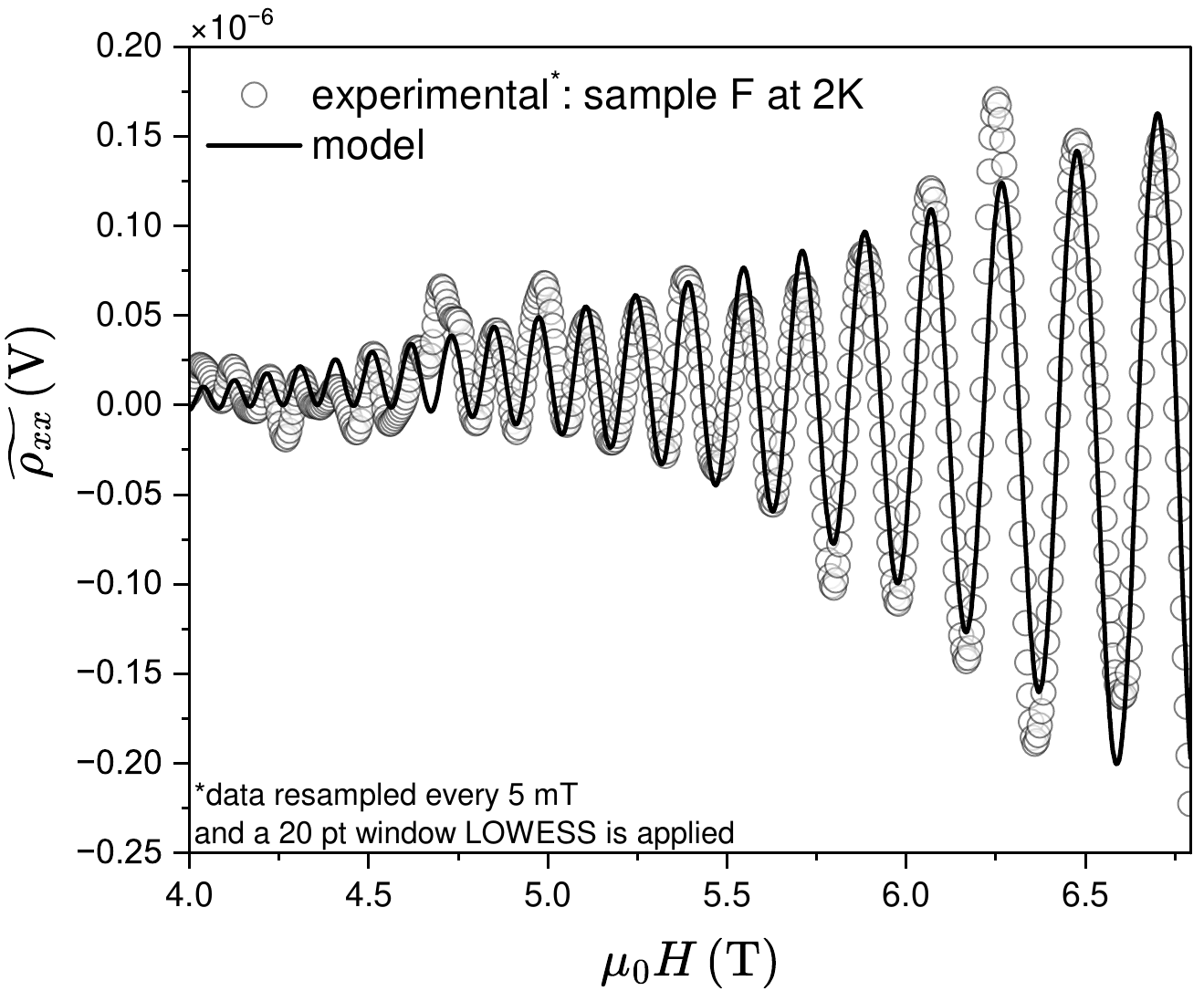}
    \caption{Upper panel: Dingle plot of $\ln\mathfrak{G}_d$ as a function of inverse magnetic field for sample F at $2\,\text{K}$. Solid line: linear fit. Lower panel: longitudinal resistance oscillations as a function of applied magnetic field for sample F at $2\,\text{K}$. Solid line: \g{lk} fit. The obtained parameters are provided in \tab \corr{S3 of the \supp}{\ref{tab_LK_parameters}}}
    \label{fig_tau_fit}
\end{figure}

In order to obtain the quantum transport time $\tau$, the extrema of $\widetilde{\rho_d}$ as a function of applied magnetic field, where $\frac{\partial^2 \widetilde{\rho_d}}{\partial H^2}=0$, are investigated. At these points, the \g{lk} formula simplifies considerably, and the quantum transport time $\tau$ can be expressed as:
\begin{equation}
\tau = \left(\frac{e}{-m_c \pi}\frac{\partial}{\partial\frac{1}{H}} \ln{\mathfrak{G}_d} \right)^{-1},
\label{eq_tau_main}
\end{equation}
with $\mathfrak{G}_d \equiv \frac{\widetilde{\rho_d}}{\rho_d^\text{(bg.)}} \sinh{\left(N M\frac{T}{H}\right)}\frac{1}{\sqrt{H}}$ and $\rho_d^\text{(bg.)}$ the smooth background of $\rho_d$. A derivation is given in Appendix C. $\tau$ can be extracted from the slope of $\ln\mathfrak{G}_d$ over $\frac{1}{H}$, leaving only $m_c$ as an unknown parameter. Since $m_c$ is already determined, $\tau$ can be obtained from $\widetilde{\rho_{xx}}(T=2\,\text{K})$ and is found to be $\tau = [( 0.26\pm 0.03 )\times 10^{-12}]\,\text{s}$. A linear fit of $\ln\mathfrak{G}_d(\frac{1}{H})$ is provided in the upper panel of \fig \ref{fig_tau_fit} (Dingle plot). The Dingle temperature $x$ is inherently linked to $\tau$ \textit{via} $x = \frac{\hbar}{2\pi k_\text{B} \tau} = (4.0\pm 0.5)\,\text{K}$. The Dingle damping is obtained by considering the bare electron mass $m_b \geq m_c$, not the cyclotron mass \cite{Shoenberg}, so the results are to be understood as an upper bound of $x$ (and lower bound of $\tau$). A direct \g{lk} fit of $\widetilde{\rho_{xx}}(T=2\,\text{K})$ results in $x_{xx}^{(\text{LK})} = (4.7\pm 0.4)\,\t{K}$, corresponding to $\tau = [(0.22 \pm 0.02)\times10^{-12}]\,\text{s}$, as depicted in the lower panel of \fig \ref{fig_tau_fit}. A similar value of $x_{xy}^{(\text{LK})} = (4.3\pm 0.3)\,\t{K}$ is found based on $\widetilde{\rho_{xy}}(T=2\,\text{K})$.

\begin{figure}[htb]
    \centering
    \includegraphics[width=0.65\linewidth]{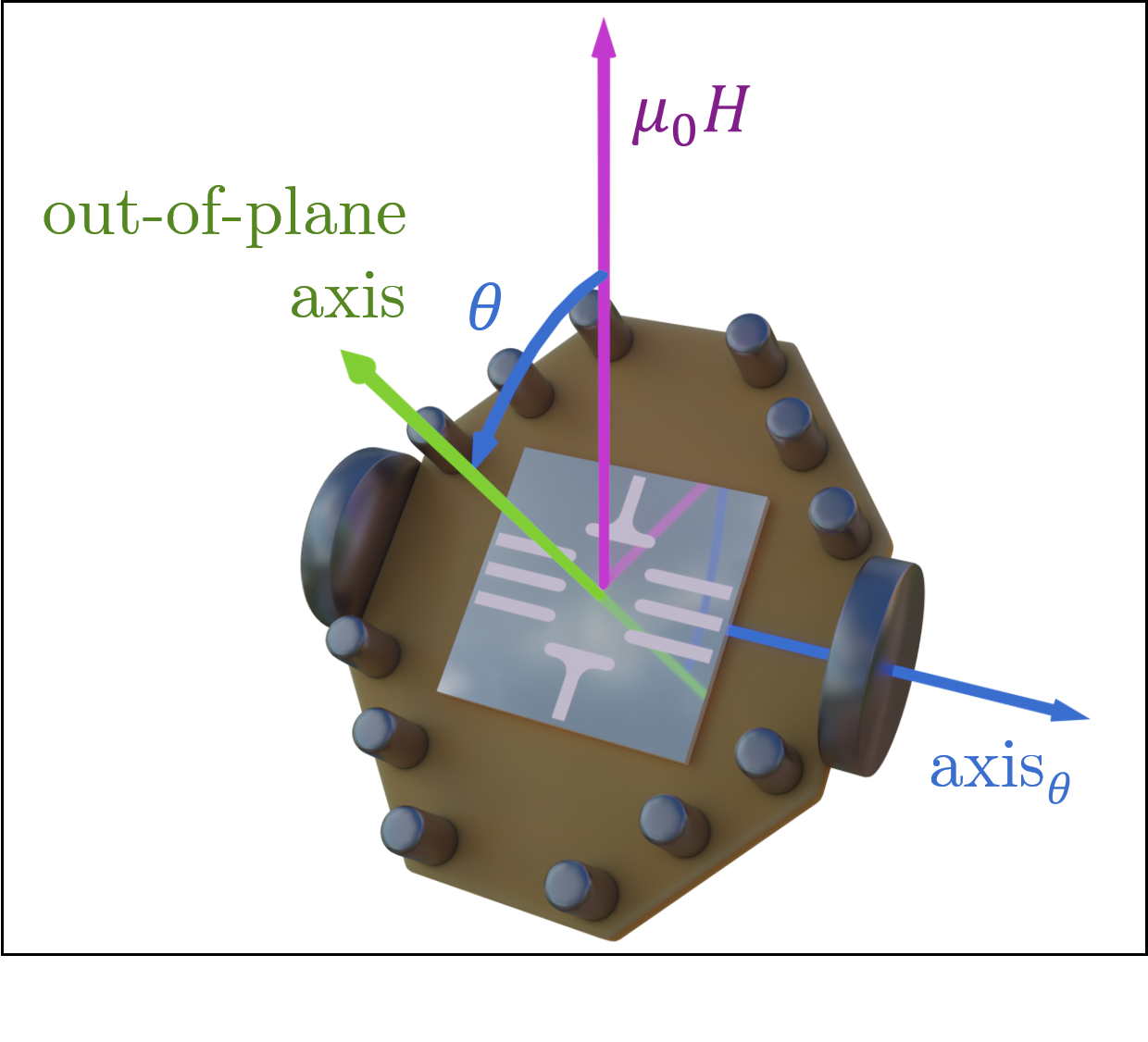}
    \caption{Visualization of the \g{sh} after a rotation along axis$_\theta$, the applied magnetic field encloses the angle $\theta$ with the out-of-plane axis of the sample.}
    \label{fig_sh}
\end{figure}

\begin{figure}[htb]
    \centering
    \includegraphics[width=1\linewidth]{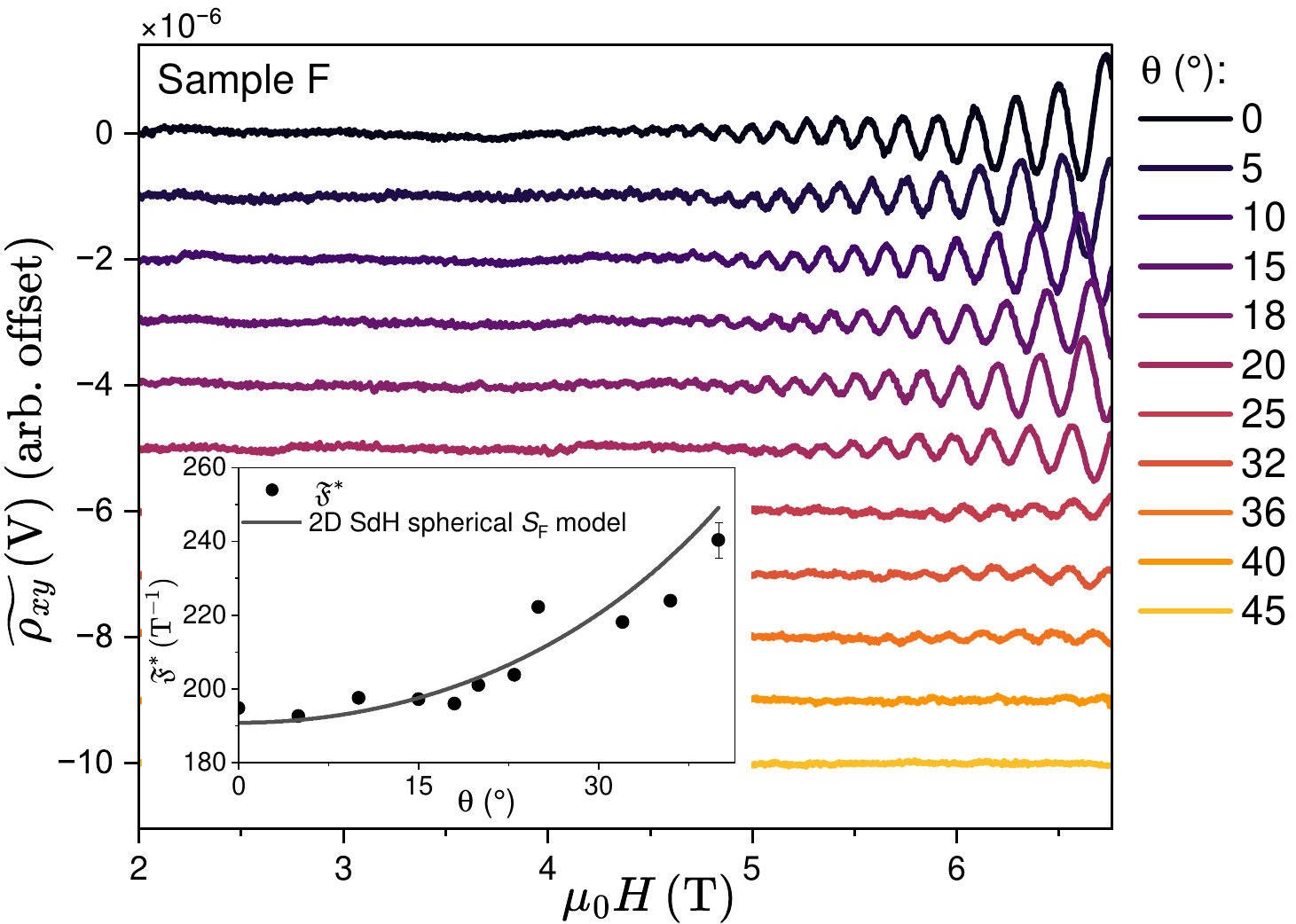}
    \caption{Oscillations in the Hall voltage $\widetilde{\rho_{xy}}$ of sample F as a function of magnetic field, applied at various angles $\theta$ to the sample surface normal. Inset: oscillation frequency $\mathfrak{F}$ as a function of $\theta$. Solid line: model resulting in $\freq(\theta=0)=(191\pm1)\,\text{T}$.}
    \label{fig_F_SdH_Hall_over_theta}
\end{figure}

Using the rotating \g{sh}, the \g{sdh} oscillations are resolved as a function of both $H$ and of the angle $\theta$ measured between $\boldsymbol{H}$ and the out-of-plane axis, as depicted in \fig \ref{fig_sh}. The $\theta$-dependent data for sample F at $2\,\text{K}$ is given in \fig \ref{fig_F_SdH_Hall_over_theta} as $\widetilde{\rho_{xy}}$ over the applied magnetic field for specific values of $\theta$. The oscillation frequency $\freq$, which is periodic in $\frac{1}{H}$, experiences an apparent increment when $\theta$ is increased: $\freq \rightarrow \frac{\freq}{\cos{\theta}}$. This conforms to a 2D \g{sdh} model, in which only the component of $\boldsymbol{H}$ along the out-of-plane axis contributes to the Landau quantization. \corr{Furthermore, the observed values of $\freq\approx200\,\text{T}$ are estimated to result in oscillation orbits larger than the thickness of the considered samples, as detailed in \secText B of the \supp.}{While it is possible, that the 2D character of the oscillations originates from an anisotropic mobility when comparing the in-plane to the out-of-plane values, geometric considerations, detailed in \secText C of the \supp, show that the real-space extent of the oscillations exceeds the flake thickness.} The uncorrected frequency $\freq^*$ is plotted in the inset to \fig \ref{fig_F_SdH_Hall_over_theta} and the solid line follows the theoretical:
\begin{equation}
    \freq^*(\theta) = \frac{\freq_0}{\cos(\theta)}
\end{equation}
where the frequency $\freq_0 = \freq(\theta=0)$ increases as a function of $\frac{1}{\cos(\theta)}$, while effects from a non-spherical Fermi surface are neglected.

\begin{figure}[htb]
    \centering
    \includegraphics[width=1\linewidth]{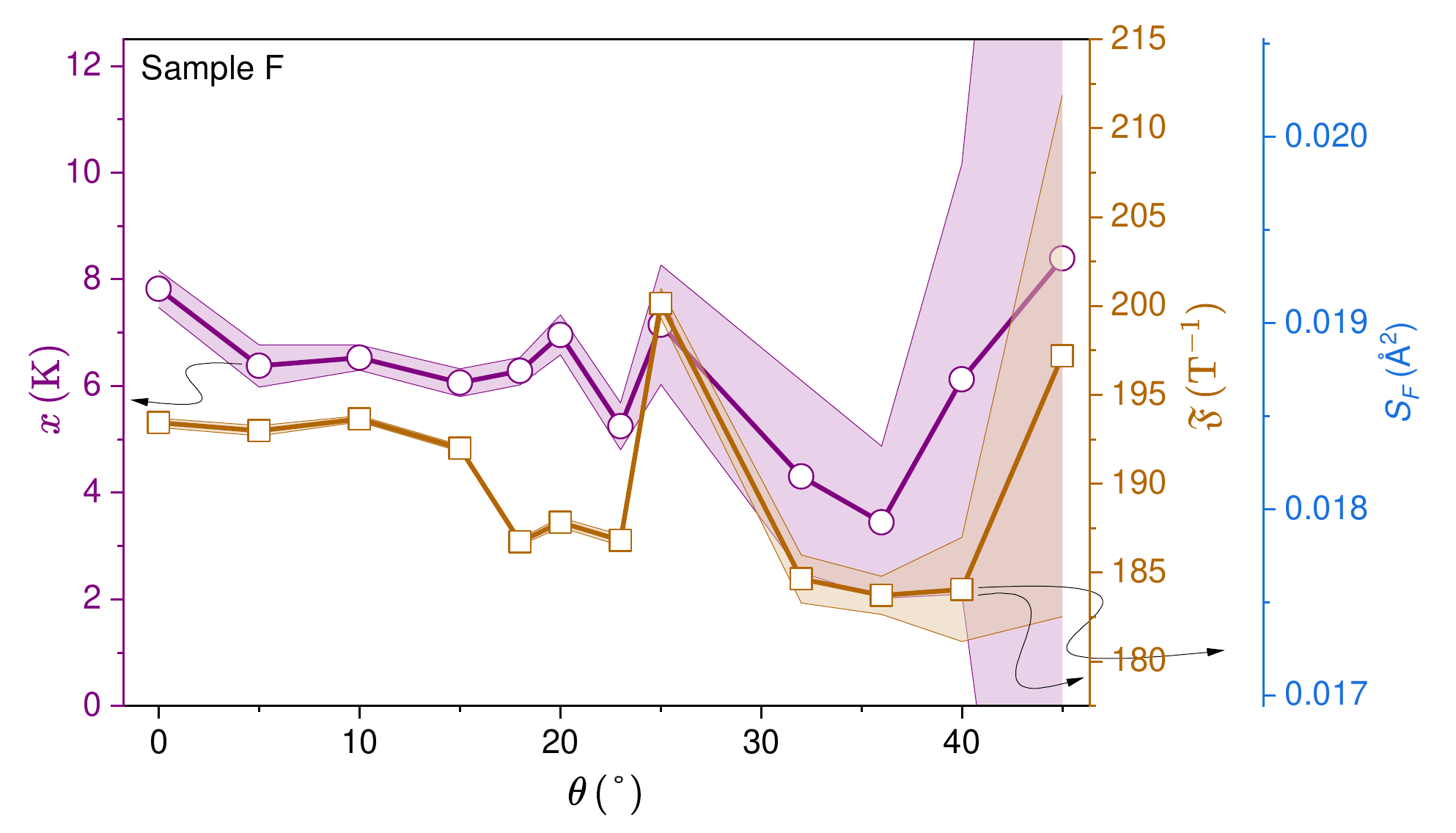}
    \caption{Dingle temperature $x$ (circles), oscillation frequency $\freq$ (squares) and extremal Fermi surface cross section $S_F$ (also squares) of sample F as a function of $\theta$. Shaded areas: error range.}
    \label{fig_F_SdH_Hall_over_theta_parameters}
\end{figure}

In order to obtain the correct values of $x$ and $\freq$, the $\widetilde{\rho_{xy}}$ data is analyzed as a function of $H_\perp = H \cos(\theta)$. The extracted parameters $x$, $\freq$ and the extremal Fermi surface cross sectional area $S_\text{F} = \freq\frac{2\pi e}{\hbar}$ as a function of $\theta$ are given in \fig \ref{fig_F_SdH_Hall_over_theta_parameters}. An equivalent plot for sample X2 is reported in \fig S12 in the \supp.

The parameters determined previously allow for the estimation of further characteristic parameters. \corr{The carrier mobility is $\mu = \frac{e \tau}{m_c} = (1200\pm100)\,\frac{\text{cm}^2}{\text{V s}}$}{The carrier mobility evaluated from $\tau$ is $\mu_q = \frac{e \tau}{m_c} = (1200\pm100)\,\frac{\text{cm}^2}{\text{V s}}$, which is significantly higher than the Hall mobility $\mu_\text{H} = \frac{l}{w\,V_{xx}} \frac{\partial V_{xy}}{\partial (\mu_0 H)} \approx 27\,\frac{\text{cm}^2}{\text{V s}}$, with $l\approx6\,\mu\text{m}$ and $w\approx1\,\mu\text{m}$ the length and width of the flake section between the contacts over which $V_{xx}$ and $V_{xy}$ are measured. This is attributed to the presence of both electron- and hole-pockets at $E_\text{F}$, which diminishes the observed Hall resistance, while the underlying charge carriers exhibit (individually) a much higher mobility.} Assuming a spherical Fermi surface, the Fermi wavevector $k_\text{F}$ is estimated from the carrier density as: $k_\text{F}^{(n)} = (3 \pi^2 n)^\frac{1}{3} \approx 0.19\,\text{Å}^{-1}$. \new{Since $n$ is derived from the Hall voltage, and assumes a single charge carrier species, $k_\text{F}^{(n)}$ may be overestimated.} Alternatively, the found value found for $S_\text{F}$ can be employed (assuming a circular Fermi surface cross section at $\theta = 0$): $k_\text{F}^{(S_\text{F})} = \left(\frac{S_\text{F}}{\pi}\right)^\frac{1}{2} \approx 0.08\,\text{Å}^{-1}$. The mean free path length is established based on $\tau$ and $m_c$ and yields $l_0^{(n)} = \frac{\hbar k_\text{F}^{(n)} \tau}{m_c} \approx 150\,\text{nm}$ or $l_0^{(S_\text{F})} = \frac{\hbar k_\text{F}^{(S_\text{F})} \tau}{m_c} \approx 80\,\text{nm}$. \new{The parameters obtained \textit{via} \g{lk} fits of samples A and F at $2\,\text{K}$ are summarized in \tab \ref{tab_LK_parameters}, and the evaluated transport parameters are tabulated in \tab \ref{tab_transport_parameters}.}  

\begin{table}
\begin{tabular}{c|c|ccc|c}
Sample & Fit                            & {$x$ (K)} & {$\freq$ (T$^{-1}$)} & {$\phi$}    & {$M$ ($m_0$)} \\ \hline
\multirow{4}{*}{F} &
  $\widetilde{\rho}_{xx}$ &
  \multirow{2}{*}{3.6(0.2)$^{[1]}$} &
  \multirow{2}{*}{192.46(0.11)$^{[1]}$} &
  0.21(0.015) &
  \multirow{5}{*}{0.321$^{[\dagger]}$} \\
       & $\widetilde{\rho}_{xy}$        &           &                      & 0.60(0.015) &               \\ \cline{2-5}
       & $\widetilde{\rho}_{xx}$        & 4.7(0.4)  & 193.1(0.3)           & 0.19(0.04)  &               \\ \cline{2-5}
       & $\widetilde{\sigma}_{xx}^{-1}$ & 3.64(0.24)& 194.19(0.18)& 0(0.024)&               \\ \cline{1-5}
A      & $\widetilde{\rho}_{xy}$        & 7(2.4)& 254.9(1.7)& -0.5(0.2)   &              
\end{tabular}
\caption{\new{Parameters obtained from \g{lk} fits at $2\,\text{K}$. $^{[\dagger]}$: Since $x$ and $M$ are highly dependent, the value $M$ is fixed to $0.321\,m_0$, which is obtained from the temperature dependence of the oscillation amplitude of sample F. $^{[1]}$: The values of $x$ and $\freq$ are constrained to be equivalent for $\widetilde{\rho}_{xx}$ and $\widetilde{\rho}_{xy}$.}}
\label{tab_LK_parameters}
\end{table}

\begin{table}[htb]
\begin{tabular}{cc|l}
\multicolumn{1}{l}{}                    & \multicolumn{1}{l|}{}                & Evaluated \textit{via}       \\ \hline
\multicolumn{1}{c|}{$\mu_q$}         & $(1200\pm100)\,\frac{\text{cm}^2}{\text{V s}}$    & $\tau$                       \\
\multicolumn{1}{c|}{$\mu_\text{H}$}                & $27\,\frac{\text{cm}^2}{\text{V s}}$                     & $V_{xx}$ and $V_{xy}$          \\
\multicolumn{1}{c|}{$n$}                & $(2\times 10^{20})\, \text{cm}^{-3}$ & Hall voltage                 \\
\multicolumn{1}{c|}{$S_\text{F}$}                & $0.02 \,\text{Å}^{-2}$                      & $\freq_0$ from $\freq(\theta)$ \\
\multicolumn{1}{c|}{$k_\text{F}^{(S_\text{F})}$} & $0.08\,\text{Å}^{-1}$                       & $S_\text{F}$                   \\
\multicolumn{1}{c|}{$k_\text{F}^{(n)}$} & $0.19\,\text{Å}^{-1}$                & $n$                          \\
\multicolumn{1}{c|}{$m_c$}              & $(0.32\pm0.02)\,m_e$                 & \eq \ref{eq_LK_T_dependence} \\
\multicolumn{1}{c|}{$\tau$}                      & $[(0.22 \pm 0.02)\times10^{-12}]\,\text{s}$ & \eq \ref{eq_tau_main}          \\
\multicolumn{1}{c|}{$l_0^{(n)}$}        & $150\,\text{nm}$                     & $k_\text{F}^{(n)}$           \\
\multicolumn{1}{c|}{$l_0^{(S_\text{F})}$}        & $80\,\text{nm}$                             & $k_\text{F}^{(S_\text{F})}$   
\end{tabular}
\caption{\new{Parameters of the charge carriers obtained by combining the analysis of the SdH oscillations and $V_\text{H}$.}}
\label{tab_transport_parameters}
\end{table}

\FloatBarrier
\subsection{Weak antilocalization}
\begin{figure}[htb]
    \centering
    \includegraphics[width=0.8\linewidth]{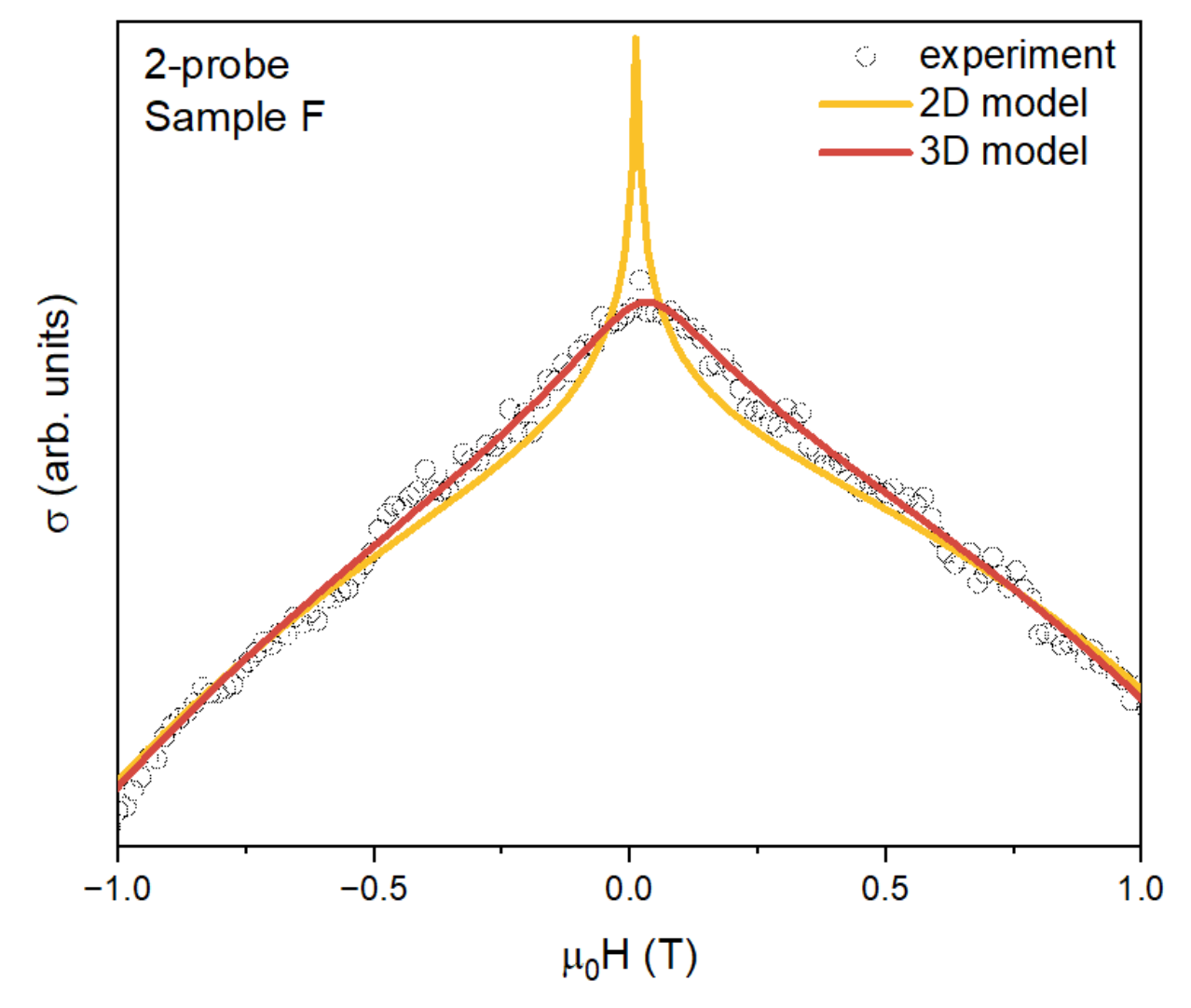}
    \caption{\new{2-terminal} conductance over applied magnetic field of sample F at $2\,\text{K}$. Models plotted as solid lines. The resulting parameters are given in \tab S3 of the \supp.}
    \label{fig_F_2terminal_3D_2D_comparison}
\end{figure}

The \new{2-terminal} conductance is obtained by measuring the voltage difference directly between source and drain. Around $H = 0$, the samples show a conductance peak, reported in \fig \ref{fig_F_2terminal_3D_2D_comparison} as conductance over applied magnetic field for sample F at $2\,\text{K}$. Similar plots are obtained for samples X1 and X2, as given in \fig S14 of the \supp. In this two-terminal configuration, the measured conductance is affected by both the \pt film and by the (metallic) Pt contacts.

The origin of the conductance peak is assigned to the \g{soc} of the charge carriers within the $12\,\text{nm}$ thin Pt contacts leading to \g{wal}: Without external magnetic field, the \g{trs} is preserved and the \g{soc} triggers a destructive interference between the backscattering paths with their time-reversed counterparts. The applied magnetic field breaks the \g{trs}, reducing the destructive backscattering interference, and thus leading to a decrease in conductance.

\begin{figure}[htb]
    \centering
    \includegraphics[width=0.85\linewidth]{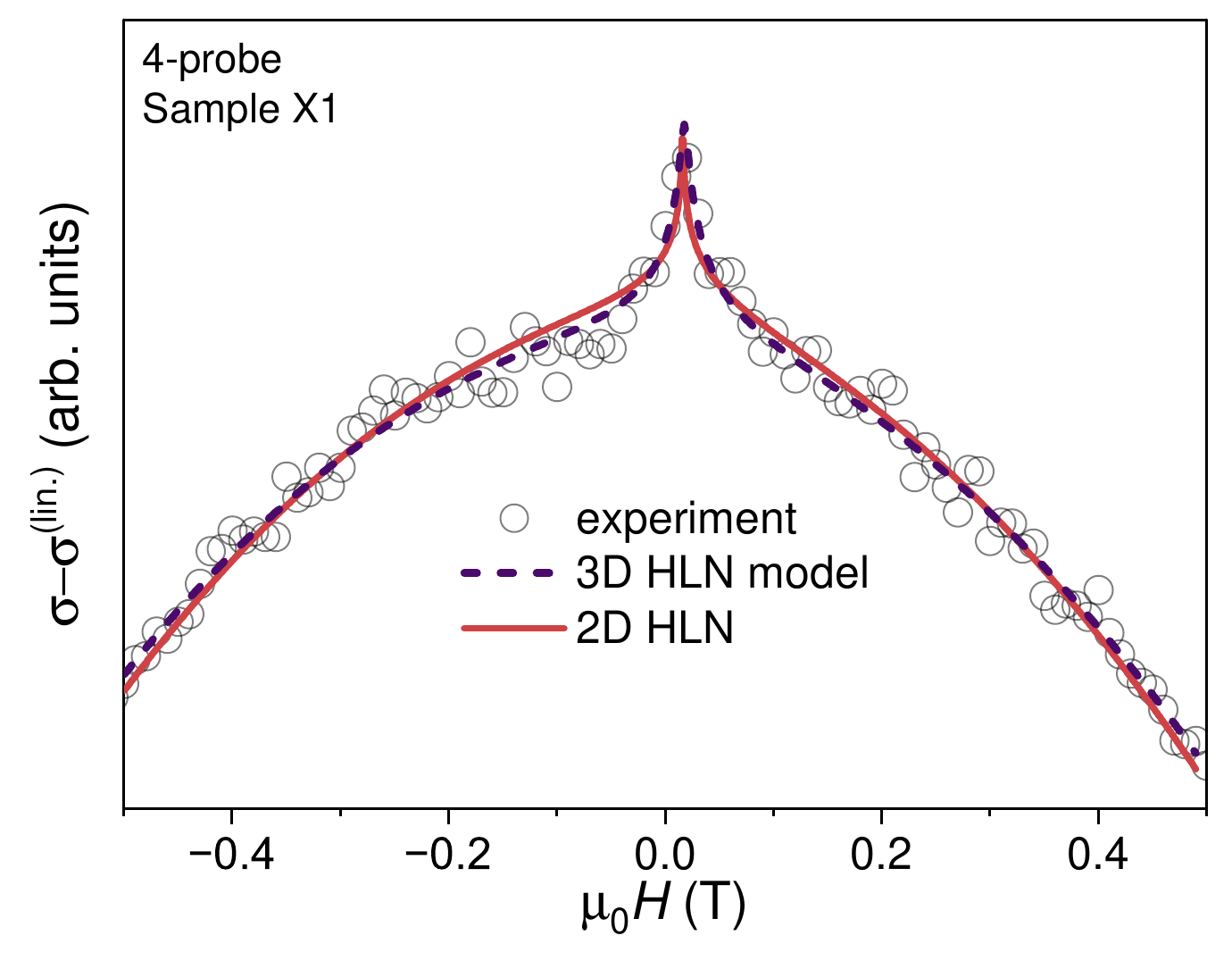}
    \caption{4--terminal conductance of sample X1 at $1.7\,\text{K}$ (Hall intermix $\sigma^\text{(lin.)}$ corrected). Comparison of the 3D \g{hln} model (dashed line) with the 2D \g{hln} model (solid line). The obtained parameters are provided in \tab S4 of the \supp.}
    \label{fig_X1_WAL_comparison}
\end{figure}

In the \new{4-terminal} conductance, sample X1 shows \g{wal}, depicted in \fig \ref{fig_X1_WAL_comparison} as conductance over applied magnetic field at $1.7\,\text{K}$. The \g{wal} remains resolvable until $T\approx 10\,\text{K}$ (\fig S15 in the \supp).

It is proposed that this \new{4-terminal} \g{wal} originates from the \g{soc} in the \pt: The Pt vacancies in \pt contribute uncompensated magnetic moments which manifest \textit{via} a Kondo effect \cite{Ouroboros}. Furthermore, it is inferred that the magnetic moments also suppress the \g{soc} in \pt. The vanishing of the Kondo effect in sample X1, correlated with the emergence of \new{4-terminal} \g{wal}, suggests that the Pt vacancy concentration varies within the bulk crystal. The Kondo effect \new{is} discussed in Appendix A, and the \g{rt} data of samples X1 and X2 are compared: \new{The Kondo effect and the \g{wal} are observed to be mutually exclusive, as one originates from the presence of Pt vacancies, while the other is inhibited by them.}\corr{Alternatively, the Pt/\pt junction may play a significant role, since it can affect the charge distribution at the Pt sites and the Pt coordination number \cite{PtPtSe2Interface}}{}.

The \g{wal} data is understood within the (2D) \g{hln} model \cite{OGHLN,Sultana2018HLN}:
\begin{equation}
    \Delta\sigma_\text{WAL}^\text{(2D)} \propto -\frac{\alpha e^2}{\pi h}\left(\ln\left(\frac{B_\phi}{H}\right)-\Psi\left(\frac{1}{2}+\frac{B_\phi}{H}\right)\right),
    \label{eq_WAL_HLN}
\end{equation}
where the magnetic field length is $B_\phi = \frac{h}{8 e \pi H l_\phi}$ ($l_\phi$ is the phase coherence length), $\Psi$ is the digamma function and $\alpha \in \left[-\frac{3}{2},-\frac{1}{2}\right]$ is a coefficient. Alternatively, a 3D extension of the \g{hln} model can be employed as in  \refsText  \cite{Nakamura_3D_HLN, SalawuWALFormula}, where it is also applied to model a Dirac system:
\begin{equation}
\begin{aligned}
    \Delta\sigma_\text{WAL}^\text{(3D)} &= A_\text{WAL} \sqrt{\frac{H}{f}}\\
    &\times\Bigg[
    2\zeta\left(\frac{1}{2},\frac{1}{2}+\frac{f}{H\, l_0^2}\right)\\
    &+\zeta\left(\frac{1}{2},\frac{1}{2}+\frac{f}{ H\, l_\phi^2}\right)\\
    &-3\zeta\left(\frac{1}{2},\frac{1}{2}+4\frac{f}{ H\, l_\text{SO}^2}+\frac{f}{H\, l_\phi^2}\right)
    \Bigg],
    \label{eq_wal_adapted}
\end{aligned}
\end{equation}
where $A_\text{WAL}$ is the magnitude of the conductance peak, $\zeta$ the Hurwitz zeta function and $f = \frac{\hbar}{4 e}=\frac{\Phi_0}{8\pi}$ a constant. There are four length scales involved: (i) The mean free path length $l_0$, (ii) the phase coherence length $l_\phi$, (iii) the spin orbit scattering length $l_\text{SO}$ and (iv) the magnetic length $l_H=\sqrt{\frac{f}{H}}$. To fit the \g{wal} in a region $|\mu_0 H| < 2\,\t{T}$, expressions for the orbital magnetoconductance and the Hall intermix are added. The \g{lk} term modeling the \g{sdh} oscillations is omitted, given that its magnitude is negligible at $|\mu_0 H| < 2\,\t{T}$. The total conductance as a function of applied magnetic field is modeled as the \g{wal} conductance taking place in parallel (additive) to the orbital magnetoconductance:
\begin{equation}
\begin{aligned}
    \sigma(H) &= \Delta\sigma_\text{WAL}^{(i)}\\
    &+ \left(\frac{1}{\sigma_0}+A_\text{MR}\,|(H +H_\text{lag})|^s+k H)\right)^{-1},
\end{aligned}
\label{eq_wal_total}
\end{equation}
with $\sigma_0$ the zero-field conductance, $A_\text{MR}$ the \g{omr} magnitude and $s$ the scaling exponent thereof. Since the highest considered temperature of $15\,\text{K}$ still lies in the range in which the system is in the Fermi liquid regime ($\approx 80\,\text{K}$), $s=2$ is chosen. The parameter $H_\text{lag}$ compensates for the lag between the nominal magnetic field and applied magnetic field, which is about $\pm16\t{mT}$ when sweeping at $8 \frac{\text{mT}}{\text{s}}$, while $k$ models the already mentioned Hall intermix. Details about the evaluation of \eq \ref{eq_wal_total} and implementation of the least-squares fit are provided in \secText D in the \supp. 

Both the 2D model and the 3D model are applied to the \new{2-terminal} conductivity of sample F at $2\,\text{K}$ (\fig \ref{fig_F_2terminal_3D_2D_comparison}). The 3D model captures the peak profile more precisely. The obtained length scales are compared in \tab S2 of the \supp. The data is comparable to the \g{wal} exhibited by Pt thin films reported in \refText \cite{PtMetalWAL}.

The 2D and the 3D model are also applied to the \new{4-terminal} \g{wal} observed in sample X1 at $1.7\,\text{K}$, depicted in \fig \ref{fig_X1_WAL_comparison} as conductance over applied magnetic field. The conductance peak shape is narrower in $H$ than the one recorded in the \new{2-terminal} measurement. The difference between the models is not as stark as for the \new{2-terminal} conductance peak. This suggests that the \new{4-terminal} \g{wal} is a distinct effect from the \new{2-terminal} \g{wal}. \new{The 3D \g{hln} fit yields $l_0 = (350\pm 160)\,\text{nm}$ and $l_\phi^\text{3D}=(350\pm120)\,\text{nm}$.}

\section{Conclusions}
Here, key aspects of the charge carriers in \pt are extracted \textit{via} the implementation of an analytic \g{lk} framework. Under application of a magnetic field, \g{sdh} oscillations, observed in the resistance measurements, occur in plane. This provides key information about the charge carriers, including the determination of the Berry phase $\Phi_\text{B} = (1.01\pm 0.05)\pi$, the cyclotron mass $m_c = (0.321\pm0.008)\,m_0$, in-plane mobility $\mu = (1200\pm100)\frac{\text{cm}^2}{\text{V s}}$ and quantum scattering time $\tau = [(0.22 \pm 0.02)\times10^{-12}]\,\text{s}$.  A spectral analysis of the oscillations yields a single species of electrons as the prevalent charge carriers, with $\freq = (194.19\pm 0.18)\,\text{T$^{-1}$}$, equating to a Fermi surface cross-section $S_\text{F} \approx 0.0185\,\text{Å$^{-2}$}$ for $t=11\,\text{nm}$. $\freq$ is reduced as the thickness is lowered, suggesting a thickness-dependence of the band structure for $8\,\text{nm}<t<20\,\text{nm}$ and potentially presenting a compelling tuning parameter for the precise electronic properties of \pt flakes and thin films. The Pt vacancies, inherently present in the grown bulk crystals, are found to play a crucial role, as they provide uncompensated magnetic moments which establish a Kondo effect, but suppress the \g{soc}, leading to a \g{wal}. While the \new{2-terminal} \g{wal} is better understood with the 3D \g{hln} model and originates from the Pt contacts, the \new{4-terminal} conductance peak also fits the 2D \g{hln} model, consistent with the 2D confinement observed from the \g{sdh} oscillations. It is concluded, that the \new{4-terminal} \g{wal} heralds substantial inherent \g{soc} in \pt. While it is plausible, that the \g{soc}, and following that, the Rashba effect leads to the observed non-zero Berry phase values, the measurements do not provide a conclusive picture of how the bulk Dirac cone influences the charge carrier properties at the Fermi level. The mechanism by which the non-trivial Berry phase emerges remains to be clarified. The employed bulk crystals provide a long-range order larger than the physical measurement area and contain Pt vacancies from growth, prompting a comparative investigation to directly grown \pt structures. The adaptation of the \g{lk} formalism is valid for a single charge carrier species with 2D oscillations. The results motivate magnetotransport studies in highly stoichiometric \pt flakes and \pt/ferromagnetic heterostructures, in order to fully harness the \g{soc} inherent in the material and demonstrate the significance of \pt for orbitronic and orbital Hall-based prospective devices.

\section*{Acknowledgments}
The authors thank J. Pe\v{s}i\'c for the fruitful discussions and her insights. This work was funded by the Austrian Science Fund (FWF) through Project No. TAI-817 and by the JKU LIT Seed funding through Project No. LIT-2022-11-SEE-119.

\section{Data availability}
The data that support the findings of this study are available from the corresponding author upon reasonable request.

\section*{Appendix A: Exposition of Kondo effect}
\begin{figure}[htb]
    \centering
    \begin{minipage}[t]{0.5\textwidth}
        \includegraphics[width=1.0\textwidth]{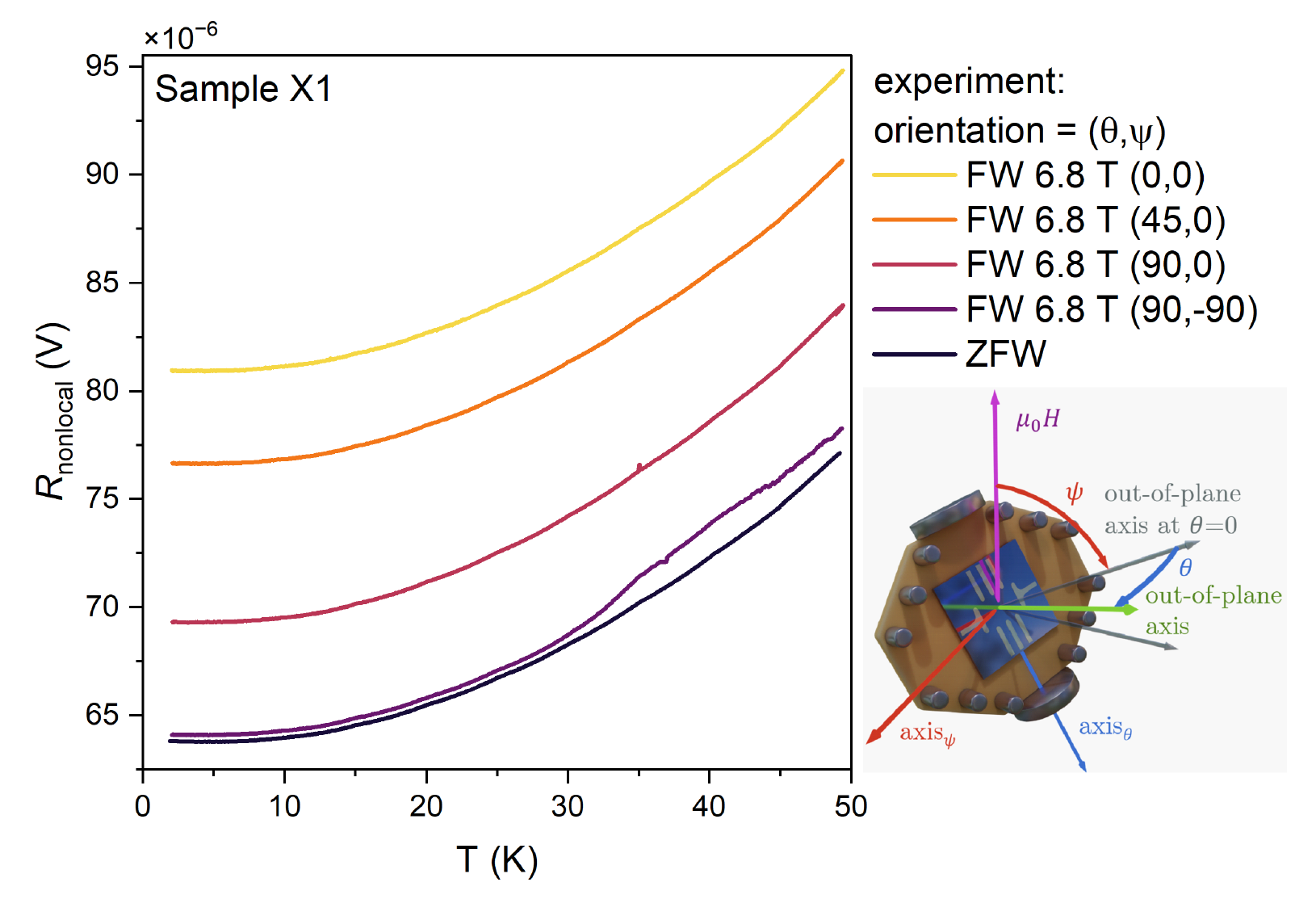}
        \caption{Left panel: \new{4-terminal} resistance over temperature of sample X1 (bulk crystal Batch 2). Right panel: sketch of how the angles $\theta$ and $\psi$ are related to the \g{sh} orientation.}
        \label{fig_X1_RT}
    \end{minipage}
\end{figure}

\begin{figure}
    \begin{minipage}[t]{0.4\textwidth}
        \includegraphics[width=1.0\textwidth]{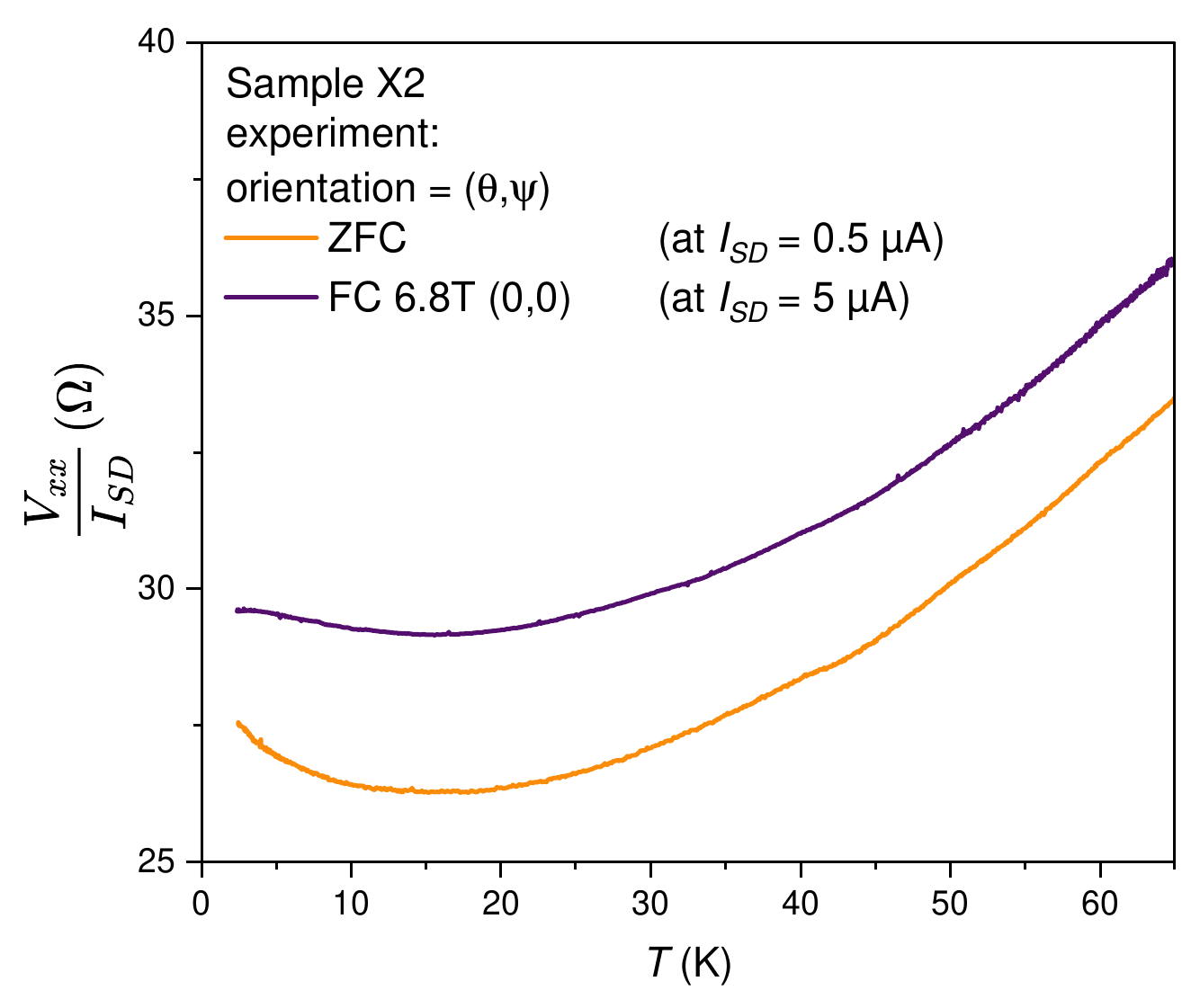}
        \caption{\corr{Nonlocal resistance over temperature of sample X2 (bulk crystal Batch 2). The ZFC signal is magnified by a factor of ten because it was recorded at one tenth the source-drain current.}{4-terminal resistance over temperature of sample X2 (bulk crystal Batch 2), normalized by the applied source-drain current.}}
        \label{fig_X2_RT}
    \end{minipage}
\end{figure}

Although both sample X1 and X2 are exfoliated from bulk crystal Batch 2, their \g{rt} behaviour is qualitatively different: The \new{4-terminal} resistance over temperature curves (\g{fw} and \g{zfw}) of sample X1 are given in \fig \ref{fig_X1_RT} and the \g{fc} and \g{zfc} curves of sample X2 are provided in \fig \ref{fig_X2_RT}. The expression $(\theta,\psi)$ describes the orientation of the \g{sh} in relation to the applied magnetic field: $\psi$ is the angle by which the \g{sh} is rotated with respect to axis$_\psi$, which is normal to $\boldsymbol{H}$; $\theta$ is the angle of orientation about the axis$_\theta$, which is normal to axis$_\psi$ and encloses an angle $(\pi/2+\psi)$ with $\boldsymbol{H}$. The angles and their respective axes are sketched in the right panel of \fig \ref{fig_X1_RT}. The curves for sample X1 are distinguished mainly by a temperature-independent offset, which is caused by the \g{omr}. The \g{omr} is strongest when the applied magnetic field is oriented along the sample surface normal, corresponding to an orientation $(0,0)$. In the $(90,0)$ orientation, the applied magnetic field lies in plane and normal to the source-drain direction. The resulting \g{omr} is reduced due to the charge carriers being confined to the sample thickness of $17\,\text{nm}$. Considering the structural anisotropy of \pt, the out-of-plane mobility can be assumed to be lower than the in-plane value, which can also account for the reduced in-plane \g{omr}. In the $(90,-90)$ orientation, the applied magnetic field lies in plane and parallel to the source-drain direction. Almost no \g{omr} is observed, since the Lorentz force is zero. The origin of the additional resistance for $T\gtrsim 35\,\text{K}$ still is unclear. The \g{zfw} shows the lowest resistance because no \g{omr} applies. As $T\to 0$, no increase in the resistance $R$ is observed.
This is in contrast to all other considered samples, which display a global minimum resistance temperature of $\approx 15\,\text{K}$ and an increase in resistance as the temperature is lowered below $T\lesssim15\,\text{K}$. This property of samples A, B and C is discussed in a previous work \cite{Ouroboros} and can be gleaned in \fig \ref{fig_X2_RT} for sample X2. The anomalous increase in resistance as $T\to 0$ is mediated by Pt vacancies, which contribute an uncompensated spin when located in the top- or bottom-most layers. At or below the Kondo temperature $T_\text{K}$, this results in a Kondo effect: The charge carriers scatter at the vacancy spin \textit{via} a double antiferromagnetic spin exchange, resulting in an increase in resistance as the temperature is lowered further.

\FloatBarrier
\onecolumngrid
\section*{Appendix B: Adaptation of the Lifshitz-Kosevich formula\label{sec_lk_derive}}
The oscillatory part of the thermodynamic potential of the electrons due to quantum oscillations $\widetilde{\Omega}$ is given by \cite{LK, Shoenberg}:
\begin{align}
    \widetilde{\Omega} &= \left(\frac{e}{2\pi c \hbar}\right)^{3/2} \frac{2kTH^{3/2}V}{(S_\text{F}'')^{1/2}}\sum_{p=1}^\infty \frac{\exp(-2\pi^2pkx/(\beta H)) \cos(\frac{1}{2}p\pi g \frac{m}{m_0})}{p^{3/2}\sinh(2\pi^2 pkT/(\beta H))}\cos\left(2\pi p\left(\frac{\freq}{H}-\frac{1}{2}\right)\pm\frac{\pi}{4}\right)\\
    &\propto H^{3/2}\sum_{p=1}^\infty R_D R_T R_S \mathfrak{O}
    \label{eq_OMEGA}
\end{align}
with $c$ the speed of light, $\hbar$ the reduced Planck quantum, $k$ the Boltzmann constant, $V$ the sample volume, $p$ the harmonic number, $x$ the Dingle temperature, $\beta = \frac{e \hbar}{m}$, $m$ the electron cyclotron mass, $g$ the Landé g-factor, $m_0$ the electron rest mass and $\freq$ the oscillation frequency. The extremal cross-sectional area of the Fermi surface perpendicular to $\boldsymbol{H}$ is $S_\text{F}$ and $\kappa$ is the component of $\boldsymbol{k}$ parallel to $\boldsymbol{H}$. The parameter $S_\text{F}''$ describes the curvature of $S_\text{F}$ with respect to $\kappa$:
\begin{equation}
    S_\text{F}'' = \frac{\text{d}^2S_\text{F}}{\text{d}\kappa^2},\quad
    \kappa = \frac{\boldsymbol{k}\cdot\boldsymbol{H}}{|\boldsymbol{H}|}.
    \label{eq_FermiSurfCrossSection}
\end{equation}
When considering the conductivity oscillations $\widetilde{\sigma}$ instead of $\widetilde{\Omega}$, the same expressions for the Dingle damping $R_D=\exp(-2\pi^2pkx/(\beta H))$, the thermal smearing $R_T=T \sinh(2\pi^2 pkT/(\beta H))^{-1}$ and the spin phase interference $R_S = \cos(\frac{1}{2}p\pi g \frac{m}{m_0})$ are applicable. The conductivity does however scale differently under application of a magnetic field:  $\widetilde{\sigma}$ is related to the variation of the density of states at the Fermi energy $\widetilde{\mathscr{D}}$  \cite{Shoenberg}:
\begin{equation}
     \frac{5}{2}\frac{\widetilde{\mathscr{D}}}{\mathscr{D}_0}+\frac{3}{2}\left(\frac{\widetilde{\mathscr{D}}}{\mathscr{D}_0}\right)^2,
    \label{sigmaFromDOSosc}
\end{equation}
where $\mathscr{D}_0$ is the steady density of states at the Fermi energy (without applied magnetic field). The ratio $\frac{\widetilde{\mathscr{D}}}{\mathscr{D}_0}$ relates to the applied magnetic field as:
\begin{align}
    \frac{\widetilde{\mathscr{D}}}{\mathscr{D}_0} &= \left(\frac{H}{2 \freq}\right)^{1/2},\\
    \frac{\widetilde{\sigma}}{\sigma} &\stackrel{\insertArrow}{\propto} \left(\frac{5}{2}\left(\frac{H}{2 \freq}\right)^\frac{1}{2}+\frac{3}{2}\left(\frac{H}{2 \freq}\right)\right).
\end{align}
The higher harmonic numbers $p>1$ are omitted since they are not observed in the considered samples (a single peak is obtained in the \g{fft}). The expression for the phase of the oscillations in $\cos\left(2\pi \left(\frac{\freq}{H}-\frac{1}{2}\right)\pm\frac{\pi}{4}\right)$ is obtained for a 3D case, wherein the offset $\delta = \frac{\pi}{4}$ results from an integration over $\kappa$ \cite{Shoenberg}. This offset is zero in the 2D case. The phase offset $\phi = \frac{1}{2}$ is expanded to also consider a potentially non-zero Berry curvature $\Phi_\text{B}$\cite{BuschBi2Se3SdH}:
\begin{align}
    2\pi \left(\frac{\freq}{H}-\frac{1}{2}\right)\pm\frac{\pi}{4} \equiv 2\pi \left(\frac{\freq}{H}-\phi\right),\\
    \phi = \left(\frac{1}{2}-\frac{\Phi_\text{B}}{2\pi}\pm\delta\right)\mod 1,\quad \delta^\text{(2D)}=0.\\
    \iff \Phi_\text{B} = 2\pi\left(\frac{1}{2}-\phi\right)\mod 2\pi
\end{align}
Thus, for $\phi=\frac{1}{2}$ the Berry phase is zero, while for $\phi=0$ the Berry phase is $\pm \pi$. The expression for the oscillation in the 2D case is $\cos\left(2\pi \left(\frac{\freq}{H}+\phi\right)\right)$.
\eq \ref{eq_OMEGA} is adapted to describe the oscillatory part of $\rho_d$ by keeping the behavior $\propto R_D R_T R_S \mathfrak{O}^\text{(2D)}$, while also considering that the origin of $\widetilde{\sigma}$ (and hence $\widetilde{\rho_d}$) lies in the modulation of $\mathscr{D}$:
\begin{align}
    \widetilde{\rho}_d &= A R_D R_T R_S \mathfrak{O}\left(\frac{5}{2}\frac{\widetilde{\mathscr{D}}}{\mathscr{D}_0}+\frac{3}{2}\left(\frac{\widetilde{\mathscr{D}}}{\mathscr{D}_0}\right)^2\right)\\
    \widetilde{\rho}_d&\stackrel{\insertArrow}{=}A T \left(\frac{5}{2}\left(\frac{H}{2 \freq}\right)^\frac{1}{2}+\frac{3}{2}\left(\frac{H}{2 \freq}\right)\right) \frac{\exp\left(-x \frac{NM}{H}\right) \cos(\pi M)}{\sinh\left(T \frac{NM}{H}\right)}\cos\left(2\pi \left(\frac{\freq}{H}+\phi\right)\right).
    \label{eq_LK_rho_inSupp}
\end{align}
To account for the \new{4-terminal} transport measurements, the amplitude $A$ is introduced. Several constants are included into $A$. The parameter $M=\frac{m_c}{m_0}$ is the fermion cyclotron mass ratio. The value of $N = 2 \pi^2\frac{k\,m_0}{\hbar\, e} \approx 14.693\,\frac{\text{K}}{\text{T}}$ is determined by substituting the values for the constants.
\section*{Appendix C: Extraction of the quantum transport time from the Lifshitz-Kosevich formula}
The quantum transport time $\tau$ is related to the Dingle temperature \textit{via} $x=\frac{\hbar}{2\pi k \tau}$, and is extracted from \eq \ref{eq_LK_rho_inSupp}: At a local extrema $\frac{\partial^2 \widetilde{\rho_d}(H)}{\partial H^2}=0$, the phase factor $|\mathfrak{O}^\text{(2D)}|=1$. A relative oscillation magnitude $\frac{\widetilde{\rho_d}}{\rho_d^\text{(bg.)}}$ is introduced to obtain a dimensionless expression (the value of $A$ changes accordingly). Let $\mathscr{H} = \frac{5}{2}\left(\frac{H}{2 \freq}\right)^\frac{1}{2}+\frac{3}{2}\left(\frac{H}{2 \freq}\right)$ be the magnetic field dependence of the oscillation amplitude. This yields:
\begin{equation}
    \frac{\widetilde{\rho_d}}{\rho_d^\text{(bg.)}} = A T \mathscr{H}\frac{\exp(-x \frac{NM}{H}) \cos(\pi M)}{\sinh(T \frac{NM}{H})},
    \label{eq_tau_derive_1}
\end{equation}
which is solved for $x$:
\begin{align}
    \iff \frac{\widetilde{\rho_d}}{\rho_d^\text{(bg.)}} \sinh\left(T \frac{NM}{H}\right) \mathscr{H}^{-1}\frac{1}{AT \cos(\pi M)}&\propto \exp\left(-x \frac{NM}{H}\right)\label{eq_tau_derive_2}\\
    \iff \ln\left(\frac{\widetilde{\rho_d}}{\rho_d^\text{(bg.)}} \sinh\left(T \frac{NM}{H}\right)\mathscr{H}^{-1}\right)-\ln(A T \cos(\pi M))&\propto -x \frac{NM}{H},
	\label{eq_tau_derive_3}
\end{align}
and taking the derivative in $\frac{1}{H}$ gives:
\begin{align}
    \frac{\partial}{\partial\frac{1}{H}} \ln\left(\frac{\widetilde{\rho_d}}{\rho_d^\text{(bg.)}} \sinh\left(T \frac{NM}{H}\right) \mathscr{H}^{-1}\right) &= -x N M.\label{eq_Dingle_temp}
\end{align}
Now putting $x=\frac{\hbar}{2\pi k \tau}$ and re-substituting $N = 2 \pi^2\frac{k\, m_0}{\hbar e}$ the right-hand term becomes:
\begin{equation}
    x N M = \frac{\pi\, m_c}{e \,\tau}.
\end{equation}
On the left-hand side of \eq \ref{eq_Dingle_temp} , the expression inside the logarithm can be summarized as:
\begin{equation}
    \mathfrak{G}_d \equiv \frac{\widetilde{\rho_d}}{\rho_d^\text{(bg.)}} \sinh{\left(N M\frac{T}{H}\right)}\mathscr{H}^{-1}.
    \label{eq_tau_G}
\end{equation}
The expression for $\tau$ becomes:
\begin{equation}
\tau = \left(\frac{e}{-m_c \pi}\frac{\partial}{\partial\frac{1}{ H}} \ln{\mathfrak{G}_d} \right)^{-1}.
\end{equation}
The second term in $\mathscr{H}$ can be neglected for $\freq\gg H \Rightarrow \left(\frac{H}{2 \freq}\right)^\frac{1}{2} \gg \left(\frac{H}{2 \freq}\right)$ , yielding a simplified expression for $\mathfrak{G}_d$:
\begin{equation}
\longrightsquigarrow{\eq\ref{eq_tau_derive_1}-\ref{eq_tau_G}}\,\mathfrak{G}_d \equiv \frac{\widetilde{\rho_d}}{\rho_d^\text{(bg.)}} \sinh{\left(N M\frac{T}{H}\right)}\frac{1}{\sqrt{H}}.
\end{equation}
\twocolumngrid
\bibliography{bib}

\onecolumngrid
\section*{Supplemental Material for \titleText}


\figureS
\tableS
\equationS


\subsection{Berry phase determination from the Lifshitz-Kosevich fit}
\new{Since, with magnetic field of up to $6.8\,\text{T}$, the lowest resolvable \g{ll} is $n_\text{LL}=29$, the linear extrapolation to $n_\text{LL}=0$ performed in the Landau fan diagram acquires a significant error. Therefore, the \g{lk} relation is employed to directly fit the data in order to substantiate the obtained value of the Berry phase. To ensure the comparability of the techniques, the inverse conductance $\sigma_{xx}^{-1}$ is fitted. The resulting value is $\Phi_\text{B}^{(\text{LK})} = (1.01\pm 0.05)\pi$.}
\begin{figure}[htb]
    \centering
    \includegraphics[width=0.45\linewidth]{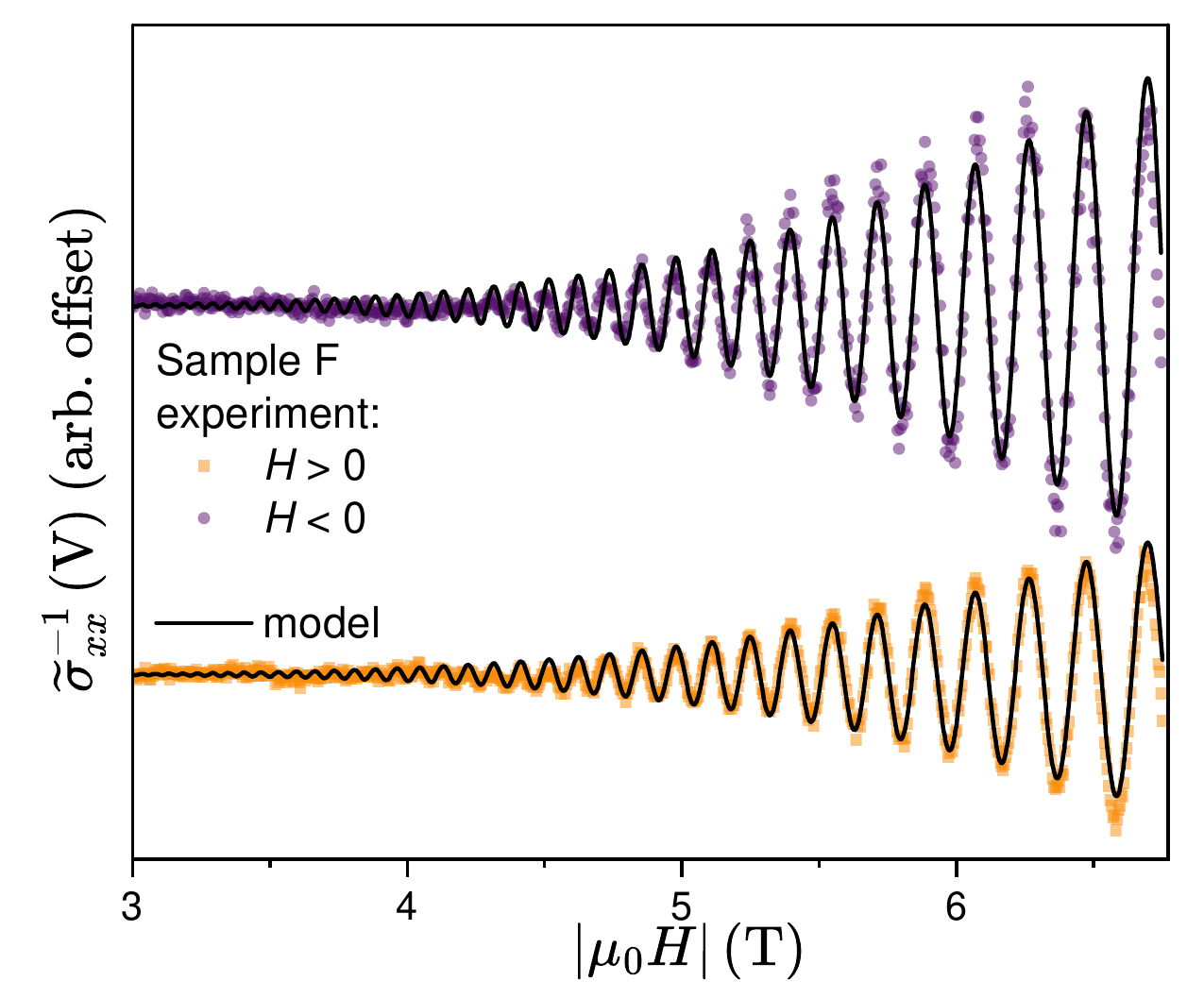}
    \caption{\new{Resistance oscillations of sample F over applied magnetic field magnitude. \g{lk} fit of the inverse conductance $\sigma_{xx}^{-1}$ to substantiate the $\Phi_B$ value. Stacked graph, showing the oscillations for both the positive and the negative field. The obtained parameters are provided in \tab II from the main text.}}
    \label{fig_F_inverseConductance_LK}
\end{figure}
\FloatBarrier

\subsection{Data treatment for the FFT\label{sec_fft_data_treatment}}
The data $\rho_d$ acquired by sweeping the applied magnetic field is almost evenly spaced in $H$. However, the \gls{lk} formula and, by extension, the data are periodic in $\frac{1}{H}$. Thus, to facilitate the \gls{fft}, evenly spaced data in $\propto\frac{1}{H}$ is needed. This is achieved by resampling the data as $R(\frac{1}{ H},T=\text{const.})$ into bins of $(2\times10^{-4})\frac{1}{\text{T}}$. The resampling procedure gives empty bins as $H\rightarrow 0$, which would limit the support of the \gls{fft} and, in turn, result in a low frequency resolution. To remedy this, values for empty bins are linearly interpolated.

\new{For sample E, the $\rho_{xy}$ data is resampled into bins of $10\,\text{mT}$ and averaged according to $R_{xy}^\text{(avg.)}(H) = \frac{R_{xy}(H)-R_{xy}(-H)}{2}$ prior to transforming the data to be evenly spaced in $\frac{1}{H}$ as described above. The resulting \g{fft} exhibits an improved signal-to-noise ratio.} 
\subsection{Estimation of the fermion cyclotron radius \label{orbitEstimation}}
In order to substantiate the 2D character of the \g{sdh} oscillations, the fermion cyclotron radius $r_c$ (in real space) is estimated by assuming that $S_\text{F}$ is the area of a circular Fermi surface cross section:
\begin{align}
    r_c = \frac{\hbar k_\text{F}}{\mu_0 H e}.
\end{align}
The Fermi wavevector $k_\text{F}$ is related to $S_\text{F}$:
\begin{align}
    k_\text{F} = \sqrt{\frac{S_\text{F}}{\pi}},\quad
    r_c \stackrel{\insertArrow}{=} \sqrt{\frac{2\hbar}{e}}\frac{\sqrt{\freq}}{\mu_0 H}.
\end{align}
The smallest value of $r_c$ in this work is achieved with $\mu_0 H = 7\,\text{T}$ and $\freq = 193\,\text{T}$. The resulting cyclotron radius is $r_c \approx 72\,\text{nm}$ and the diameter is $d_c \approx 144\,\text{nm}$. This is significantly larger than the thickness of the considered samples, which range from $8\,\text{nm}$ to $26\,\text{nm}$. Therefore, the obtained values of $\freq$ cannot result from oscillations around an in-plane axis.
\subsection{Implementation of the weak antilocalization fitting}
\label{sec_WAL_implementation}
\begin{figure}[htb]
    \centering
    \includegraphics[width=0.4\linewidth]{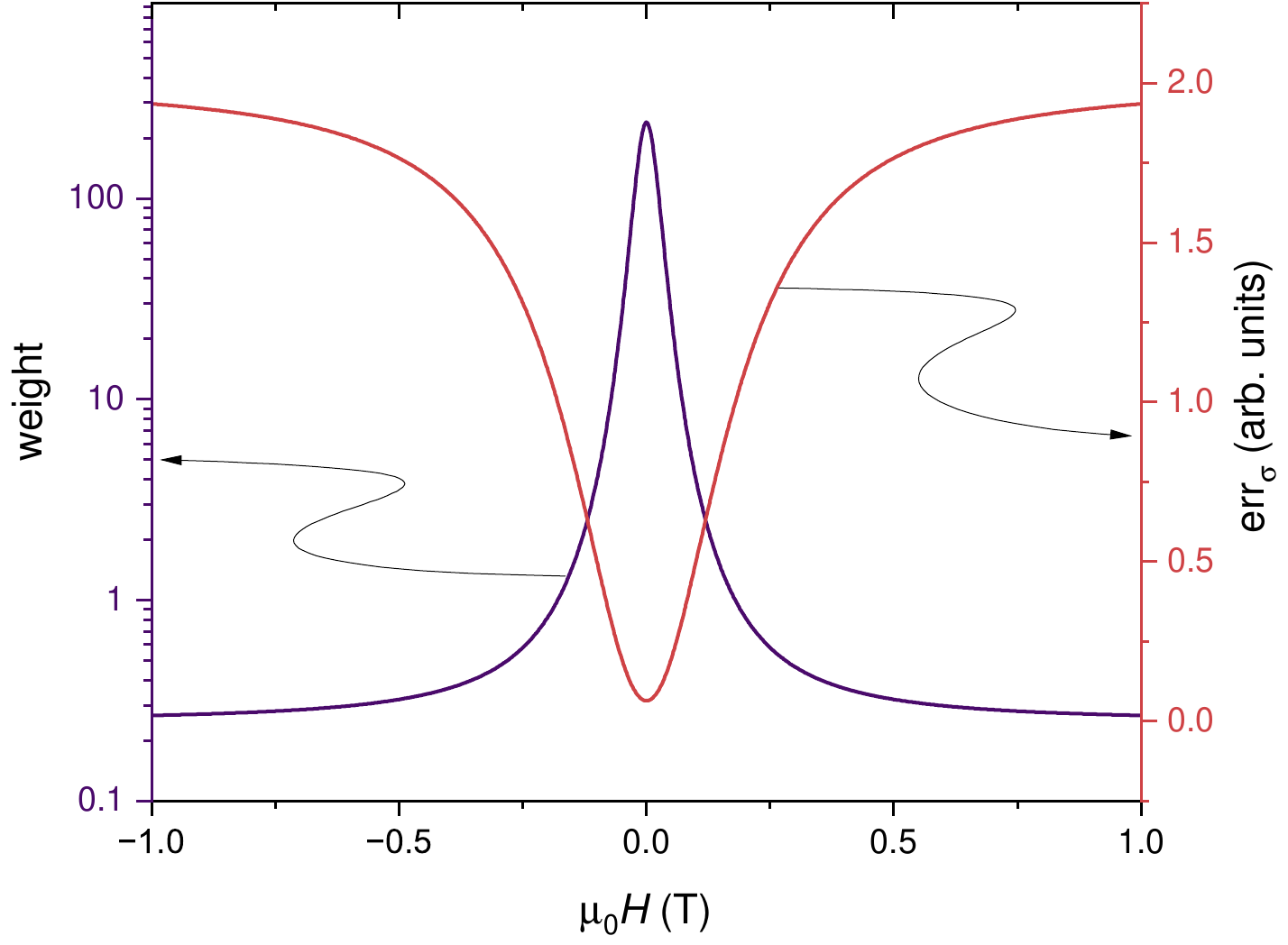}
    \caption{Assigned weight $w$ (log-scale) and $\text{err}_\sigma$
as function of applied magnetic field for $W=30\,\text{T$^{-2}$}$ for the least-squares fitting of data containing a \g{wal} peak around $H=0$.}
\label{fig_wal_weights}
\end{figure}

The implementation of the Hurwitz $\zeta$ function $\zeta\left(s,a\right)= \sum_{n=0}^\infty (n+a)^{-s}$ is challenging, due to the fact, that it cannot be evaluated analytically, since the sum $\sum_{n=0}^\infty (n+a)^{-s}$ diverges to $+\infty$ for $s=\frac{1}{2}$ and $a\geq\frac{1}{2}$. Mathematica \cite{Mathematica} calculates the sum \textit{via} analytic continuation and is used to generate $1000$ precomputed values for $s=\frac{1}{2}$ and $a\in[0.1;1000]$ (logarithmically spaced). The fitting function interpolates between these precomputed values to evaluate $\zeta$. The least-squares fit of the \g{wal} is hindered by the fact that the \g{wal} peak occurs only over a limited range of $H$ and the sum of squares is not significantly affected by an improper \g{wal} peak fit. This is remedied by assigning lower error $\text{err}_\sigma$ to the data points around $H=0$: 
\begin{equation}
    \text{err}_\sigma(H)\propto\frac{1}{1+\frac{W}{(W\,H)^2+1}},\quad W\in\mathbb{R}_+
\end{equation}
The error is related to the weighting $w$ of the data points during the least-squares fit \textit{via}:
\begin{align}
    w(H) = (\text{err}_\sigma(H))^{-2},\quad
\end{align}
where $\text{err}_\sigma(H)$ is the error in $\sigma$. The error $\text{err}_\sigma$ and corresponding weight $w$ for $W=30\,\text{T$^{-2}$}$ are depicted in \fig \ref{fig_wal_weights} as a function of the applied magnetic field.
\FloatBarrier
\subsection{X-ray photoemission spectroscopy}
\begin{figure}[htb]
    \centering
    \includegraphics[width=0.35\linewidth]{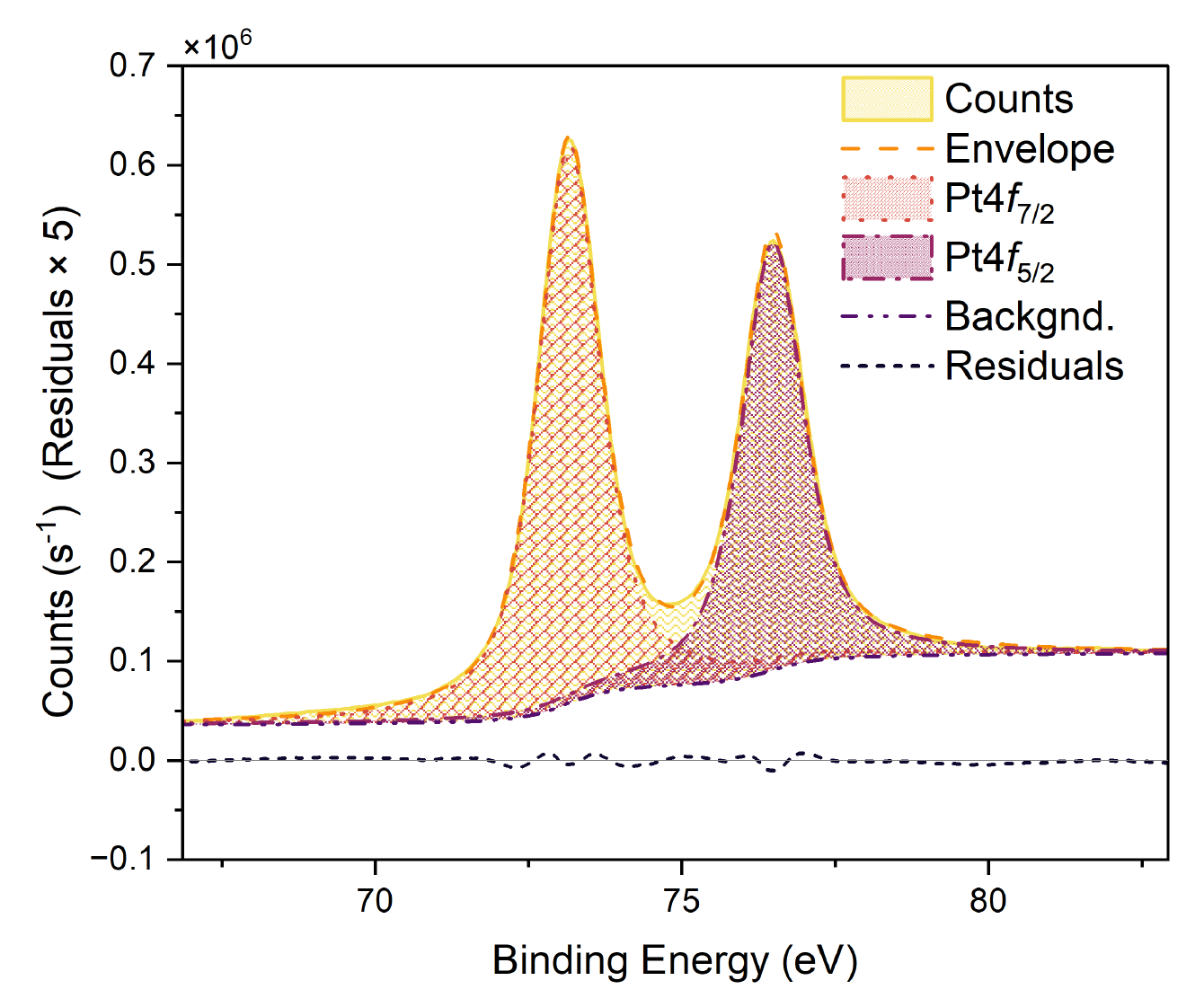}
     \includegraphics[width=0.35\linewidth]{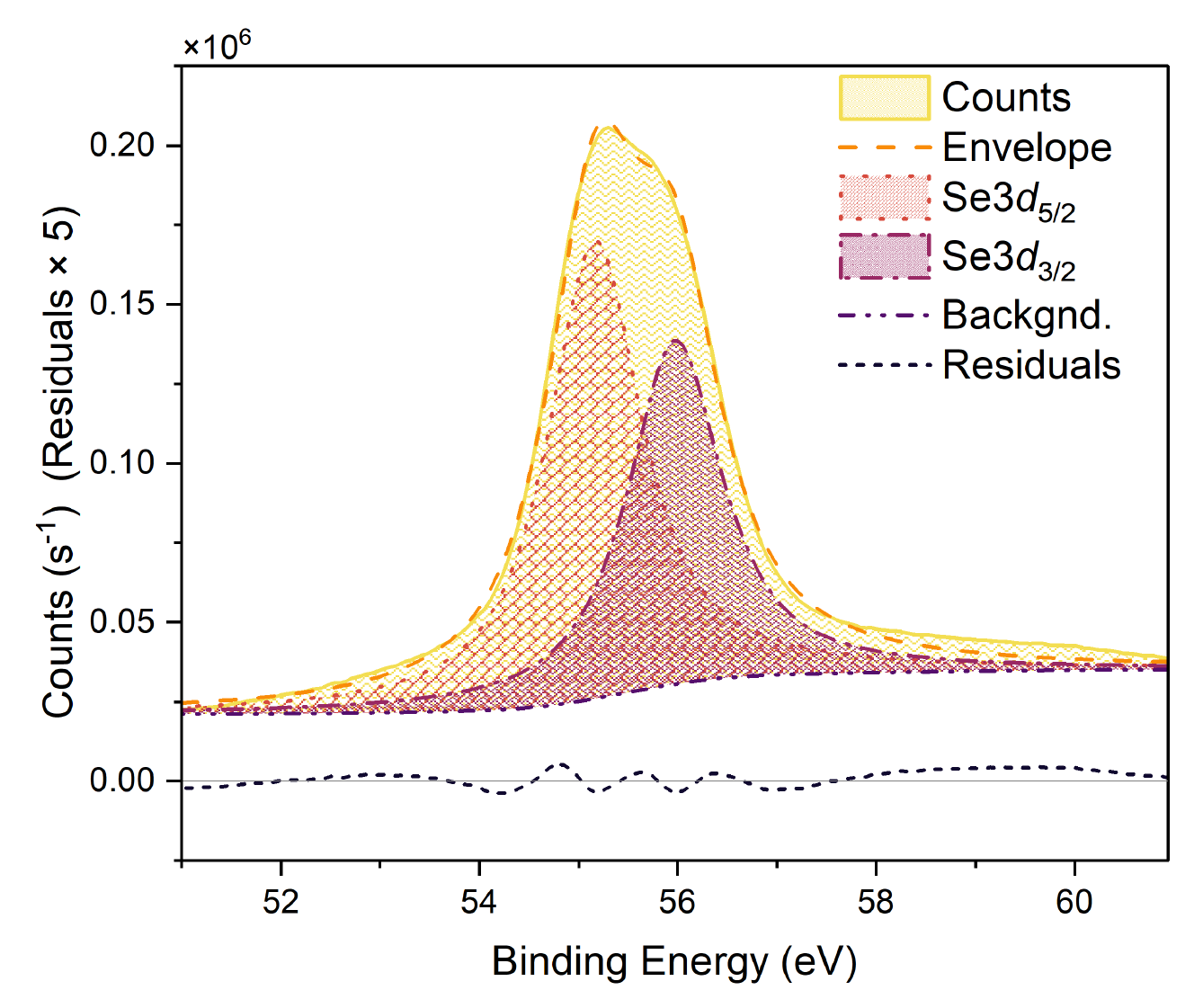}
    \caption{\g{xps} data as counts over binding energy for Pt4\textit{f} (left) and Se3\textit{d} (right). The obtained parameters are given in \tab \ref{tab_XPS}.}
    \label{fig_XPS}
\end{figure}

Flakes from Batch 2 of \pt are measured \textit{via} \g{xps} in order to investigate their chemical composition. The obtained spectra of the Pt4\textit{f} and Se3\textit{d} ranges are depicted in \fig \ref{fig_XPS} as \g{cps} over \g{be} and are comparable to those of the \corr{first Batch}{Batch 1} of \pt \cite{Ouroboros}. The peak parameters are reported in \tab \ref{tab_XPS}, where L/G mix refers to the Lorentzian/Gauss ratio of the Voigt peak shape. A constraint is in place which sets the \g{fwhm} and the L/G mix of spin-split peaks to the same value. The height ratio is given with respect to the doublet peak with lower binding energy. 
\FloatBarrier
\subsection{Supplemental figures\label{sec_supp_fig}}
\begin{figure}[htb]
    \centering
    \includegraphics[width=8cm]{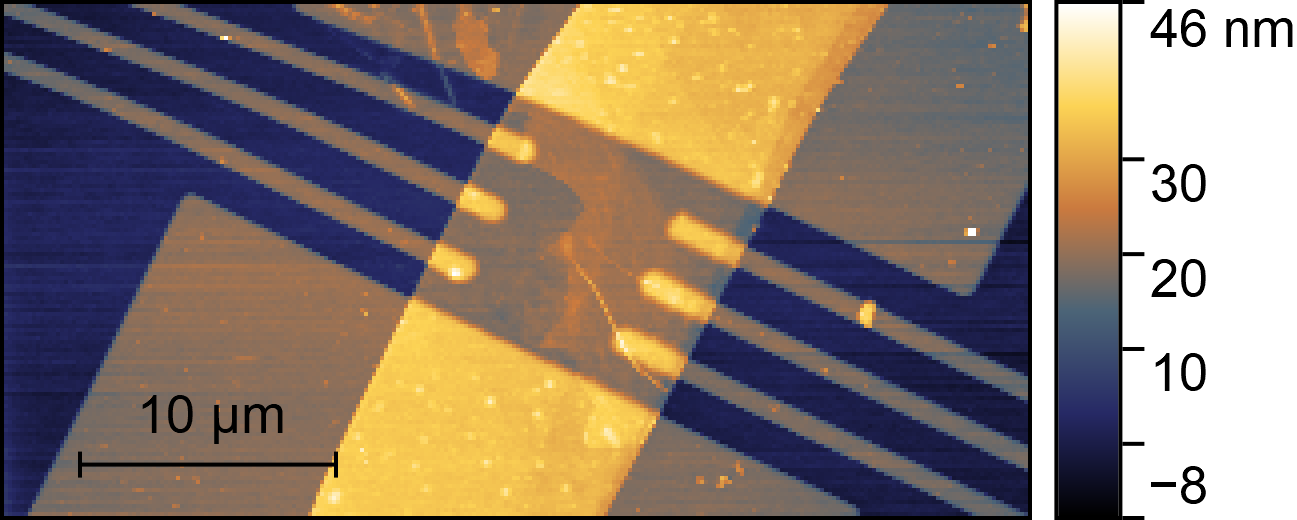}
    \caption{\g{afm} image of a flake exfoliated from bulk crystal Batch 2 on Pt contacts.}
    \label{fig_AFM}
\end{figure}

\begin{figure}[h]
        \centering
        \includegraphics[height=4.5cm]{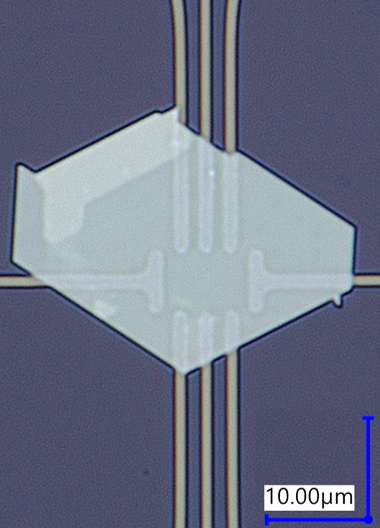}
        \includegraphics[height=4.5cm]{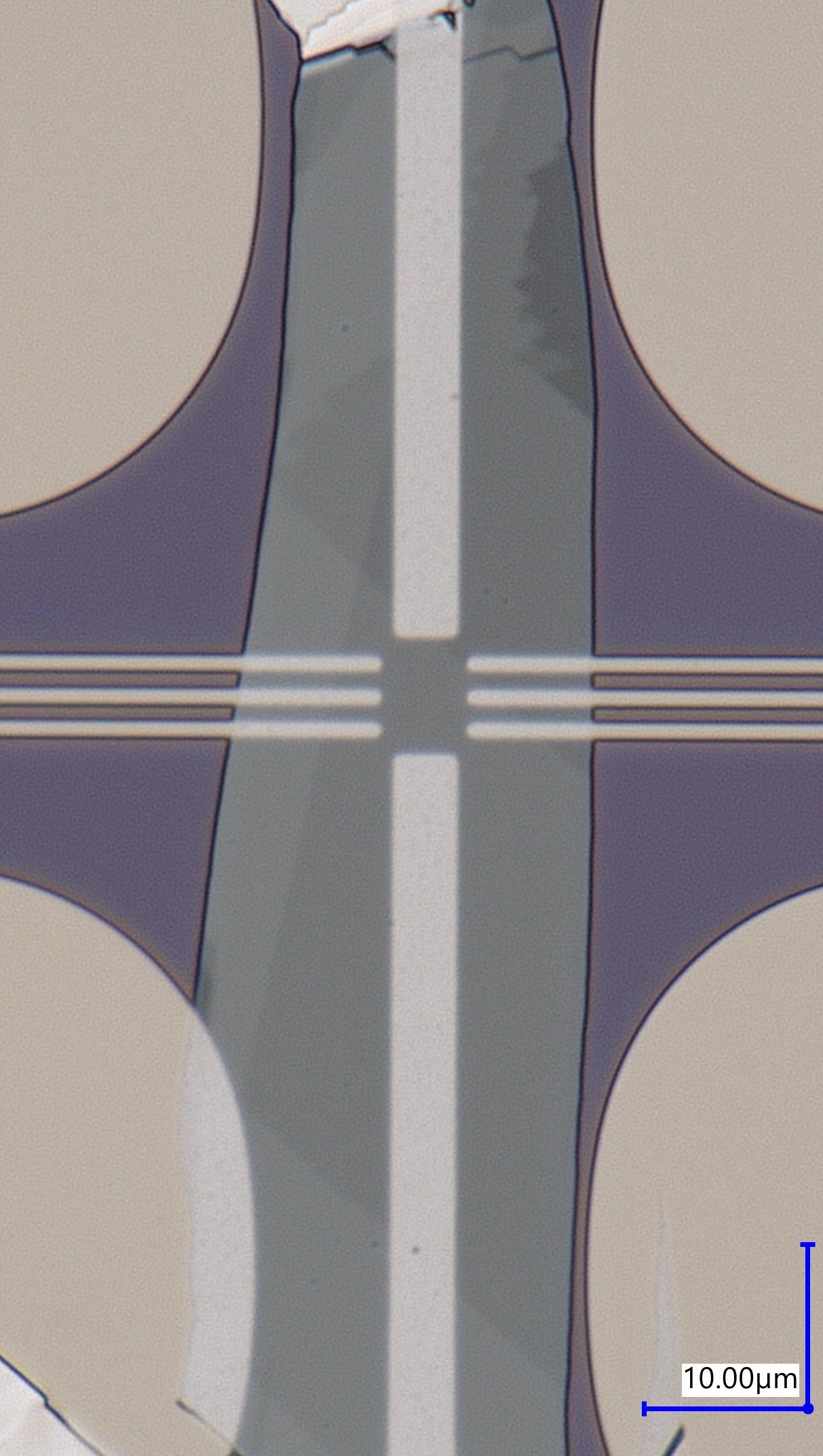}
        \includegraphics[height=4.5cm]{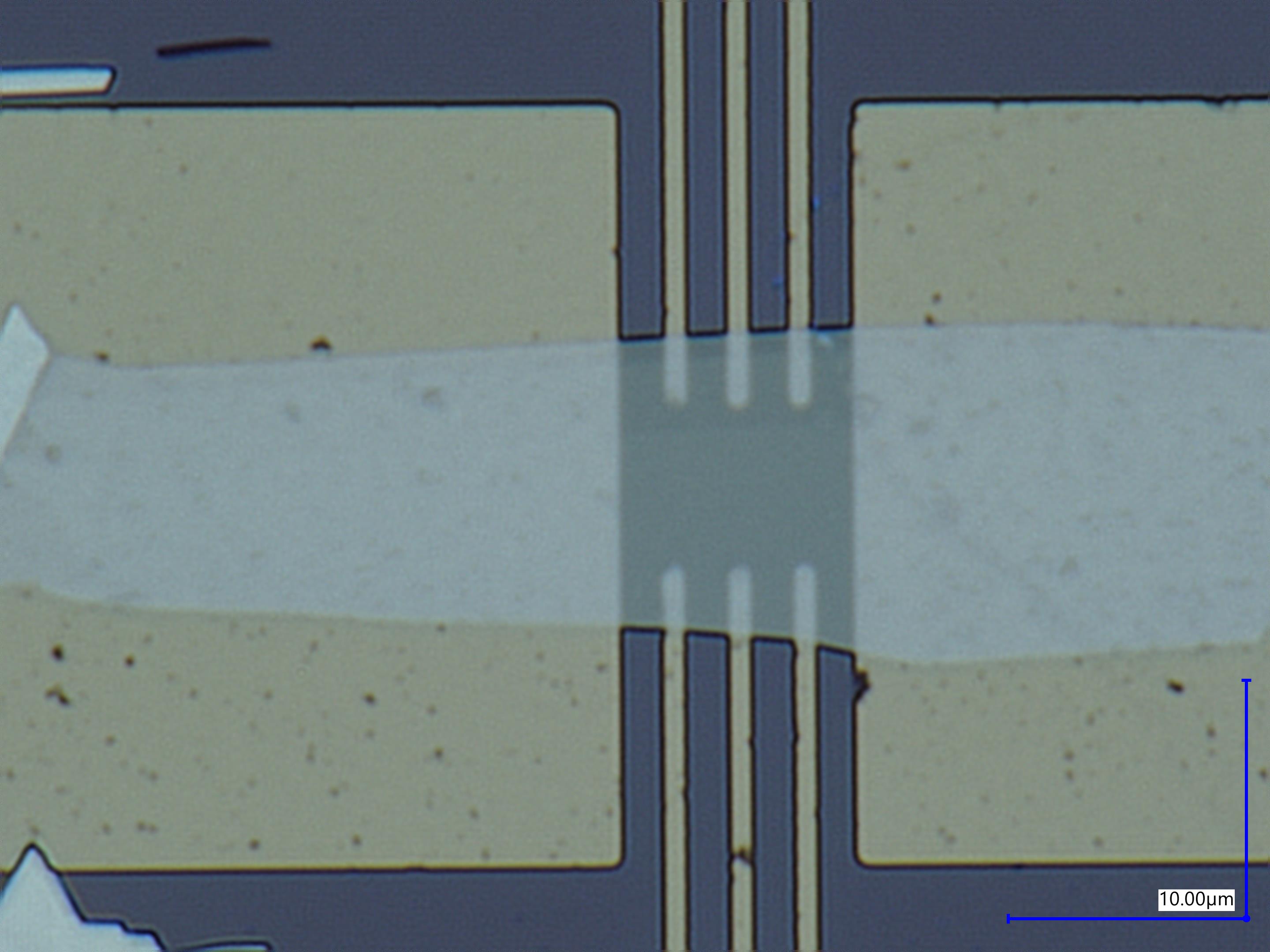}
        \caption{\new{Representative optical images of considered samples. Left to right: Samples B, D and X2.}}
        \label{fig_OM_supp}
\end{figure}

The topography of a \pt flake placed onto the Pt contacts is established by \g{afm} and depicted in \fig \ref{fig_AFM}. 
\FloatBarrier
\new{Representative optical images of samples B, F, and X2 are provided in \fig \ref{fig_OM_supp}. The illumination conditions vary across the samples.}

\begin{figure}[htb]
    \centering
    \includegraphics[height=6cm]{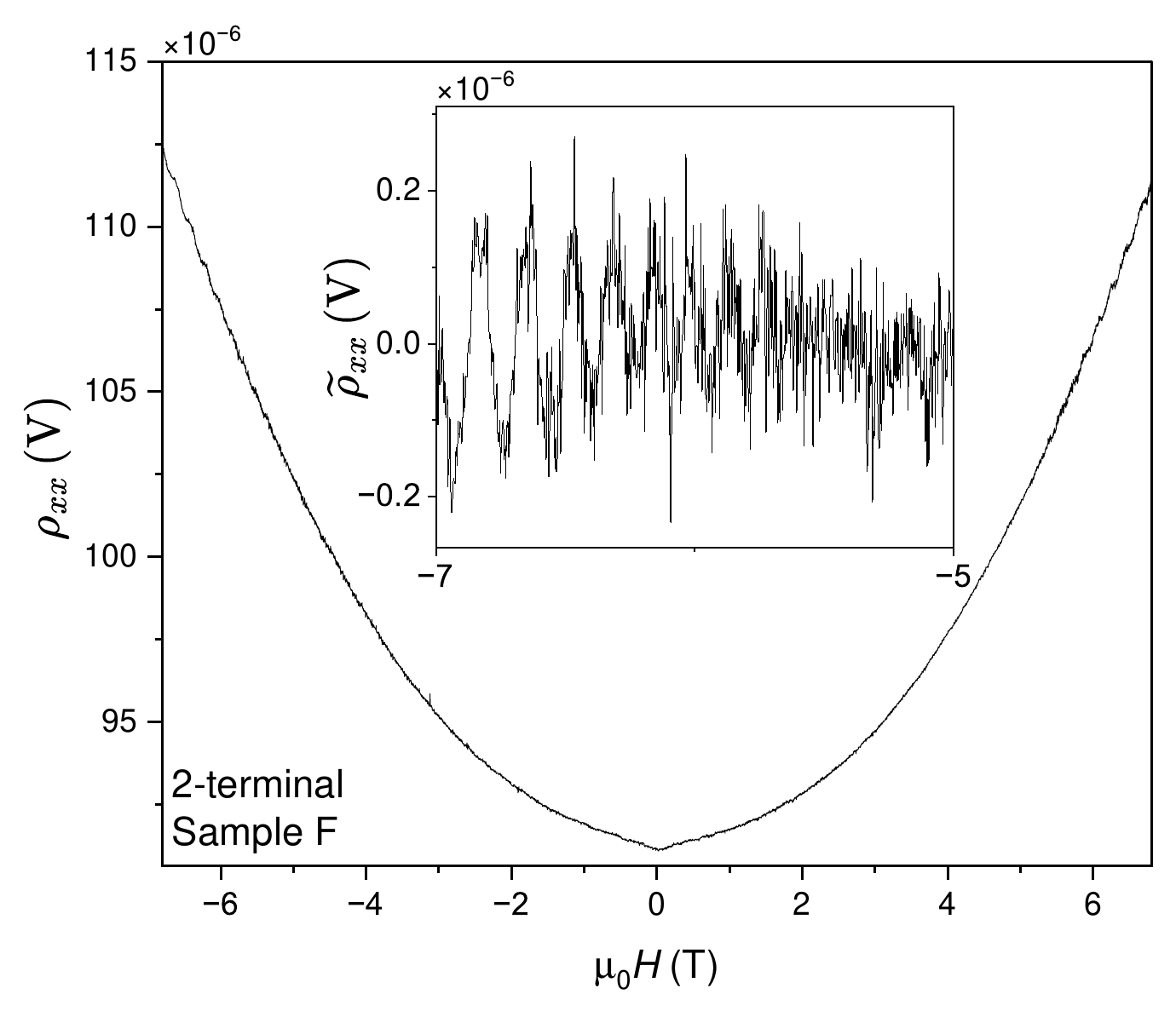}
    \caption{\new{2-terminal} $\rho_{xx}$ as a function of $H$ for sample F at $2\,\text{K}$. Inset: \gls{sdh} oscillations $\tilde{\rho}_{xx}$ for $\mu_0 H \leq 5\,\text{T}$.}
    \label{fig_Rasa_MR_2terminal}
\end{figure}

\begin{figure}[htb]
    \centering
    \includegraphics[height=6cm]{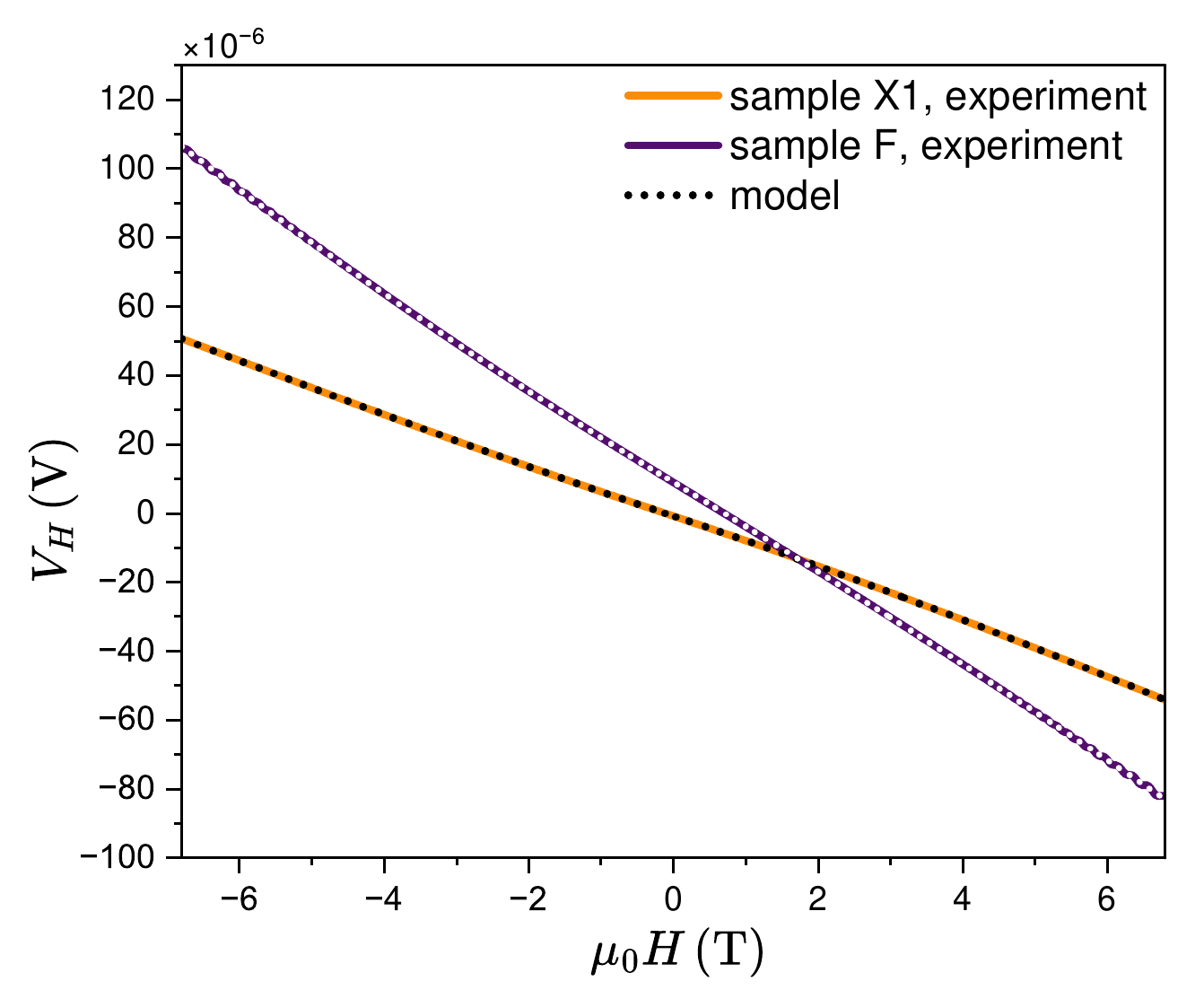}
    \caption{Hall voltage as a function of applied magnetic field for samples F and X1 at $2\,\text{K}$. Dotted lines: model \new{according to \eq \ref{eq_Hall}}. The resulting parameters are reported in \tab \ref{tab_Hall}.}
    \label{fig_Hall_supp}
\end{figure}

To exclude the \new{4-terminal} setup of the measurements as source of the observed oscillatory behavior, \fig \ref{fig_Rasa_MR_2terminal} exhibits the \new{2-terminal} resistivity $\rho_{xx}$ (source = high voltage and drain = low voltage) as a function of the applied magnetic field for sample F at $2\,\text{K}$. The inset shows the oscillations $\tilde{\rho}_{xx}$ for $\mu_0 H \leq 5\,\text{T}$. A \gls{fft} spectrum thereof puts $F\approx 195\,\text{T}$, consistent with the \new{4-terminal} value.
In \fig \ref{fig_Hall_supp}, the Hall voltage as a function of the applied magnetic field for samples F and X1 at $2\,\text{K}$ is reported. The employed fit is:
\begin{equation}
    A_0+A_1 \mu_0 H+A_2 |\mu_0 H|^2+A_m \text{sgn}(H) |\mu_0 H|^m, \quad \{A_i,m\} \in \mathbb{R}^+, \,m > 1,
    \label{eq_Hall}
\end{equation}
with $A_i$ positive prefactors and $\text{sgn}(.)$ denoting the signum function. The linear slope component $A_1$ is relevant for the extraction of $n$. The parameters $A_0$ and $A_2$ account for the longitudinal transport intermix. $A_m$ models a general non-linear Hall mechanism with power $m$. \g{sdh} oscillations are not considered, as they are not relevant for establishing the proper $A_1$ value. The extracted values for $n$ are $n_\text{F} = [(2.243\pm0.002)\times10^{20}]\,\text{cm}^{-3}$ and $n_\text{X1} = [(2.670\pm0.001)\times10^{20}]\,\text{cm}^{-3}$ respectively. The given errors are solely the ones originating from the fit, neglecting errors made by assuming \eq \ref{eq_Hall}.
\FloatBarrier
\begin{figure}[htb]
    \centering
    \includegraphics[height=6cm]{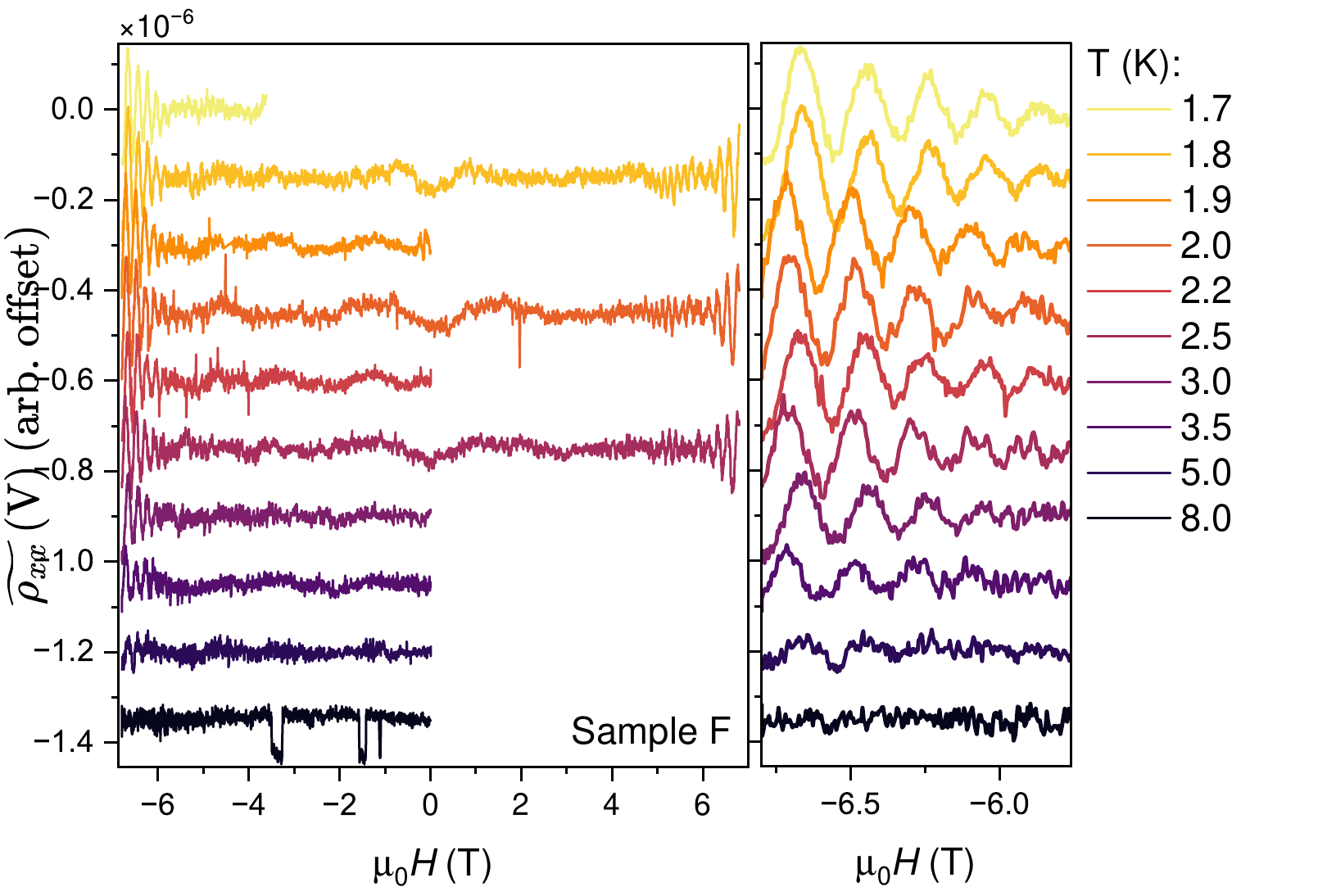}
    \caption{Oscillation in the longitudinal voltage $\widetilde{\rho_{xx}}$ of sample F as a function of applied magnetic field at specific temperatures. The lag of the applied magnetic field $H_\text{lag}$ is not corrected. Left panel: full range. Right panel: $H<-5.5\,\text{T}$.}
    \label{fig_F_osc_xx}
\end{figure}
\begin{figure}[htb]
    \centering
    \includegraphics[height=6cm]{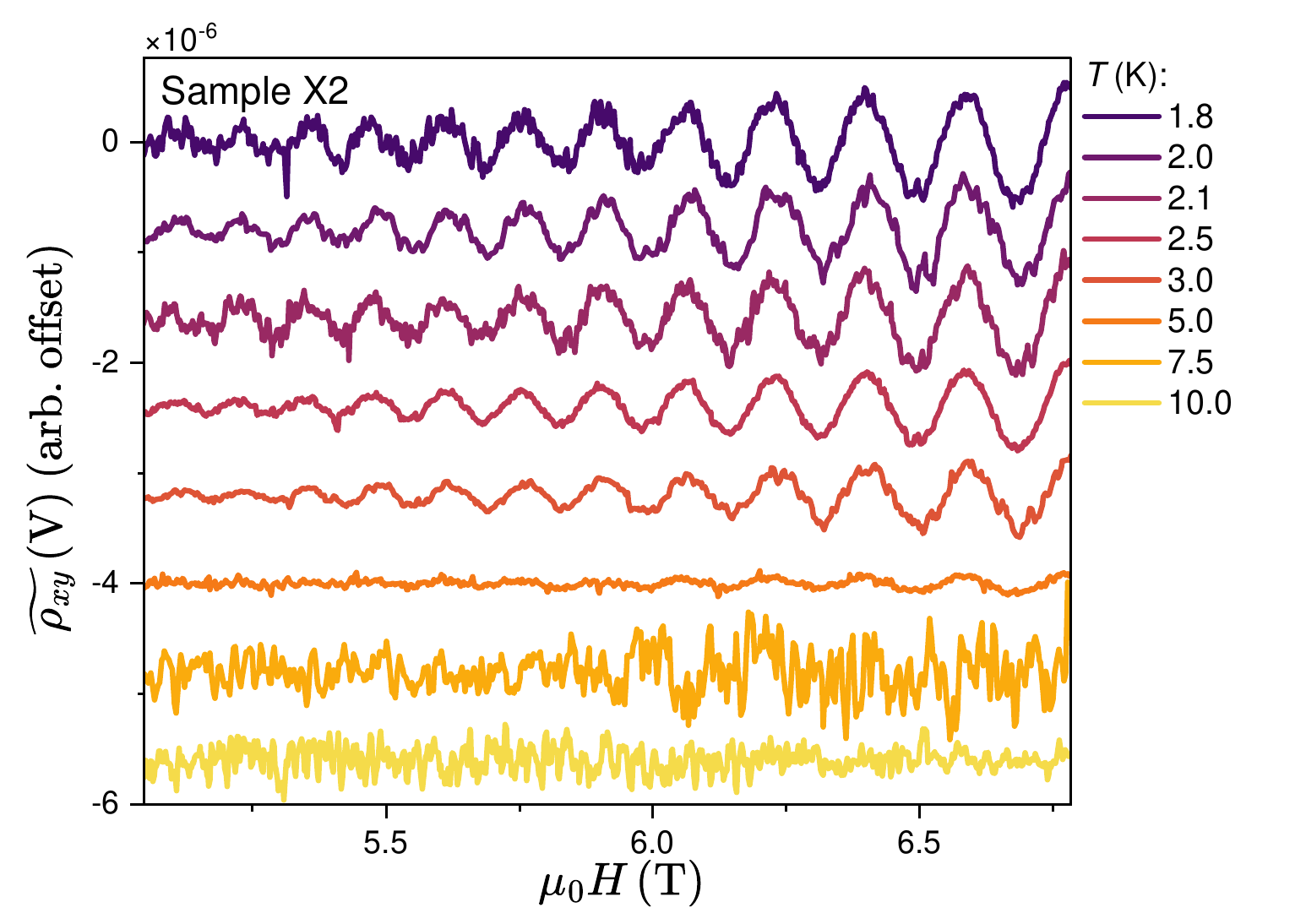}
    \caption{Oscillation in the Hall voltage $\widetilde{\rho_{xy}}$ of sample X2 as a function of applied magnetic field at specific temperatures.}
    \label{fig_X2_osc_xy}
\end{figure}
\begin{figure}[htb]
    \centering
    \includegraphics[height=6cm]{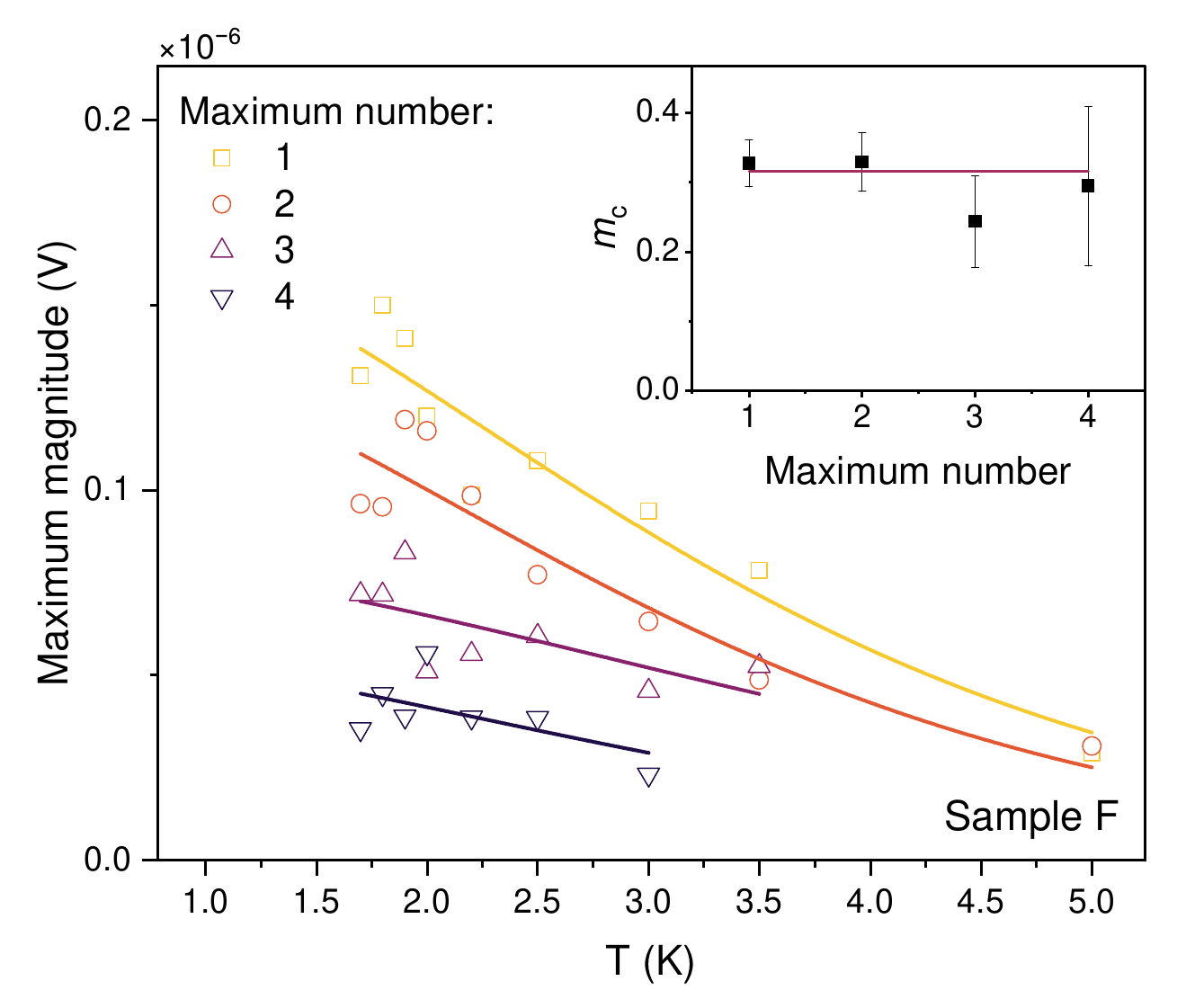}
    \caption{Determination of $m_c$ \textit{via} $\widetilde{\rho_{xx}}$ of sample F: fit of maxima at magnetic fields corresponding to integer \g{ll} over $T$. Solid lines: model. Inset: obtained values of $m_c^{(i,\,xx)}$ and weighted average (solid line). The magnitude of the first maximum (number 1) refers to the maximum in $\widetilde{\rho_{xx}}$ at the highest applied magnetic field amplitude $|\mu_0H|\approx 6.7\,\text{T}$. The subsequent maxima magnitudes (numbers 2 to 4) are found at reduced $|\mu_0H|$.}
    \label{fig_F_lambda_and_mc_xx}
\end{figure}
\begin{figure}[htb]
    \centering
    \includegraphics[height=6cm]{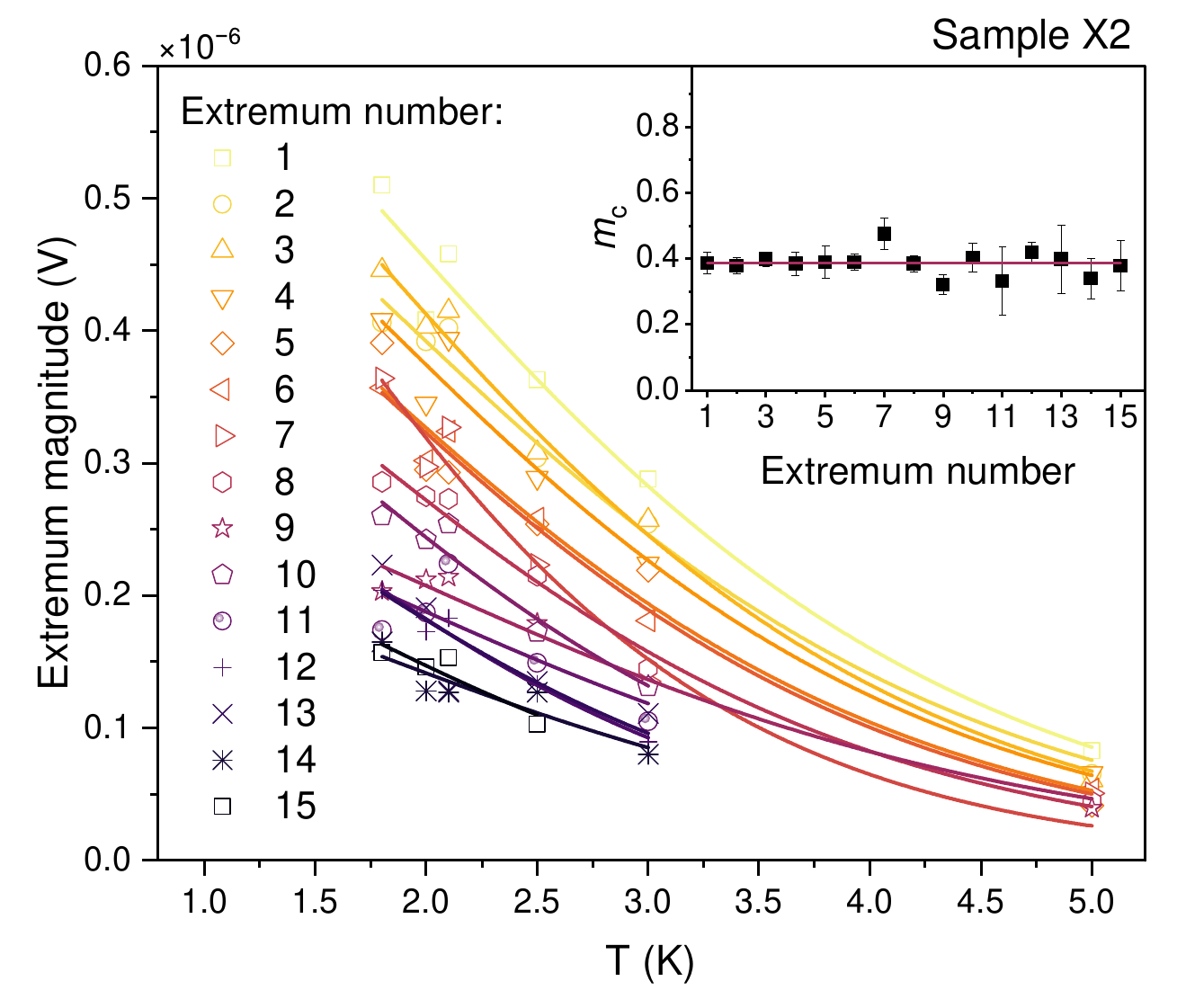}
    \caption{Determination of $m_c$ \textit{via} $\widetilde{\rho_{xy}}$ of sample X2. Extrema magnitudes over $T$. Solid lines: model. Inset: obtained values of $m_c^{(i,\,xx)}$ and weighted average (solid line). To enhance the determination of $m_c$, both minima and maxima are considered. 
The magnitude of the first extremum (number 1) refers to the extremum in $\widetilde{\rho_{xx}}$ at the highest applied magnetic field amplitude $|\mu_0H|\approx 6.7\,\text{T}$. The subsequent extrema magnitudes (number 2 to 15) are found at reduced $|\mu_0H|$. }
    \label{fig_X2_lambda_and_mc_xy}
\end{figure}
\begin{figure}[htb]
    \centering
    \includegraphics[height=6cm]{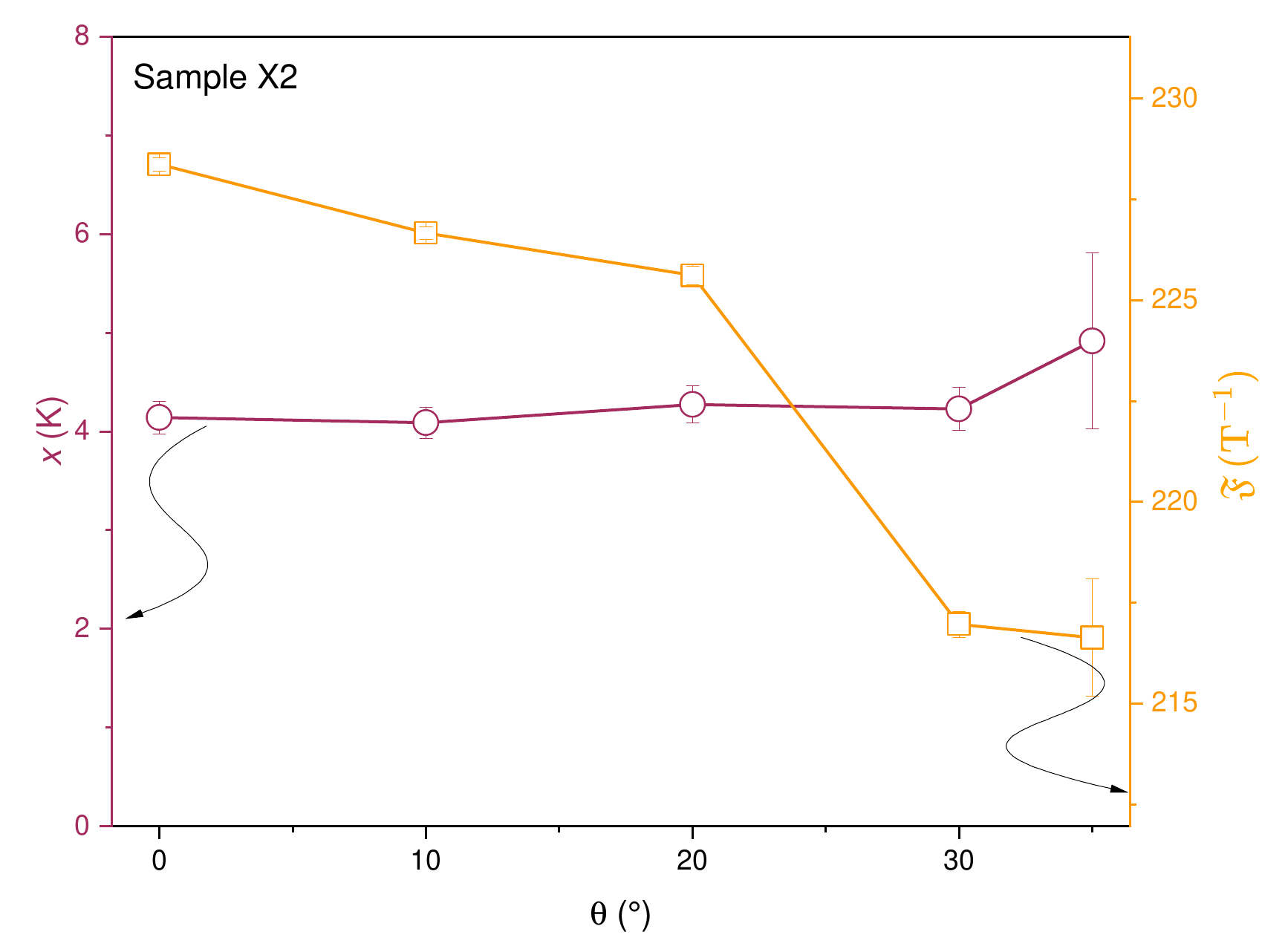}
    \caption{Dingle temperature $x$ (circles) and oscillation frequency $\freq$ (squares) of sample X2 as a function of magnetic field angle $\theta$.}
    \label{fig_X2_SdH_Hall_over_theta_parameters}
\end{figure}
\begin{figure}[htb]
    \centering
    \includegraphics[height=6cm]{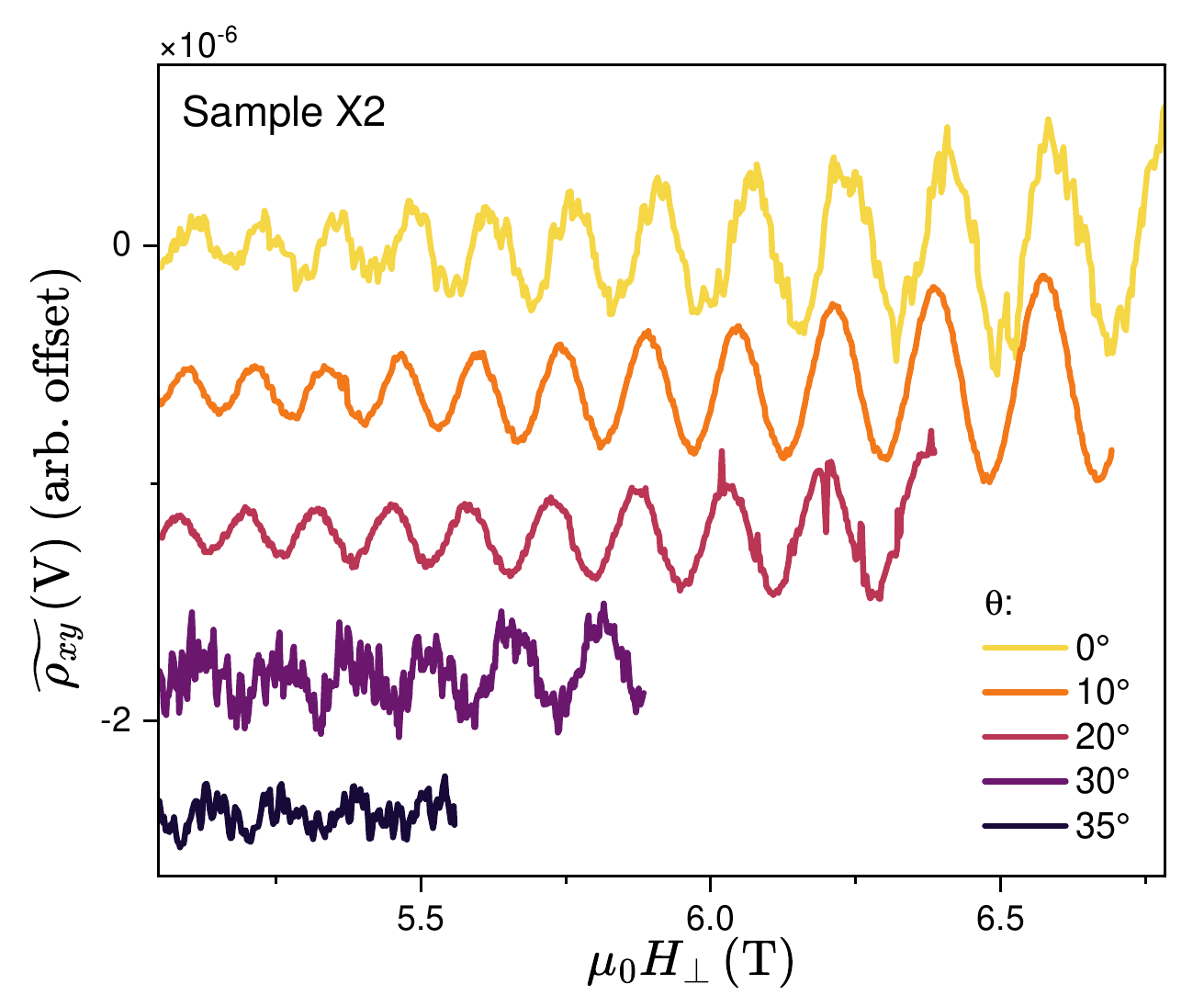}
    \caption{Oscillations in the Hall voltage as a function of the normal component of the applied magnetic field for sample X2 at $2\,\text{K}$ at specific magnetic field angles from the surface normal $\theta$.}
    \label{fig_X2_SdH_Hall_over_theta}
\end{figure}
\FloatBarrier
Samples F and X1 show \g{wal} in the \new{2-terminal} resistance measurement: \fig \ref{fig_wal_2terminal} gives the \new{2-terminal} conductance as a function of applied magnetic field, fitted by \eq 8 from the main text. In this two-terminal configuration, the measured conductance is influenced by both the \pt film and the (metallic) Pt contacts. The \g{wal} shape is broader in the magnetic field direction than the \g{wal} peak observed in the \new{4-terminal} data. Sample F, exfoliated from bulk crystal Batch 1, also shows the \new{2-terminal} \g{wal} and the data can be fitted with comparable length scales to sample X1 (Batch 2). The data are comparable to those of metallic thin films \cite{PtMetalWAL}.
\begin{figure}[htb]
    \centering
    \includegraphics[height=6cm]{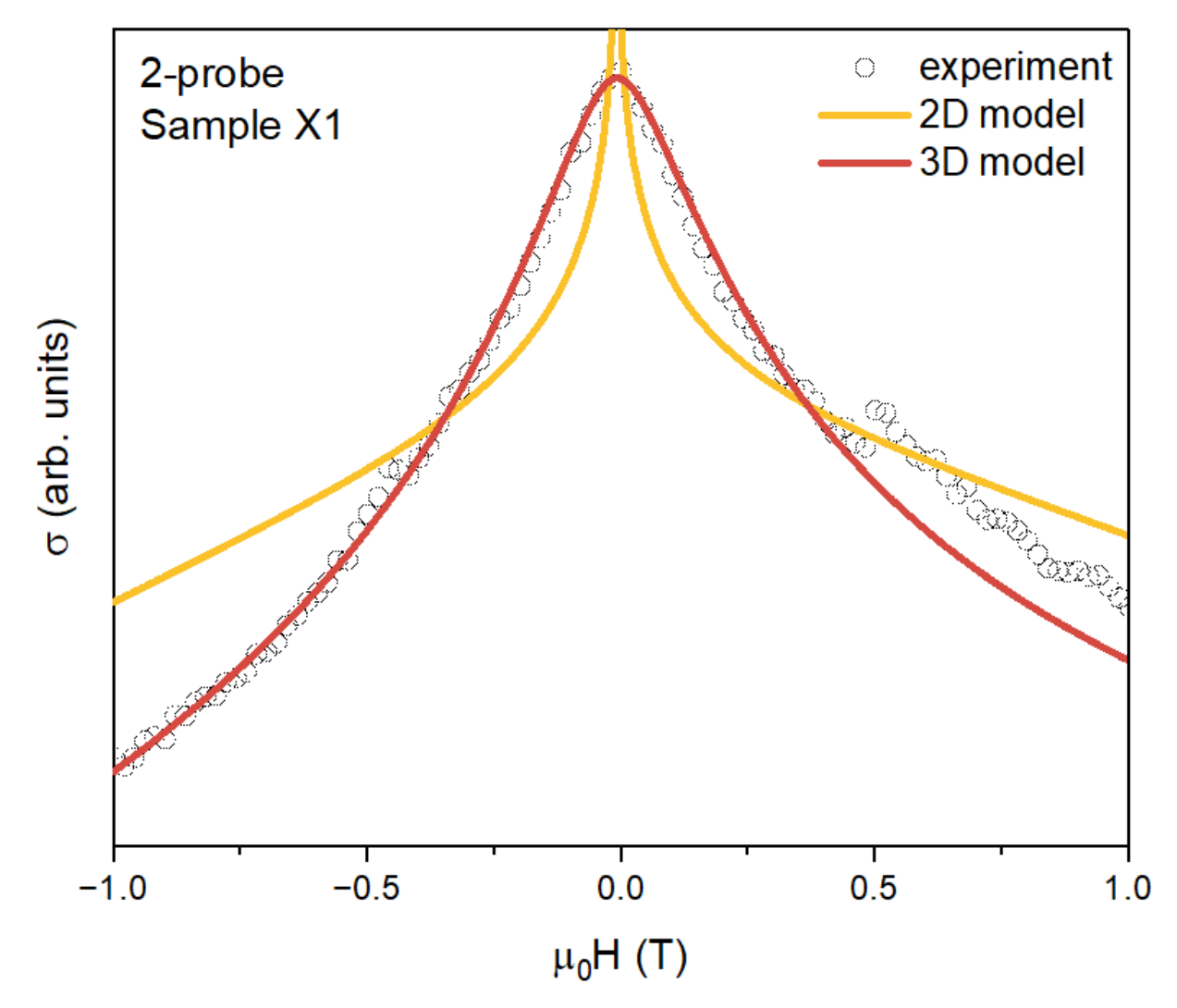}
    \includegraphics[height=6cm]{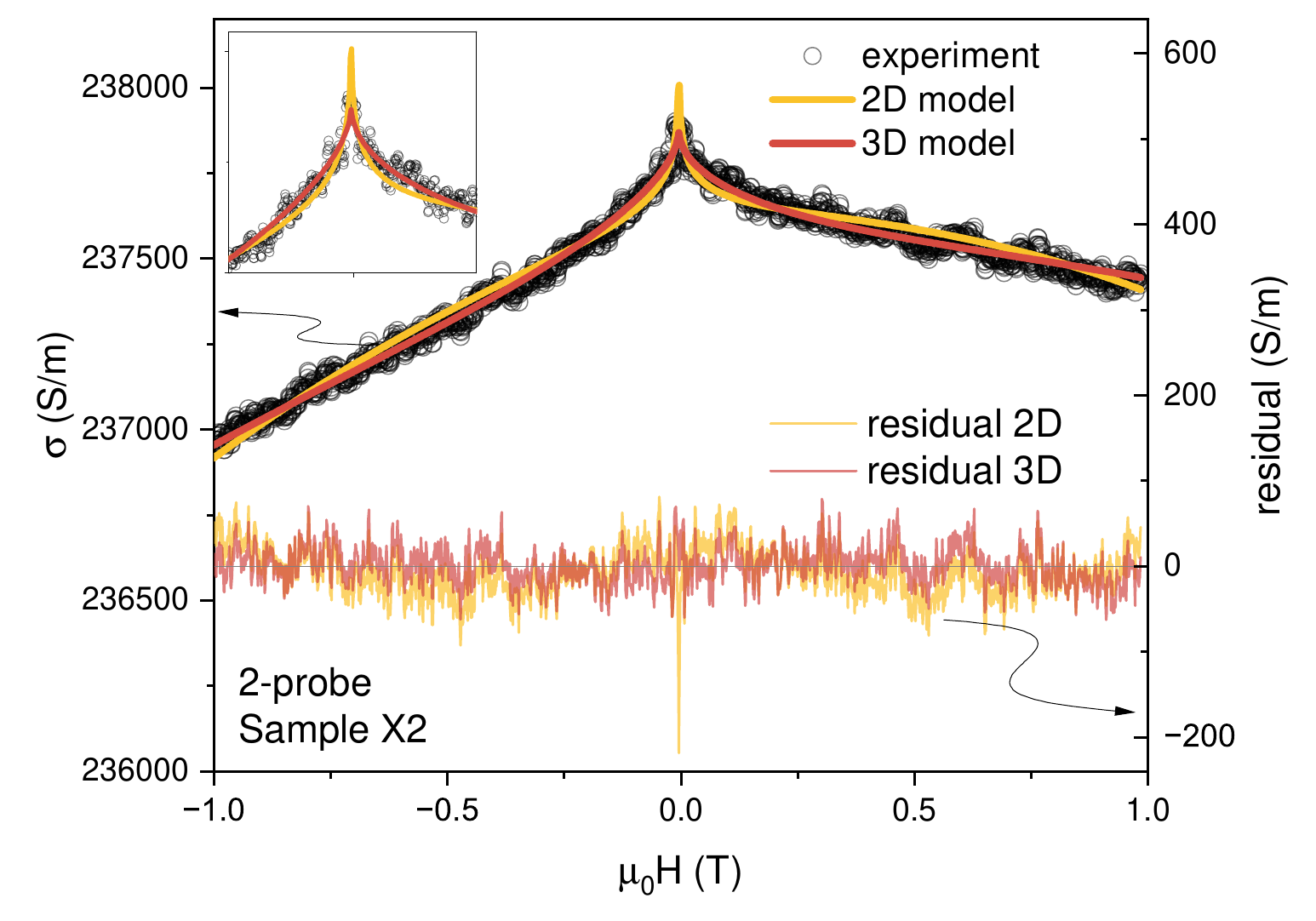}
    \caption{\g{wal} in the \new{2-terminal} conductivity over applied magnetic field. Left panel: sample X1 at $1.7\,\text{K}$. Right panel: \new{2-terminal} conductivity (left axis) and residuals (right axis) over applied magnetic field of Sample X2 at $2\,\text{K}$. The respective obtained parameters are provided in \tab \ref{tab_WAL_2terminal}.}
    \label{fig_wal_2terminal}
\end{figure}
\FloatBarrier
The \new{4-terminal} conductance as a function of applied magnetic field for specific temperatures $T \in(2,15)\,\text{K}$ is reported in \fig \ref{fig_X1_WAL_fit}. The employed model adequately follows the conductance peaks near $H = 0$. For $T \geq 3.5\,\text{K}$, the peak flattens out and it remains resolvable for $T\lesssim 10\,\text{K}$. Since the parameter $A_\text{WAL}$ contains no temperature dependence, the same value is used to fit all temperatures. Due to the analytic nature of the model, the parameters $l_i$ are not fully independent, and no significant values were found. It can be estimated that the length scales are on the order of $\sim 500\,\text{nm}$.
\begin{figure}[htb]
    \centering
    \includegraphics[height=6.5cm]{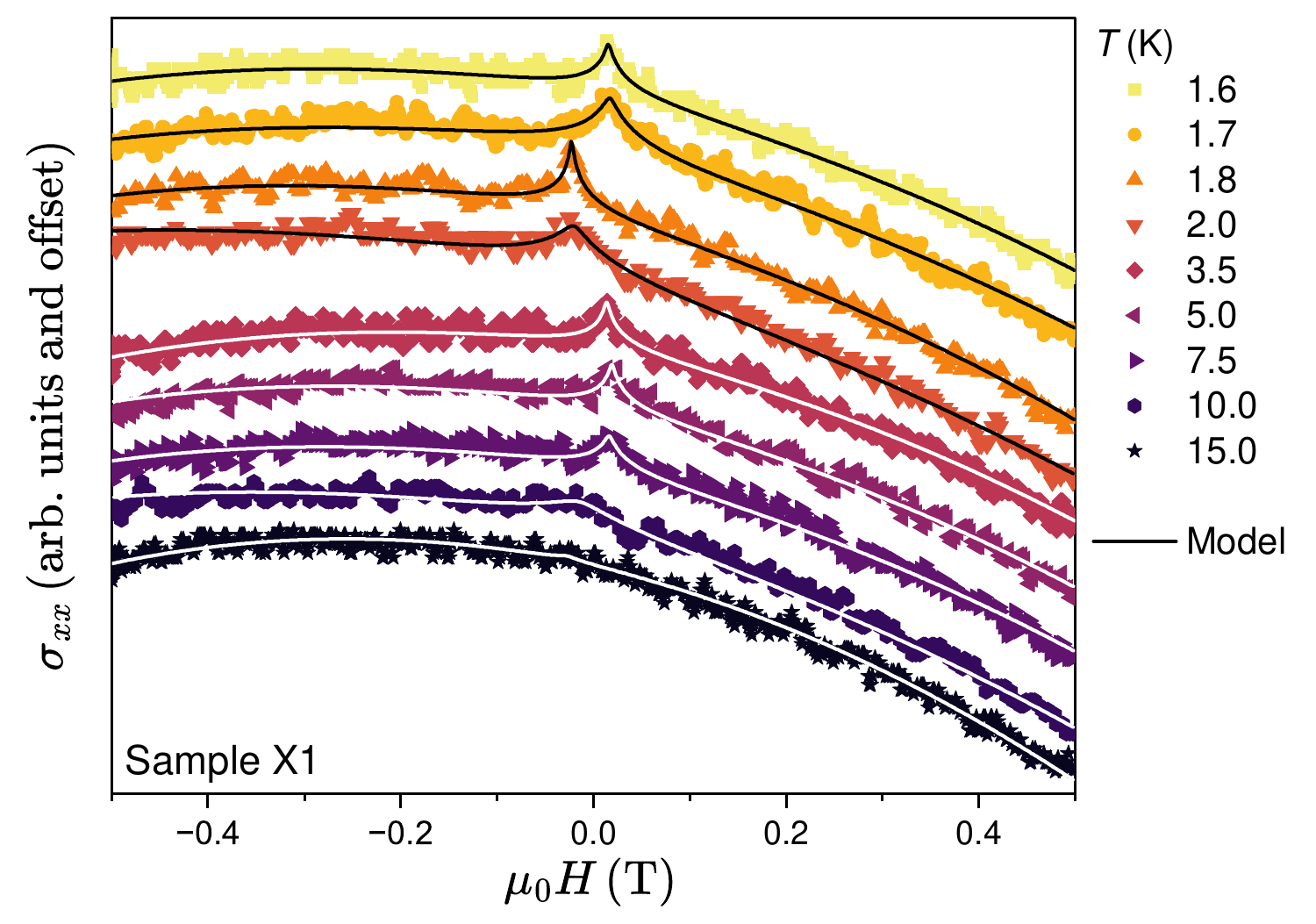}
    \caption{\new{4-terminal} conductance of sample X1 as a function of applied magnetic field for various temperatures. Solid lines: 3D model. The resulting parameters are given in \tab \ref{tab_X1_wal_Ti}.}
    \label{fig_X1_WAL_fit}
\end{figure}
\FloatBarrier
\clearpage
\subsection{Supplemental tables}

\begin{table}[H]
\centering
\begin{tabular}{l|S[table-format=3.2(2), table-number-alignment = center]}
$\freq$ (\corr{$\frac{n_\text{LL}}{\text{T}}$}{$\frac{n_\text{LL}}{1/\text{T}}$})& 193.6(0.3)\\ \hline
$\phi$& 0.09(0.04)\\
\end{tabular}
\caption{Parameters resulting from the linear fit of the Landau fan diagram of $\widetilde{\sigma_{xx}}$ of sample F at $2\,\text{K}$. $\freq$ is given in (\corr{$\frac{n_\text{LL}}{\text{T}}$}{$\frac{n_\text{LL}}{1/\text{T}}$}) in order to highlight the relation to $n_\text{LL}$. (\corr{$\frac{n_\text{LL}}{\text{T}}$}{$\frac{n_\text{LL}}{1/\text{T}}$} is of equal dimension as T).}
\label{tab_LandauFan}
\end{table}

\begin{table}[H]
\centering
\begin{tabular}{
    l|
    S[table-format=1.3(1)e-3, table-number-alignment = center] 
    S[table-format=1.1(1)e-3, table-number-alignment = center]
    S[table-format=1.1(1)e-3, table-number-alignment = center]|
    S[table-format=1.1(3)e-3, table-number-alignment = center]
}
sample & {$l_0$ (m)} & {$l_\phi^\text{3D}$ (m)} & {$l_\text{SO}$ (m)} & {$l_\phi^\text{2D}$ (m)} \\ \hline
F & 6.0(4)e-8   & 5.8(4)e-8  & 2.1(3)e-7  & 1.7(3)e-8 \\
X1 & 5.9(2)e-8   & 5.6(2)e-8  & 2.1(2)e-7  & 4(300)e-9   \\
X2 & 4.044(2)e-4 & 5(1)e-8    & 8.2(6)e-8  & 2.3(6)e-8 \\
\end{tabular}
\caption{\corr{Table of length scales of the local \g{wal}.}{Table of length scales obtained from fitting the 2-terminal conductance of sample F at $2\,\text{K}$ with \eq 9 from the main text. The resulting curves are presented in \fig 12 in the main text.}}
\label{tab_WAL_2terminal}
\end{table}

\begin{table}[H]
\centering
\begin{tabular}{
    l|
    S[table-format=1.1(1)e-3] 
    S[table-format=1.1(1)e-3]
    S[table-format=1.0(2)e-3]|
    S[table-format=1.0(3)e-3]
}
sample & {$l_0$ (m)} & {$l_\phi^\text{3D}$ (m)} & {$l_\text{SO}$ (m)} & {$l_\phi^\text{2D}$ (m)} \\ \hline
X1 & 3.5(1.6)e-7& 3.5(1.2)e-7& 4(15)e-6& 2(500)e-8\\
\end{tabular}
\caption{\corr{Table of length scales of the nonlocal \g{wal}.}{Table of length scales obtained from fitting the 4-terminal conductance $\sigma-\sigma^\text{(lin.)}$ of sample X1 at $1.7\,\text{K}$ with \eq 9 from the main text. The resulting curves are presented in \fig 13 in the main text. The linear conductance contribution due to Hall intermix is $\sigma^\text{(lin.)}=-28.5\,\frac{1/\text{V}}{\text{T}}$ and the zero-field conductance is $\sigma_0 = 14468\, \frac{1}{\text{V}}$}}
\label{tab_WAL}
\end{table}

\begin{table}[H]
\centering
\begin{tabular}{
    S[table-format=2.1]|
    S[table-format=2.2(2)e-3]
    S[table-format=1.2(4)e-3]
    S[table-format=5.1(2)]
    c
    c
}
{$T$ (K)}   & {$A_\text{MR}$ (V/T$^2$)} & {$k$ (V/T)}   &{$\sigma_0$ (1/V)} & {$A_\text{WAL}^{(*)}$} & {$s^{[\dagger]}$}     \\ \hline
1.6         &2.38(0.11)e-7               &1.59(3)e-7     &14449.4(0.4)& \multirow{9}{*}{\rotatebox[origin=c]{-90}{$(9\pm6)\E{-7}$}}   &\multirow{9}{*}{2}\\
1.7         &2.27(0.16)e-7               &1.59(4)e-7     &14466.2(0.6)&    &                                                  \\
1.8         &2.57(0.20)e-7               &1.67(5)e-7     &14466.4(0.6)&    &                                                  \\
2.0         &2.01(0.24)e-7               &1.87(5)e-7     &14416  (1.1)&    &                                                  \\
3.5         &2.69(0.10)e-7               &1.39(3)e-7     &14442.4(0.3)&    &                                                  \\
5.0         &2.50(0.20)e-7               &1.58(6)e-7     &14479.2(0.6)&    &                                                  \\
7.5         &2.50(0.14)e-7               &1.61(4)e-7     &14451.8(0.4)&    &                                                  \\
10.0        &  2.3(0.4)e-7               &1.78(6)e-7     &14437  (2  )&    &                                                  \\
15.0        &3.41(0.15)e-7               &1.59(5)e-7     &14295.9(0.6)&    &                                                  
\end{tabular}
\begin{tabular}{
    S[table-format=2.1]|
    S[table-format=2.2(5)e-3]
    S[table-format=1.2(4)e-3]
    S[table-format=5.1(3)e-3]
    S[table-format=2.3(2)e-3]
}
{$T$ (K)} & {$l_0$ (m)}& {$l_\phi^\text{3D}$ (m)} & {$l_\text{SO}$ (m)} & {$H_\text{lag}$ (T)}  \\ \hline
1.6  &4(4)e-7       &   6(9)e-7&    5.3(1.8)e-7&  -1.57(0.023)e-2\\
1.7  &3(4)e-7       &	4(8)e-7&	4(4)e-7&	-1.71(0.03)e-2\\
1.8  &10(210)e-7    &	9(400)e-7&	6(20)e-7&	 2.26(0.026)e-2\\
2.0  &2.4(2.4)e-7   &	2(5)e-7&	4(10)e-7&	 2.04(0.08)e-2\\
3.5  &5(8)e-7       &	7(18)e-7&	6(3)e-7&	-1.40(0.021)e-2\\
5.0  &5(8)e-7       &	7(21)e-7&	6(4)e-7&	-1.94(0.03)e-2\\
7.5  &4(4)e-7       &	4(8)e-7&	6(6)e-7&	-1.64(0.03)e-2\\
10.0 &1.45(2.8)e-7  &	1.4(0.8)e-7&	4(4)e-7&	 1.44(0.019)e-2\\
15.0 &1.8(0.8)e-7   &	2.6(0.9)e-7&	6(4)e-7&	 2.73(0.04)e-2\\
\end{tabular}
\caption{Parameters obtained from fitting the \new{4-terminal} conductivity of sample X1 over applied magnetic field with \eq 8 from the main text at various temperatures. \new{The corresponding fit curves are presented in \fig \ref{fig_X1_WAL_fit}.} $^{(*)}$: value is constrained to be equivalent at all temperatures. $^{[\dagger]}$: The \g{omr} power is fixed to $2$ since the range of $|\mu_0 H| \leq 1\,\text{T}$ is insufficient to resolve the curvature.}
\label{tab_X1_wal_Ti}
\end{table}

\begin{table}[H]
\centering
\begin{tabular}{l
    S[table-format=2.2]|
    S[table-format=6.2]
    S[table-format=1.2]
    S[table-format=6.2]
    S[table-format=1.2]
    S[table-format=2.2]
}
{Name} & {Peak BE (eV)}& {Height (CPS)}& {Height Ratio} & {Area (CPS$\times$eV)}& {FWHM fit param (eV)} & {L/G Mix (\%)}\\ \hline
{Pt4\textit{f}$_{7/2}$} & 73.15              & 563285.91  & 1            & 929682.81   & 1.2                 & 67.58           \\
{Pt4\textit{f}$_{5/2}$} & 76.48              & 431326.87  & 0.77         & 708657.71   & 1.2                 & 67.58           \\ \hline
{Se3\textit{d}$_{5/2}$} & 55.16              & 144037.56  & 1            & 234490.27   & 1.14                & 87.25           \\
{Se3\textit{d}$_{3/2}$} & 55.97              & 108205.12  & 0.75         & 176400.01   & 1.14                & 87.25           \\
\end{tabular}
\caption{\corr{Peak fit table for the Pt4\textit{f} range and the Se3\textit{d} range.}{Table of parameters from fitting the Pt4\textit{f} range and the Se3\textit{d} range of the \g{xps} data discussed in \secText E and displayed in \fig \ref{fig_XPS}.}}
\label{tab_XPS}
\end{table}

\begin{table}[H]
\centering
\begin{tabular}{l|
    S[table-format=3.3(1)e-3]
    S[table-format=3.3(1)e-3]
    S[table-format=3.3(1)e-3]
}
{sample} & {$A_0$ (V)} &  {$A_1$ (V/T)} & {$A_2$ (V/T$^2$)} \\ \hline
F   &  8.982(3)e-6    &-1.265(1)e-5 &5.92(2)e-8\\
X1  & -8.289(7)e-7    &-6.875(3)e-6 &-2.030(3)e-8\\
\multicolumn{4}{c}{}\\[-.5ex]
{sample} & {$A_m$ (V/T$^m$)} & {$m$}&\\ \hline
F   &-2.23(7)e-7    &1.91(1)&\\
X1  &-2.08(2)e-7    &1.727(3)&\\
\end{tabular}
\caption{Parameters obtained from fitting the Hall voltage over applied magnetic field of samples F and X1 with \eq \ref{eq_Hall}\new{, as plotted in \fig \ref{fig_Hall_supp}}.}
\label{tab_Hall}
\end{table}

\FloatBarrier
\clearpage

\subsection{Potential sources of confusion}
\renewcommand{\descriptionlabel}[1]{\hspace\labelsep \normalfont #1\quad}
\begin{description}
    \item[$a_{xx}$ or $a^{(xx)}$] longitudinal component or parameter following from evaluation of longitudinal components
    \item[$a_{xy}$ or $a^{(xy)}$] transversal (Hall) component or parameter following from evaluation of transversal (Hall) components
    \item[$\widetilde{a}$] oscillatory part of $a$
    \item[$A$] positive real constant, which may change after every step or sentence
    \item[$f$] constant, equivalent to $\frac{\hbar}{4e}$
    \item[$F$] sample F
    \item[$\boldsymbol{H}$] vector-valued applied magnetic field, always given in Tesla
    \item[$H$] equivalent to $\boldsymbol{H}\cdot\hat{e}_H$
    \item[$I$] Drain-source current (scalar-valued)
    \item[$\boldsymbol{j}$] current density direction (vector-valued)
    \item[$\mathfrak{F}$] \g{sdh} oscillation frequency
    \item[$\mathcal{F}$] Fourier transform
    \item[$\text{FFT}$] fast Fourier transform
    \item[$m_0$] free electron rest mass
    \item[$m_c$] fermion cyclotron mass, related to the extremal Fermi surface cross section
    \item[$m^*$] effective fermion mass, related to the band curvature
    \item[$m_b$] bare electron mass
    \item[$m$] anomalous Hall power
    \item[$M=\frac{m_c}{m_0}$] fermion cyclotron mass ratio
    \item[$\sim$] on the same order of magnitude as
    \item[$\gtrsim$] larger or on the same order of magnitude as
    \item[$\approx$] approximately
    \item[$\cong$] cognate with  
    \item[$\propto$] "directly proportional to" or "uniquely identifies a related parameter"
    \item[$\insertArrow$] previous results are inserted or expanded
    \item[$\longrightsquigarrow{ref.}\,a$] performing steps similar to ref. yields $a$
    \item[$\phi$] \g{sdh} oscillation phase
    \item[$\Phi_B$] Berry phase
    \item[$\Phi_0$] Magnetic flux quantum
    \item[$\theta$ and $\psi$] angles of $\boldsymbol{H}$ given in degrees. The substrate is mounted onto the \gls{sh}, such that, in the $\theta = 0$ and $\psi=0$ orientation, the current density direction coincides with $\text{axis}_\psi$. An additional rotation along $\text{axis}_\theta$ is available, with $\text{axis}_\theta$ being normal to the $\text{axis}_\psi$ and rotating with the \gls{sh}. The resulting angle $\theta$ is measured between the out-of-plane axis of the \gls{sh} and the plane spanned by $(\boldsymbol{H}\wedge\text{axis}_\theta)$.
    \item[$\zeta(.,.)$] The Hurwitz Zeta function $\zeta(s,a) = \sum_{n=0}^\infty (n+a)^{-s}$, where, in this work, $s=1/2$ and $a \in \mathbb{R}$ and $a > 1/2$. The sum is divergent if evaluated numerically and is thus evaluated \textit{via} analytic continuation.
    \item[\new{2-terminal}] measuring the voltage difference over the leads, Pt contacts and the \pt flake
    \item[\new{4-terminal}] measuring only the voltage difference over the \pt flake
\end{description}

\end{document}